\documentclass[12pt,a4paper]{article}

\DeclareMathSizes{12}{12}{7}{5}
\usepackage{amsfonts}
\usepackage[dvips]{graphicx}
\usepackage{latexsym,amsmath,amssymb,graphics,stmaryrd}
\usepackage{cite}
\usepackage{array}
\usepackage{subfigure}
\usepackage{multirow}
\usepackage{epsfig}
\usepackage[dvips]{graphicx}
\usepackage[dvips]{color}
\usepackage{pstricks}
\usepackage{pst-node}
\usepackage{pst-plot}
\usepackage{dsfont}
\usepackage{amsthm}
\usepackage{graphics}
\usepackage{subfigure}
\usepackage{fancyhdr}

\usepackage{rotating}
\newlength{\arlength}

\usepackage{feynmp}

\usepackage{Dalgebra}

\usepackage{feynmp}

\fancypagestyle{plain}{
\fancyhf{}
\newcommand{\fpage}{\iffloatpage{}{\thepage}}
\fancyfoot[C]{\fpage}

}

\newcommand{\col}{~,}
\newcommand{\pnt}{~.}
\newcommand{\AdS}{\text{AdS}}
\newcommand{\CFT}{\text{CFT}}

\newcommand{\twob}{{\text{II}\,\text{B}}}

\newcommand{\FP}{\mathrm{FP}}

\newcommand{\YM}{\text{YM}}

\newcommand{\unitmatrix}{\mathds{1}}

\newcommand{\comm}[2]{\left[#1\smash[b]{\mathbin{,}}#2\right]}
\newcommand{\acomm}[2]{\left\{#1\smash[b]{\mathbin{,}}#2\right\}}

\newcommand{\de}{\operatorname{d}\!}

\newcommand{\e}{\operatorname{e}}

\newlength{\neglength}
\newlength{\diameter}

\newcommand{\svertex}[3][0.5]{%
\fmfiequ{#2}{point #1*length(#3) of #3}
}
\newcommand{\dvertex}[3]{%
\fmfiequ{#1}{point length(#3)/3 of #3}
\fmfiequ{#2}{point 2length(#3)/3 of #3}
}
\newcommand{\vvertex}[3]{%
\fmfipath{px}
\fmfiequ{px}{(0,ypart(#2))..(100,ypart(#2))}
\fmfiequ{#1}{point xpart(#3 intersectiontimes px) of #3}
}

\newcommand{\chione}[1][black]{%
\fmftop{v1}
\fmfbottom{v3}
\fmfforce{(0.125w,h)}{v1}
\fmfforce{(0.125w,0)}{v3}
\fmffixed{(0.25w,0)}{v1,v2}
\fmffixed{(0.25w,0)}{v3,v4}
\fmf{plain,tension=0.5,right=0.25,fore=#1}{v1,vc1}
\fmf{plain,tension=0.5,left=0.25,fore=#1}{v2,vc1}
\fmf{plain,tension=1.25,fore=#1}{vc1,vc2}
\fmf{plain,tension=0.5,left=0.125,fore=#1}{vc3,vc2}
\fmf{plain,tension=0.5,left=0.25,fore=#1}{v3,vc2}
\fmf{plain,tension=0.5,right=0.25,fore=#1}{v4,vc2}
\fmf{plain,tension=0.5,right=0,width=1mm,fore=#1}{v3,v4}
\fmfposition
\fmfipath{p[]}
\fmfipair{vd[],vm[],vu[]}
\fmfiset{p1}{vpath(__v1,__vc1)}
\fmfiset{p2}{vpath(__v2,__vc1)}
\fmfiset{p3}{vpath(__vc1,__vc2)}
\fmfiset{p4}{vpath(__v3,__vc2)}
\fmfiset{p5}{vpath(__v4,__vc2)}
\svertex{vm1}{p1}
\dvertex{vu1}{vd1}{p1}
\svertex{vm2}{p2}
\dvertex{vu2}{vd2}{p2}
\svertex{vm3}{p3}
\dvertex{vu3}{vd3}{p3}
\svertex{vm4}{p4}
\dvertex{vd4}{vu4}{p4}
\svertex{vm5}{p5}
\dvertex{vd5}{vu5}{p5}
}

\newcommand{\chioneg}[1][black]{%
\fmftop{v1}
\fmfbottom{v4}
\fmfforce{(0.125w,h)}{v1}
\fmfforce{(0.125w,0)}{v4}
\fmffixed{(0.25w,0)}{v1,v2}
\fmffixed{(0.25w,0)}{v2,v3}
\fmffixed{(0.25w,0)}{v4,v5}
\fmffixed{(0.25w,0)}{v5,v6}
\fmf{plain,tension=0.5,right=0.25,fore=#1}{v1,vc1}
\fmf{plain,tension=0.5,left=0.25,fore=#1}{v2,vc1}
  \fmf{plain,tension=1.25,fore=#1}{vc1,vc2}
\fmf{plain,tension=0.5,left=0.25,fore=#1}{v4,vc2}
\fmf{plain,tension=0.5,right=0.25,fore=#1}{v5,vc2}
\fmf{plain,fore=#1}{v3,v6}
\fmf{plain,tension=0.5,right=0,width=1mm,fore=#1}{v4,v6}
\fmfposition
\fmfipath{p[],pg}
\fmfipair{vd[],vm[],vu[],vgd[],vgm[],vgu[],vg[]}
\fmfiset{p1}{vpath(__v1,__vc1)}
\fmfiset{p2}{vpath(__v2,__vc1)}
\fmfiset{p3}{vpath(__vc1,__vc2)}
\fmfiset{p5}{vpath(__v5,__vc2)}
\fmfiset{p4}{vpath(__v4,__vc2)}
\fmfiset{pg}{vpath(__v3,__v6)}
\svertex{vm1}{p1}
\dvertex{vu1}{vd1}{p1}
\svertex{vm2}{p2}
\dvertex{vu2}{vd2}{p2}
\svertex{vm3}{p3}
\dvertex{vu3}{vd3}{p3}
\svertex{vm4}{p4}
\dvertex{vd4}{vu4}{p4}
\svertex{vm5}{p5}
\dvertex{vd5}{vu5}{p5}
\vvertex{vgu2}{vu2}{pg}
\vvertex{vgm2}{vm2}{pg}
\vvertex{vgd2}{vd2}{pg}
\vvertex{vgu3}{vu3}{pg}
\vvertex{vgm3}{vm3}{pg}
\vvertex{vgd3}{vd3}{pg}
\vvertex{vgu5}{vu5}{pg}
\vvertex{vgm5}{vm5}{pg}
\vvertex{vgd5}{vd5}{pg}
\vvertex{vg1}{vloc(__vc1)}{pg}
\vvertex{vg2}{vloc(__vc2)}{pg}
}

\newcommand{\gchioneg}[1][black]{%
\fmftop{v1}
\fmfbottom{v5}
\fmfforce{(0.125w,h)}{v1}
\fmfforce{(0.125w,0)}{v5}
\fmffixed{(0.25w,0)}{v1,v2}
\fmffixed{(0.25w,0)}{v2,v3}
\fmffixed{(0.25w,0)}{v3,v4}
\fmffixed{(0.25w,0)}{v5,v6}
\fmffixed{(0.25w,0)}{v6,v7}
\fmffixed{(0.25w,0)}{v7,v8}
\fmf{plain,fore=#1}{v1,v5}
\fmf{plain,tension=0.5,right=0.25,fore=#1}{v2,vc1}
\fmf{plain,tension=0.5,left=0.25,fore=#1}{v3,vc1}
  \fmf{plain,tension=1.25,fore=#1}{vc1,vc2}
\fmf{plain,tension=0.5,left=0.25,fore=#1}{v6,vc2}
\fmf{plain,tension=0.5,right=0.25,fore=#1}{v7,vc2}
\fmf{plain,fore=#1}{v4,v8}
\fmf{plain,tension=0.5,right=0,width=1mm,fore=#1}{v5,v8}
\fmfposition
\fmfipath{p[],pgl,pgr}
\fmfipair{vd[],vm[],vu[],vgd[],vglm[],vglu[],vgld[],vgl[],vgrm[],vgru[],vgrd[],vgr[]}
\fmfiset{p1}{vpath(__v2,__vc1)}
\fmfiset{p2}{vpath(__v3,__vc1)}
\fmfiset{p3}{vpath(__vc1,__vc2)}
\fmfiset{p5}{vpath(__v7,__vc2)}
\fmfiset{p4}{vpath(__v6,__vc2)}
\fmfiset{pgl}{vpath(__v1,__v5)}
\fmfiset{pgr}{vpath(__v4,__v8)}
\svertex{vm1}{p1}
\dvertex{vu1}{vd1}{p1}
\svertex{vm2}{p2}
\dvertex{vu2}{vd2}{p2}
\svertex{vm3}{p3}
\dvertex{vu3}{vd3}{p3}
\svertex{vm4}{p4}
\dvertex{vd4}{vu4}{p4}
\svertex{vm5}{p5}
\dvertex{vd5}{vu5}{p5}
\vvertex{vglu1}{vu1}{pgl}
\vvertex{vglm1}{vm1}{pgl}
\vvertex{vgld1}{vd1}{pgl}
\vvertex{vglu3}{vu3}{pgl}
\vvertex{vglm3}{vm3}{pgl}
\vvertex{vgld3}{vd3}{pgl}
\vvertex{vglu5}{vu5}{pgl}
\vvertex{vglm5}{vm5}{pgl}
\vvertex{vgld5}{vd5}{pgl}
\vvertex{vgl1}{vloc(__vc1)}{pgl}
\vvertex{vgl2}{vloc(__vc2)}{pgl}
\vvertex{vgru2}{vu2}{pgr}
\vvertex{vgrm2}{vm2}{pgr}
\vvertex{vgrd2}{vd2}{pgr}
\vvertex{vgru3}{vu3}{pgr}
\vvertex{vgrm3}{vm3}{pgr}
\vvertex{vgrd3}{vd3}{pgr}
\vvertex{vgru5}{vu5}{pgr}
\vvertex{vgrm5}{vm5}{pgr}
\vvertex{vgrd5}{vd5}{pgr}
\vvertex{vgr1}{vloc(__vc1)}{pgr}
\vvertex{vgr2}{vloc(__vc2)}{pgr}
}

\newcommand{\chionetwo}[1][black]{%
\fmftop{v1}
\fmfbottom{v4}
\fmfforce{(0.125w,h)}{v1}
\fmfforce{(0.125w,0)}{v4}
\fmffixed{(0.25w,0)}{v1,v2}
\fmffixed{(0.25w,0)}{v2,v3}
\fmffixed{(0.25w,0)}{v4,v5}
\fmffixed{(0.25w,0)}{v5,v6}
\fmffixed{(0,whatever)}{vc1,vc3}
\fmffixed{(0,whatever)}{vc2,vc4}
\fmf{plain,tension=0.5,right=0.25}{v1,vc1}
\fmf{plain,tension=0.5,left=0.25}{v2,vc1}
\fmf{phantom,tension=0.5,right=0.25}{v2,vc2}
\fmf{plain,tension=0.5,left=0.25}{v3,vc2}
\fmf{plain,tension=0.5,left=0.25}{v4,vc3}
\fmf{phantom,tension=0.5,right=0.25}{v5,vc3}
\fmf{plain,tension=0.5,left=0.25}{v5,vc4}
\fmf{plain,tension=0.5,right=0.25}{v6,vc4}
\fmf{plain,tension=1.25,left=0}{vc1,vc3}
\fmf{plain,tension=1.25,left=0}{vc2,vc4}
\fmffreeze
\fmf{plain,tension=1,left=0}{vc2,vc3}
\fmf{plain,tension=0.5,right=0,width=1mm}{v4,v6}
\fmffreeze
\fmfposition
\fmfipath{p[]}
\fmfipair{vd[],vm[],vu[]}
\fmfiset{p1}{vpath(__v1,__vc1)}
\fmfiset{p2}{vpath(__v2,__vc1)}
\fmfiset{p6}{vpath(__v3,__vc2)}
\fmfiset{p4}{vpath(__v4,__vc3)}
\fmfiset{p8}{vpath(__v5,__vc4)}
\fmfiset{p9}{vpath(__v6,__vc4)}
\fmfiset{p3}{vpath(__vc1,__vc3)}
\fmfiset{p7}{vpath(__vc2,__vc4)}
\fmfiset{p5}{vpath(__vc2,__vc3)}
\svertex{vm1}{p1}
\svertex{vm2}{p2}
\svertex{vm3}{p3}
\svertex{vm4}{p4}
\svertex{vm5}{p5}
\svertex{vm6}{p6}
\svertex{vm7}{p7}
\svertex{vm8}{p8}
\svertex{vm9}{p9}
}

\newcommand{\chionetwoone}[1][black]{%
\fmftop{v1}
\fmfbottom{v4}
\fmfforce{(0.125w,h)}{v1}
\fmfforce{(0.125w,0)}{v4}
\fmffixed{(0.25w,0)}{v1,v2}
\fmffixed{(0.25w,0)}{v2,v3}
\fmffixed{(0.25w,0)}{v4,v5}
\fmffixed{(0.25w,0)}{v5,v6}
\fmffixed{(0,whatever)}{vc1,vc3}
\fmffixed{(0,whatever)}{vb2,vb4}
\fmffixed{(0,whatever)}{vc1,vb1}
\fmffixed{(0,whatever)}{vc1,vb3}
\fmffixed{(whatever,0)}{vb1,vb2}
\fmffixed{(whatever,0)}{vb3,vb4}
\fmf{plain,tension=0.5,right=0.25}{v1,vc1}
\fmf{plain,tension=0.5,left=0.25}{v2,vc1}
\fmf{phantom,tension=0.5,right=0.25}{v2,vb2}
\fmf{plain,tension=0.5,left=0.25}{v3,vb2}
\fmf{plain,tension=0.5,left=0.25}{v4,vc3}
\fmf{plain,tension=0.5,right=0.25}{v5,vc3}
\fmf{phantom,tension=0.5,left=0.25}{v5,vb4}
\fmf{plain,tension=0.5,right=0.25}{v6,vb4}
\fmf{plain,tension=1.25,left=0}{vc1,vb1}
\fmf{plain,tension=1.25,left=0}{vb1,vb3}
\fmf{plain,tension=1.25,left=0}{vb3,vc3}
\fmf{plain,tension=1.25,left=0}{vb2,vb4}
\fmffreeze
\fmf{plain,tension=1,left=0}{vb1,vb2}
\fmf{plain,tension=1,left=0}{vb3,vb4}
\fmf{plain,tension=0.5,right=0,width=1mm}{v4,v6}
\fmffreeze
\fmfposition
}

\newcommand{\chionethree}[1][black]{%
\fmftop{v1}
\fmfbottom{v5}
\fmfforce{(0.125w,h)}{v1}
\fmfforce{(0.125w,0)}{v5}
\fmffixed{(0.25w,0)}{v1,v2}
\fmffixed{(0.25w,0)}{v2,v3}
\fmffixed{(0.25w,0)}{v3,v4}
\fmffixed{(0.25w,0)}{v5,v6}
\fmffixed{(0.25w,0)}{v6,v7}
\fmffixed{(0.25w,0)}{v7,v8}
\fmf{plain,tension=0.5,right=0.25,fore=#1}{v1,vc1}
\fmf{plain,tension=0.5,left=0.25,fore=#1}{v2,vc1}
  \fmf{plain,tension=1.25,fore=#1}{vc1,vc2}
\fmf{plain,tension=0.5,left=0.25,fore=#1}{v5,vc2}
\fmf{plain,tension=0.5,right=0.25,fore=#1}{v6,vc2}
\fmf{plain,tension=0.5,right=0.25,fore=#1}{v3,vc3}
\fmf{plain,tension=0.5,left=0.25,fore=#1}{v4,vc3}
  \fmf{plain,tension=1.25,fore=#1}{vc3,vc4}
\fmf{plain,tension=0.5,left=0.25,fore=#1}{v7,vc4}
\fmf{plain,tension=0.5,right=0.25,fore=#1}{v8,vc4}
\fmf{plain,tension=0.5,right=0,width=1mm,fore=#1}{v5,v8}
\fmfposition
\fmfipath{pl[],pr[]}
\fmfipair{vld[],vlm[],vlu[],vrd[],vrm[],vru[]}
\fmfiset{pl1}{vpath(__v1,__vc1)}
\fmfiset{pl2}{vpath(__v2,__vc1)}
\fmfiset{pl3}{vpath(__vc1,__vc2)}
\fmfiset{pl5}{vpath(__v6,__vc2)}
\fmfiset{pl4}{vpath(__v5,__vc2)}
\fmfiset{pr1}{vpath(__v3,__vc3)}
\fmfiset{pr2}{vpath(__v4,__vc3)}
\fmfiset{pr3}{vpath(__vc3,__vc4)}
\fmfiset{pr5}{vpath(__v8,__vc4)}
\fmfiset{pr4}{vpath(__v7,__vc4)}
\svertex{vlm1}{pl1}
\dvertex{vlu1}{vld1}{pl1}
\svertex{vlm2}{pl2}
\dvertex{vlu2}{vld2}{pl2}
\svertex{vlm3}{pl3}
\dvertex{vlu3}{vld3}{pl3}
\svertex{vlm4}{pl4}
\dvertex{vld4}{vlu4}{pl4}
\svertex{vlm5}{pl5}
\dvertex{vld5}{vlu5}{pl5}
\svertex{vrm1}{pr1}
\dvertex{vru1}{vrd1}{pr1}
\svertex{vrm2}{pr2}
\dvertex{vru2}{vrd2}{pr2}
\svertex{vrm3}{pr3}
\dvertex{vru3}{vrd3}{pr3}
\svertex{vrm4}{pr4}
\dvertex{vrd4}{vru4}{pr4}
\svertex{vrm5}{pr5}
\dvertex{vrd5}{vru5}{pr5}

}

\newcommand{\chionetwothree}[1][black]{%
\fmftop{v1}
\fmfbottom{v5}
\fmfforce{(0.125w,h)}{v1}
\fmfforce{(0.125w,0)}{v5}
\fmffixed{(0.25w,0)}{v1,v2}
\fmffixed{(0.25w,0)}{v2,v3}
\fmffixed{(0.25w,0)}{v3,v4}
\fmffixed{(0.25w,0)}{v5,v6}
\fmffixed{(0.25w,0)}{v6,v7}
\fmffixed{(0.25w,0)}{v7,v8}
\fmffixed{(0,whatever)}{vc1,vc4}
\fmffixed{(0,whatever)}{vc2,vc5}
\fmffixed{(0,whatever)}{vc3,vc6}

\fmf{plain,tension=0.5,right=0.25}{v1,vc1}
\fmf{plain,tension=0.5,left=0.25}{v2,vc1}
\fmf{phantom,tension=0.5,right=0.25}{v2,vc2}
\fmf{plain,tension=0.5,left=0.25}{v3,vc2}
\fmf{phantom,tension=0.5,right=0.25}{v3,vc3}
\fmf{plain,tension=0.5,left=0.25}{v4,vc3}
\fmf{plain,tension=0.5,left=0.25}{v5,vc4}
\fmf{phantom,tension=0.5,right=0.25}{v6,vc4}
\fmf{plain,tension=0.5,left=0.25}{v6,vc5}
\fmf{phantom,tension=0.5,right=0.25}{v7,vc5}
\fmf{plain,tension=0.5,left=0.25}{v7,vc6}
\fmf{plain,tension=0.5,right=0.25}{v8,vc6}
\fmf{plain,tension=1.25,left=0}{vc1,vc4}
\fmf{plain,tension=1.25,left=0}{vc2,vc5}
\fmf{plain,tension=1.25,left=0}{vc3,vc6}
\fmffreeze
\fmf{plain,tension=1,left=0}{vc4,vc2}
\fmf{plain,tension=1,left=0}{vc5,vc3}
\fmf{plain,tension=0.5,right=0,width=1mm}{v5,v8}
\fmffreeze
\fmfposition
}

\newcommand{\chitwoonethree}[1][black]{%
\fmftop{v1}
\fmfbottom{v5}
\fmfforce{(0.125w,h)}{v1}
\fmfforce{(0.125w,0)}{v5}
\fmffixed{(0.25w,0)}{v1,v2}
\fmffixed{(0.25w,0)}{v2,v3}
\fmffixed{(0.25w,0)}{v3,v4}
\fmffixed{(0.25w,0)}{v5,v6}
\fmffixed{(0.25w,0)}{v6,v7}
\fmffixed{(0.25w,0)}{v7,v8}
\fmffixed{(whatever,0.5h)}{v5,vc1}
\fmffixed{(0,whatever)}{vc1,vc4}
\fmffixed{(0,whatever)}{vc2,vc5}
\fmffixed{(0,whatever)}{vc3,vc6}
\fmffixed{(whatever,0)}{vc1,vc3}
\fmffixed{(whatever,0)}{vc3,vc5}

\fmf{plain,tension=0.5,right=0.125}{v1,vc1}
\fmf{phantom,tension=0.5,left=0.25}{v2,vc1}
\fmf{plain,tension=0.5,right=0.25}{v2,vc2}
\fmf{plain,tension=0.5,left=0.25}{v3,vc2}
\fmf{phantom,tension=0.5,right=0.25}{v3,vc3}
\fmf{plain,tension=0.5,left=0.125}{v4,vc3}
\fmf{plain,tension=0.5,left=0.25}{v5,vc4}
\fmf{plain,tension=0.5,right=0.25}{v6,vc4}
\fmf{phantom,tension=0.5,left=0.25}{v6,vc5}
\fmf{phantom,tension=0.5,right=0.25}{v7,vc5}
\fmf{plain,tension=0.5,left=0.25}{v7,vc6}
\fmf{plain,tension=0.5,right=0.25}{v8,vc6}
\fmf{plain,tension=1.25,left=0}{vc1,vc4}
\fmf{plain,tension=1.25,left=0}{vc2,vc5}
\fmf{plain,tension=1.25,left=0}{vc3,vc6}
\fmffreeze
\fmf{plain,tension=1,left=0}{vc1,vc5}
\fmf{plain,tension=1,left=0}{vc5,vc3}
\fmf{plain,tension=0.5,right=0,width=1mm}{v5,v8}
\fmffreeze
\fmfposition
}

\newcommand{\chionethreetwo}[1][black]{%
\fmftop{v1}
\fmfbottom{v5}
\fmfforce{(0.125w,h)}{v1}
\fmfforce{(0.125w,0)}{v5}
\fmffixed{(0.25w,0)}{v1,v2}
\fmffixed{(0.25w,0)}{v2,v3}
\fmffixed{(0.25w,0)}{v3,v4}
\fmffixed{(0.25w,0)}{v5,v6}
\fmffixed{(0.25w,0)}{v6,v7}
\fmffixed{(0.25w,0)}{v7,v8}
\fmffixed{(whatever,0.5h)}{v5,vc2}
\fmffixed{(0,whatever)}{vc1,vc4}
\fmffixed{(0,whatever)}{vc2,vc5}
\fmffixed{(0,whatever)}{vc3,vc6}
\fmffixed{(whatever,0)}{vc2,vc4}
\fmffixed{(whatever,0)}{vc4,vc6}
\fmf{plain,tension=0.5,right=0.25}{v1,vc1}
\fmf{plain,tension=0.5,left=0.25}{v2,vc1}
\fmf{phantom,tension=0.5,right=0.25}{v2,vc2}
\fmf{phantom,tension=0.5,left=0.25}{v3,vc2}
\fmf{plain,tension=0.5,right=0.25}{v3,vc3}
\fmf{plain,tension=0.5,left=0.25}{v4,vc3}
\fmf{plain,tension=0.5,left=0.125}{v5,vc4}
\fmf{phantom,tension=0.5,right=0.25}{v6,vc4}
\fmf{plain,tension=0.5,left=0.25}{v6,vc5}
\fmf{plain,tension=0.5,right=0.25}{v7,vc5}
\fmf{phantom,tension=0.5,left=0.25}{v7,vc6}
\fmf{plain,tension=0.5,right=0.125}{v8,vc6}
\fmf{plain,tension=1.25,left=0}{vc1,vc4}
\fmf{plain,tension=1.25,left=0}{vc2,vc5}
\fmf{plain,tension=1.25,left=0}{vc3,vc6}
\fmffreeze
\fmf{plain,tension=1,left=0}{vc2,vc4}
\fmf{plain,tension=1,left=0}{vc2,vc6}
\fmf{plain,tension=0.5,right=0,width=1mm}{v5,v8}
\fmffreeze
\fmfposition
}

\newcommand{\cvert}[7][]{%
\settoheight{\eqoff}{$\times$}%
\setlength{\eqoff}{0.5\eqoff}%
\addtolength{\eqoff}{-8\unitlength}%
\raisebox{\eqoff}{%
\fmfframe(1,2)(1,2){%
\begin{fmfchar*}(12,12)
\fmfright{v3,v2}
\fmfpoly{phantom}{v1,v3,v2}
\fmf{#2}{v1,vc1}
\fmf{#3}{vc1,v2}
\fmf{#4}{vc1,v3}
\fmffreeze
\fmfposition
\fmfipath{p[]}
\fmfiset{p1}{vpath(__v1,__vc1)}
\fmfiset{p2}{vpath(__vc1,__v2)}
\fmfiset{p3}{vpath(__vc1,__v3)}
{#1}
\fmfis{phantom,ptext.clen=6,ptext.hout=3,ptext.oout=12,ptext.out=#5,ptext.sep=;}{p1}
\fmfis{phantom,ptext.clen=6,ptext.hin=3,ptext.oin=14,ptext.in=#6,ptext.sep=;}{p2}
\fmfis{phantom,ptext.side=right,ptext.clen=6,ptext.hin=-3,ptext.oin=14,ptext.in=#7,ptext.sep=;}{p3}
\end{fmfchar*}}}
}

\newcommand{\qvert}[8]{%
\settoheight{\eqoff}{$\times$}%
\setlength{\eqoff}{0.5\eqoff}%
\addtolength{\eqoff}{-8\unitlength}%
\raisebox{\eqoff}{%
\fmfframe(1,2)(1,2){%
\begin{fmfchar*}(12,12)
\fmfleft{v2,v1}
\fmfright{v3,v4}
\fmfforce{(0,0)}{v1}
\fmfforce{(0,h)}{v2}
\fmfforce{(w,h)}{v3}
\fmfforce{(w,0)}{v4}
\fmf{#1}{v1,vc1}
\fmf{#2}{v2,vc1}
\fmf{#3}{vc1,v3}
\fmf{#4}{vc1,v4}
\fmffreeze
\fmfposition
\fmfipath{p[]}
\fmfiset{p1}{vpath(__v1,__vc1)}
\fmfiset{p2}{vpath(__v2,__vc1)}
\fmfiset{p3}{vpath(__vc1,__v3)}
\fmfiset{p4}{vpath(__vc1,__v4)}
\fmfis{phantom,ptext.clen=6,ptext.hout=3,ptext.oout=12,ptext.out=#6,ptext.sep=;}{p2}
\fmfis{phantom,ptext.clen=6,ptext.hout=3,ptext.oout=12,ptext.out=#5,ptext.sep=;}{p1}
\fmfis{phantom,ptext.side=right,ptext.clen=6,ptext.hin=-3,ptext.oin=12,ptext.in=#8,ptext.sep=;}{p4}
\fmfis{phantom,ptext.clen=6,ptext.hin=3,ptext.oin=12,ptext.in=#7,ptext.sep=;}{p3}
\end{fmfchar*}}}
}

\newcommand{\cVat}[7][]{%
\settoheight{\eqoff}{$\times$}%
\setlength{\eqoff}{0.5\eqoff}%
\addtolength{\eqoff}{-13\unitlength}%
\raisebox{\eqoff}{%
\fmfframe(2,1)(2,1){%
\begin{fmfchar*}(21,24)
\fmftop{v2}
\fmfbottom{v3}
\fmfforce{(w,h)}{v2}
\fmfforce{(w,0)}{v3}
\fmfpoly{phantom}{v1,v3,v2}
\fmf{phantom,tension=2}{v1,vg1}
\fmf{phantom}{v2,vg1}
\fmf{phantom}{v3,vg1}
\fmffreeze
\fmf{phantom,tension=1}{vg2,v2}
\fmf{phantom,tension=1}{vg3,v3}
\fmf{phantom}{vg1,v1}
\fmf{phantom,tension=0.5}{vg1,vg2}
\fmf{phantom,tension=0.5}{vg1,vg3}
\fmffreeze
\fmfposition
\fmf{phantom}{vg2,vg3}
\fmfipath{p[],pca}
\fmfipair{vm[],vo[],vi[]}
\fmfiset{p1}{vpath(__vg1,__v1)}
\fmfiset{p2}{vpath(__vg2,__v2)}
\fmfiset{p3}{vpath(__vg3,__v3)}
\fmfiset{p4}{vpath(__vg2,__vg3)}
\fmfiset{p6}{vpath(__vg3,__vg1)}
\fmfiset{p5}{vpath(__vg1,__vg2)}
{#1}
\fmfis{#2,ptext.clen=7,ptext.hin=3,ptext.hout=3,ptext.oin=6,ptext.oout=6,ptext.sep=;}{reverse(p1)}
\fmfis{#3,ptext.clen=7,ptext.hin=3,ptext.hout=3,ptext.oin=6,ptext.oout=6,ptext.sep=;}{p5}
\fmfis{#4,ptext.clen=7,ptext.hin=-9,ptext.hout=-9,ptext.oin=6,ptext.oout=6,ptext.sep=;}{p6}
\fmfis{#5,ptext.clen=7,ptext.hin=3,ptext.hout=3,ptext.oin=6,ptext.oout=6,ptext.sep=;}{p4}
\fmfis{#6,ptext.clen=7,ptext.hin=3,ptext.hout=3,ptext.oin=6,ptext.oout=6,ptext.sep=;}{p2}
\fmfis{#7,ptext.clen=7,ptext.hin=-9,ptext.hout=-9,ptext.oin=6,ptext.oout=6,ptext.sep=;}{p3}
\end{fmfchar*}}}
}

\newcommand{\cVatg}[6][]{%
\settoheight{\eqoff}{$\times$}%
\setlength{\eqoff}{0.5\eqoff}%
\addtolength{\eqoff}{-13\unitlength}%
\raisebox{\eqoff}{%
\fmfframe(2,1)(2,1){%
\begin{fmfchar*}(21,24)
\fmftop{v3}
\fmfbottom{v2}
\fmfforce{(w,h)}{v3}
\fmfforce{(w,0)}{v2}
\fmfpoly{phantom}{v1,v2,v3}
\fmf{phantom,tension=2}{v1,vg1}
\fmf{phantom}{v2,vg1}
\fmf{phantom}{v3,vg1}
\fmffreeze
\fmf{phantom,tension=1}{vgo2,v2}
\fmf{phantom,tension=1}{vgo3,v3}
\fmf{phantom}{vg1,vgm1}
\fmf{phantom}{vgm1,v1}
\fmf{#3,tension=0.5}{vg1,vgi2}
\fmf{#4,tension=0.5}{vg1,vgi3}
\fmf{#3,tension=1}{vgi2,vgm2}
\fmf{#3,tension=1}{vgm2,vgo2}
\fmf{#4,tension=1}{vgi3,vgm3}
\fmf{#4,tension=1}{vgm3,vgo3}
\fmffreeze
\fmfposition
\fmfipath{p[]}
\fmfipair{vm[],vo[],vi[]}
\fmfiset{p1}{vpath(__vg1,__v1)}
\fmfiset{p2}{vpath(__vgo2,__v2)}
\fmfiset{p3}{vpath(__vgo3,__v3)}
\fmfiset{vm1}{vloc(__vgm1)}
\fmfiset{vi2}{vloc(__vgi2)}
\fmfiset{vm2}{vloc(__vgm2)}
\fmfiset{vo2}{vloc(__vgo2)}
\fmfiset{vi3}{vloc(__vgi3)}
\fmfiset{vm3}{vloc(__vgm3)}
\fmfiset{vo3}{vloc(__vgo3)}
{#1}
\fmfis{#2,ptext.clen=7,ptext.hout=3,ptext.oin=0,ptext.oout=6,ptext.sep=;}{reverse(p1)}
\fmfis{#5,ptext.clen=7,ptext.hin=3,ptext.oin=6,ptext.oout=0,ptext.sep=;}{p2}
\fmfis{#6,ptext.clen=7,ptext.hin=-8,ptext.oin=6,ptext.oout=0,ptext.sep=;}{p3}
\end{fmfchar*}}}
}

\newcommand{\cVatt}[9][]{%
\settoheight{\eqoff}{$\times$}%
\setlength{\eqoff}{0.5\eqoff}%
\addtolength{\eqoff}{-14\unitlength}%
\raisebox{\eqoff}{%
\fmfframe(2,2)(2,2){%
\begin{fmfchar*}(21,24)
\fmftop{v3}
\fmfbottom{v2}
\fmfforce{(w,h)}{v2}
\fmfforce{(w,0)}{v3}
\fmfpoly{phantom}{v1,v3,v2}
\fmf{phantom,tension=2}{v1,vg1}
\fmf{phantom}{v2,vg1}
\fmf{phantom}{v3,vg1}
\fmffreeze
\fmf{phantom,tension=1}{vgo2,v2}
\fmf{phantom,tension=1}{vgo3,v3}
\fmf{phantom}{vg1,v1}
\fmf{#3,tension=0.5}{vg1,vgi2}
\fmf{#4,tension=0.5}{vg1,vgi3}
\fmf{#6,tension=0.5}{vgi2,vgo2}
\fmf{#7,tension=0.5}{vgi3,vgo3}
\fmffreeze
\fmfposition
\fmf{#5,label=$ $}{vgi2,vgi3}
\fmf{#5}{vgo2,vgo3}
\fmfipath{p[],pca}
\fmfipair{vm[],vo[],vi[]}
\fmfiset{p1}{vpath(__vg1,__v1)}
\fmfiset{p2}{vpath(__vgo2,__v2)}
\fmfiset{p3}{vpath(__vgo3,__v3)}
\fmfiset{p4}{vpath(__vgi2,__vgi3)}
\fmfiset{p5}{vpath(__vgi3,__vg1)}
\fmfiset{p6}{vpath(__vg1,__vgi2)}
\fmfiset{p7}{vpath(__vgo2,__vgo3)}
\fmfiset{p8}{vpath(__vgo3,__vgi3)}
\fmfiset{p9}{vpath(__vgi2,__vgo2)}
{#1}
\fmfis{#2,ptext.clen=7,ptext.hout=3,ptext.oin=0,ptext.oout=6,ptext.sep=;}{reverse(p1)}
\fmfis{#8,ptext.clen=7,ptext.hin=3,ptext.oin=6,ptext.oout=0,ptext.sep=;}{p2}
\fmfis{#9,ptext.clen=7,ptext.hin=-8,ptext.oin=6,ptext.oout=0,ptext.sep=;}{p3}
\end{fmfchar*}}}
}

\newcommand{\cVab}[7][plain]{%
\settoheight{\eqoff}{$\times$}%
\setlength{\eqoff}{0.5\eqoff}%
\addtolength{\eqoff}{-13\unitlength}%
\raisebox{\eqoff}{%
\fmfframe(2,1)(6,1){%
\begin{fmfchar*}(21,24)
\fmftop{v3}
\fmfbottom{v2}
\fmfforce{(w,h)}{v3}
\fmfforce{(w,0)}{v2}
\fmffixed{(0,0.75h)}{v2,vg3}
\fmffixed{(0,0.25h)}{v2,vg2}
\fmfpoly{phantom}{v1,v2,v3}
\fmf{#1}{vg2,vg3}
\fmf{#1}{vg2,v2}
\fmf{#1}{vg3,v3}
\fmf{photon}{vg #2,v1}
\fmf{photon,left=#3}{vg3,vg2}
\fmffreeze
\fmfposition
\fmfipath{p[],pca}
\fmfipair{vm[],vo[],vi[]}
\fmfiset{p1}{vpath(__vg #2,__v1)}
\fmfiset{p2}{vpath(__vg2,__v2)}
\fmfiset{p3}{vpath(__vg3,__v3)}
\fmfiset{p4}{vpath(__vg2,__vg3)}
\fmfiset{p5}{reverse(vpath(__vg3,__vg2))}
\fmfis{phantom,ptext.clen=6,ptext.hin=3,ptext.hout=3,ptext.oin=6,ptext.oout=6,ptext.out=#4,ptext.sep=;}{reverse(p1)}
\fmfis{phantom,ptext.clen=6,ptext.hin=3,ptext.hout=3,ptext.oin=6,ptext.oout=6,ptext.out=#5,ptext.sep=;}{p3}
\fmfis{phantom,ptext.clen=6,ptext.hin=-9,ptext.hout=-9,ptext.oin=6,ptext.oout=6,ptext.out=#6,ptext.sep=;}{p2}
\fmfis{phantom,ptext.clen=6,ptext.hin=-9,ptext.hout=-9,ptext.oin=8,ptext.oout=8,#7,ptext.sep=;}{p4}
\end{fmfchar*}}}
}

\newcommand{\cccV}[9][]{%
\settoheight{\eqoff}{$\times$}%
\setlength{\eqoff}{0.5\eqoff}%
\addtolength{\eqoff}{-13\unitlength}%
\raisebox{\eqoff}{%
\fmfframe(2,1)(2,1){%
\begin{fmfchar*}(24,24)
\fmftop{v4}
\fmfleft{v1}
\fmfbottom{v2}
\fmfright{v2}
\fmfforce{(0.5w,h)}{v4}
\fmfforce{(0,0.5h)}{v1}
\fmfforce{(w,0.5h)}{v3}
\fmfforce{(0.5w,0)}{v2}
\fmffixed{(0.5w,0h)}{vs1,vs3}
\fmf{phantom}{vs1,v1}
\fmf{phantom}{vs2,v2}
\fmf{phantom}{vs3,v3}
\fmf{phantom}{vs4,v4}
\fmfpoly{phantom}{vs1,vs2,vs3,vs4}
\fmffreeze
\fmfposition
\fmfipath{p[],pca}
\fmfipair{vm[],vo[],vi[]}
\fmfiset{p1}{vpath(__vs1,__v1)}
\fmfiset{p2}{vpath(__vs2,__v2)}
\fmfiset{p3}{vpath(__vs3,__v3)}
\fmfiset{p4}{vpath(__vs4,__v4)}
\fmfiset{p5}{vpath(__vs1,__vs2)}
\fmfiset{p6}{vpath(__vs2,__vs3)}
\fmfiset{p7}{vpath(__vs3,__vs4)}
\fmfiset{p8}{vpath(__vs4,__vs1)}
{#1}
\fmfis{#2,ptext.clen=7,ptext.hin=3,ptext.hout=3,ptext.oin=5,ptext.oout=3,ptext.sep=;}{reverse(p1)}
\fmfis{#3,ptext.clen=7,ptext.hin=3,ptext.hout=3,ptext.oin=5,ptext.oout=3,ptext.sep=;}{p2}
\fmfis{#4,ptext.clen=7,ptext.hin=3,ptext.hout=3,ptext.oin=5,ptext.oout=3,ptext.sep=;}{p3}
\fmfis{#5,ptext.clen=7,ptext.hin=-8,ptext.hout=-8,ptext.oin=5,ptext.oout=3,ptext.sep=;}{p4}
\fmfis{#6,ptext.clen=7,ptext.hin=-8,ptext.hout=-8,ptext.oin=5,ptext.oout=5,ptext.sep=;}{p5}
\fmfis{#7,ptext.clen=7,ptext.hin=-8,ptext.hout=-8,ptext.oin=5,ptext.oout=5,ptext.sep=;}{p6}
\fmfis{#8,ptext.clen=7,ptext.hin=-8,ptext.hout=-8,ptext.oin=5,ptext.oout=5,ptext.sep=;}{p7}
\fmfis{#9,ptext.clen=7,ptext.hin=3,ptext.hout=3,ptext.oin=5,ptext.oout=5,ptext.sep=;}{reverse(p8)}
\end{fmfchar*}}}
}

\newcommand{\swftwoone}{%
\settoheight{\eqoff}{$\times$}%
\setlength{\eqoff}{0.5\eqoff}%
\addtolength{\eqoff}{-7.5\unitlength}%
\raisebox{\eqoff}{%
\fmfframe(1,0)(1,0){%
\begin{fmfchar*}(20,15)
\fmfleft{v1}
\fmfright{v2}
\fmffixed{(0.66w,0)}{vc1,vc2}
\fmf{plain}{v1,vc1}
\fmf{plain}{vc2,v2}
\fmf{plain,left=1}{vc1,vc2}
\fmf{plain,left=1}{vc2,vc1}
\fmffreeze
\fmfposition
\fmfipair{vm[]}
\svertex{vm1}{vpath(__vc1,__vc2)}
\svertex{vm2}{vpath(__vc2,__vc1)}
\fmfi{photon}{vm1--vm2}
\end{fmfchar*}}}}

\newcommand{\swftwotwo}[3][plain]{%
\settoheight{\eqoff}{$\times$}%
\setlength{\eqoff}{0.5\eqoff}%
\addtolength{\eqoff}{-7.5\unitlength}%
\raisebox{\eqoff}{%
\fmfframe(1,0)(1,0){%
\begin{fmfchar*}(20,15)
\fmfleft{v1}
\fmfright{v2}
\fmffixed{(0.33w,0)}{vc1,vc2}
\fmffixed{(0.33w,0)}{vc2,vc3}
\fmf{#1}{v1,vc1}
\fmf{#1}{vc1,vc2}
\fmf{#1}{vc2,vc3}
\fmf{#1}{vc3,v2}
\fmf{photon,tension=0.5,left=#3}{vc1,vc2}
\fmf{photon,tension=0.5,left=#2}{vc1,vc3}
\fmffreeze
\fmfposition
\fmfipath{p[]}
\end{fmfchar*}}}}

\newcommand{\swftwofour}[3][plain]{%
\settoheight{\eqoff}{$\times$}%
\setlength{\eqoff}{0.5\eqoff}%
\addtolength{\eqoff}{-7.5\unitlength}%
\raisebox{\eqoff}{%
\fmfframe(1,0)(1,0){%
\begin{fmfchar*}(20,15)
\fmfleft{v1}
\fmfright{v2}
\fmffixed{(0.33w,0)}{vc1,vc2}
\fmffixed{(0.33w,0)}{vc2,vc3}
\fmf{#1}{v1,vc1}
\fmf{#1}{vc1,vc2}
\fmf{#1}{vc2,vc3}
\fmf{#1}{vc3,v2}
\fmf{photon,tension=0.5,left=#2}{vc1,vc2}
\fmf{photon,tension=0.5,left=#3}{vc2,vc3}
\fmffreeze
\fmfposition
\fmfipath{p[]}
\end{fmfchar*}}}}

\newcommand{\swftwofive}[2][plain]{%
\settoheight{\eqoff}{$\times$}%
\setlength{\eqoff}{0.5\eqoff}%
\addtolength{\eqoff}{-7.5\unitlength}%
\raisebox{\eqoff}{%
\fmfframe(1,0)(1,0){%
\begin{fmfchar*}(20,15)
\fmfleft{v1}
\fmfright{v2}
\fmffixed{(0.22w,0)}{vc1,vc2}
\fmffixed{(0.22w,0)}{vc2,vc3}
\fmffixed{(0.22w,0)}{vc3,vc4}
\fmf{#1}{v1,vc1}
\fmf{#1}{vc1,vc2}
\fmf{#1}{vc2,vc3}
\fmf{#1}{vc3,vc4}
\fmf{#1}{vc4,v2}
\fmf{photon,tension=0.5,left=#2}{vc1,vc3}
\fmf{photon,tension=0.5,left=-(#2)}{vc2,vc4}
\fmffreeze
\fmfposition
\fmfipath{p[]}
\fmfiset{p1}{vpath(__v1,__vc1)}
\fmfiset{p2}{vpath(__vc1,__vc2)}
\fmfiset{p3}{vpath(__vc2,__vc3)}
\fmfiset{p4}{vpath(__vc3,__vc4)}
\fmfiset{p5}{vpath(__vc4,__v2)}
\end{fmfchar*}}}}

\newcommand{\swftwosix}[3]{%
\settoheight{\eqoff}{$\times$}%
\setlength{\eqoff}{0.5\eqoff}%
\addtolength{\eqoff}{-7.5\unitlength}%
\raisebox{\eqoff}{%
\fmfframe(1,0)(1,0){%
\begin{fmfchar*}(20,15)
\fmfleft{v1}
\fmfright{v2}
\fmffixed{(0.33w,0)}{vc1,vc2}
\fmffixed{(0.33w,0)}{vc2,vc3}
\fmffixed{(0,0.33w)}{vc2,vc}
\fmf{plain}{v1,vc1}
\fmf{plain}{vc1,vc2}
\fmf{plain}{vc2,vc3}
\fmf{plain}{vc3,v2}
\fmf{#1,tension=0.5,left=0.5}{vc1,vc}
\fmf{#2,tension=0.5,left=0.5}{vc,vc3}
\fmffreeze
\fmf{#3,tension=0.5}{vc2,vc}
\fmffreeze
\fmfposition
\fmfipath{p[]}
\fmfiset{p1}{vpath(__v1,__vc1)}
\fmfiset{p2}{vpath(__vc1,__vc2)}
\fmfiset{p3}{vpath(__vc2,__vc3)}
\fmfiset{p4}{vpath(__vc3,__v2)}
\fmfiset{p5}{vpath(__vc1,__vc)}
\fmfiset{p6}{vpath(__vc2,__vc)}
\fmfiset{p7}{vpath(__vc,__vc3)}
\end{fmfchar*}}}}

\newcommand{\nvml}[3][1]{%
\fmfcmd{%
begingroup;
save a, vp, tvp, nvp, tv, nv, ip, ts, tt, is, it, n, m, scale, t, r, s, ttpr, tnpr, ep, mm;
path lcirc;
pair vp[][], tvp[][], tv[][], nvp[][], nv[][], ip[][], ts[], is[], tt[], it[], ep[], mid;
n := #2;
m:=3;
for i=1 upto n:
for j=1 upto m:
a[i][j] := arctime ((j-1)/(m-1)*arclength pm[i]) of pm[i];
vp[i][j] := point a[i][j] of pm[i];
tvp[i][j] := unitvector direction a[i][j] of pm[i];
nvp[i][j] := tvp[i][j] rotated -90;
endfor;
endfor;
if(vp[1][1]=vp[n][m]):
vp[n+1][1] := vp[1][1];
tvp[0][m] := tvp[n][m];
nvp[0][m] := nvp[n][m];
tvp[n+1][1] :=tvp[1][1];
nvp[n+1][1] :=nvp[1][1];
else:
vp[n+1][1] := vp[n][m];
tvp[0][m] := (0,0);
nvp[0][m] := (0,0);
tvp[n+1][1] :=tvp[n][m];
nvp[n+1][1] :=nvp[n][m];
fi;
s := 1;
for i=1 upto n:
for j=1 upto m:
if (j=1):
tv[i][1] := (tvp[i-1][m]+tvp[i][1]);
nv[i][1] := (nvp[i-1][m]+nvp[i][1]);
if (not(tv[i][1]=(0,0))):
tv[i][1] := unitvector tv[i][1];
fi;
if (not(nv[i][1]=(0,0))):
nv[i][1] := unitvector nv[i][1];
fi;
ttpr := tvp[i][1] dotprod tvp[i-1][m];
tnpr := tvp[i][1] dotprod nvp[i-1][m];
elseif (j=m):
tv[i][m] := (tvp[i][m]+tvp[i+1][1]);
nv[i][m] := (nvp[i][m]+nvp[i+1][1]);
if (not(tv[i][m]=(0,0))):
tv[i][m] := unitvector tv[i][m];
fi;
if (not(nv[i][m]=(0,0))):
nv[i][m] := unitvector nv[i][m];
fi;
ttpr := tvp[i][m] dotprod tvp[i+1][1];
tnpr := -tvp[i][m] dotprod nvp[i+1][1];
else:
nv[i][j] :=nvp[i][j];
tv[i][j] :=tvp[i][j];
fi;
scale := 25;
if ((j=1) or (j=m)):
 if ((tnpr<=0) and not((tv[i][j]=(0,0)) or (nv[i][j]=(0,0)))):
  ip[i][j] := vp[i][j] shifted(0.15*scale*nvp[i][j]);
  ts[s] := tvp[i][j];
  is[s] := ip[i][j];
  s:=s+1;
 else:
  if ((j=1) and (ttpr>0)):
  fi;
 fi;
else:
 ip[i][j] := vp[i][j] shifted(0.15*scale*nv[i][j]);
 ts[s] := tv[i][j];
 is[s] := ip[i][j];
 s:=s+1;
fi;
endfor;
endfor;
if(vp[1][1]=vp[n][m]):
ts[s] := ts[1];
is[s] := is[1];
else:
tv[n+1][1] := unitvector (tvp[n][m]+tvp[n+1][1]);
nv[n+1][1] := unitvector (nvp[n][m]+nvp[n+1][1]);
ip[n+1][1] := vp[n+1][1] shifted(0.15*scale*nv[n+1][1]);
ts[s] := tv[n+1][1];
is[s] := ip[n+1][1];
fi;
t=#1;
lcirc:=is[1];
for k=2 upto s:
lcirc := lcirc{ts[k-1]}..tension t..{ts[k]}is[k];
endfor;
mm := arctime (0.5* arclength lcirc) of lcirc;
if(vp[1][1]=vp[n][m]):
ep1 := point arctime (0* arclength lcirc) of lcirc of lcirc;
ep2 := point mm of lcirc;
mid := 1/2[ep1,ep2];
else:
ep1 := point mm of lcirc;
ep2 :=unitvector direction mm of lcirc rotated -90;
mid:= ep1 shifted(0.2*scale*ep2);
fi;
draw(lcirc) withpen pencircle scaled 0.25;
drawarrow(subpath(mm*0.8,mm*1.1) of lcirc) withpen pencircle scaled 0.25;
endgroup;
}
\fmfiv{label=#3,l.dist=0}{mid}
}

\DeclareMathOperator{\tr}{tr}

\DeclareMathOperator{\perm}{P}

\DeclareMathOperator{\Kop}{K}
\DeclareMathOperator{\Rop}{R}

\DeclareMathOperator{\D}{D}
\DeclareMathOperator{\barD}{\vphantom{\D}\smash[t]{\bar{\mathrm{D}}}}
\DeclareMathOperator{\Ld}{L}

\newlength{\eqoff}

\newlength{\unit}
\psset{xunit=\unit,yunit=\unit,runit=\unit}
\newlength{\linew}
\psset{linewidth=\linew}

\numberwithin{equation}{section}

\newcommand{\mympostgrey}{0.75 white}

\begin{document}
\begin{fmffile}{graphs}
\fmfcmd{%
input Dalgebra
}

\fmfcmd{%
def getmid(suffix p) =
  pair p.mid[], p.off[], p.dir[];
  for i=0 upto 36:
    p.dir[i] = dir(5*i);
    p.mid[i]+p.off[i] = directionpoint p.dir[i] of p;
    p.mid[i]-p.off[i] = directionpoint -p.dir[i] of p;
  endfor
enddef;
}

\fmfcmd{%
marksize=2mm;
def draw_mark(expr p,a) =
  begingroup
    save t,tip,dma,dmb; pair tip,dma,dmb;
    t=arctime a of p;
    tip =marksize*unitvector direction t of p;
    dma =marksize*unitvector direction t of p rotated -45;
    dmb =marksize*unitvector direction t of p rotated 45;
    linejoin:=beveled;
    draw (-.5dma.. .5tip-- -.5dmb) shifted point t of p;
  endgroup
enddef;
style_def derplain expr p =
    save amid;
    amid=.5*arclength p;
    draw_mark(p, amid);
    draw p;
enddef;
style_def derphoton expr p =
    save amid;
    amid=.5*arclength p;
    draw_mark(p, amid);
    draw wiggly p;
enddef;
def draw_marks(expr p,a) =
  begingroup
    save t,tip,dma,dmb,dmo; pair tip,dma,dmb,dmo;
    t=arctime a of p;
    tip =marksize*unitvector direction t of p;
    dma =marksize*unitvector direction t of p rotated -45;
    dmb =marksize*unitvector direction t of p rotated 45;
    dmo =marksize*unitvector direction t of p rotated 90;
    linejoin:=beveled;
    draw (-.5dma.. .5tip-- -.5dmb) shifted point t of p withcolor 0white;
    draw (-.5dmo.. .5dmo) shifted point t of p;
  endgroup
enddef;
style_def derplains expr p =
    save amid;
    amid=.5*arclength p;
    draw_marks(p, amid);
    draw p;
enddef;
def draw_markss(expr p,a) =
  begingroup
    save t,tip,dma,dmb,dmo; pair tip,dma,dmb,dmo;
    t=arctime a of p;
    tip =marksize*unitvector direction t of p;
    dma =marksize*unitvector direction t of p rotated -45;
    dmb =marksize*unitvector direction t of p rotated 45;
    dmo =marksize*unitvector direction t of p rotated 90;
    linejoin:=beveled;
    draw (-.5dma.. .5tip-- -.5dmb) shifted point t of p withcolor 0white;
    draw (-.5dmo.. .5dmo) shifted point arctime a+0.25 mm of p of p;
    draw (-.5dmo.. .5dmo) shifted point arctime a-0.25 mm of p of p;
  endgroup
enddef;
style_def derplainss expr p =
    save amid;
    amid=.5*arclength p;
    draw_markss(p, amid);
    draw p;
enddef;
style_def dblderplains expr p =
    save amidm;
    save amidp;
    amidm=.5*arclength p-0.75mm;
    amidp=.5*arclength p+0.75mm;
    draw_mark(p, amidm);
    draw_marks(p, amidp);
    draw p;
enddef;
style_def dblderplainss expr p =
    save amidm;
    save amidp;
    amidm=.5*arclength p-0.75mm;
    amidp=.5*arclength p+0.75mm;
    draw_mark(p, amidm);
    draw_markss(p, amidp);
    draw p;
enddef;
style_def dblderplainsss expr p =
    save amidm;
    save amidp;
    amidm=.5*arclength p-0.75mm;
    amidp=.5*arclength p+0.75mm;
    draw_marks(p, amidm);
    draw_markss(p, amidp);
    draw p;
enddef;
}

%
%

\fmfcmd{%
thin := 1pt; 
thick := 2thin;
arrow_len := 4mm;
arrow_ang := 15;
curly_len := 3mm;
dash_len := 1.5mm; 
dot_len := 1mm; 
wiggly_len := 2mm; 
wiggly_slope := 60;
zigzag_len := 2mm;
zigzag_width := 2thick;
decor_size := 5mm;
dot_size := 2thick;
}


\begingroup\parindent0pt
\vspace*{2em}
\begingroup\LARGE
Superspace calculation of the three-loop dilatation operator
of $\mathcal{N}=4$ SYM theory
\par\endgroup
\vspace{1.5em}
\begingroup\large
Christoph Sieg
\par\endgroup
\vspace{1em}
\begingroup\itshape
Niels Bohr International Academy \\
Niels Bohr Institute \\
Blegdamsvej 17 \\
2100 Copenhagen \\
Denmark
\par\endgroup
\vspace{1em}
\begingroup\ttfamily
csieg@nbi.dk
\par\endgroup
\vspace{1.5em}
\endgroup

\paragraph{Abstract.}
We derive the three-loop dilatation operator of the flavor $SU(2)$
subsector of $\mathcal{N}=4$ supersymmetric Yang-Mills 
theory in the planar limit 
by a direct Feynman diagram
calculation in $\mathcal{N}=1$ superspace.
The transcendentality three contributions which appear 
in intermediate steps cancel among each other, leaving a rational
result which confirms the predictions from integrability.
We derive finiteness conditions that allow us to avoid the 
explicit evaluation of entire classes of Feynman graphs 
and also yield constraints on the $\D$-algebra manipulations. 
Based on these results, we discover universal 
cancellation mechanisms.
As a check for the consistency of our result, we verify the cancellation
of all higher-order poles.



\newpage


\section{Introduction}

\label{sec:introduction}

The hints of integrability found in
type $\twob$ string theory in $\AdS_5\times\text{S}^5$ and in
$\mathcal{N}=4$ supersymmetric Yang-Mills (SYM) theory with gauge group 
$SU(N)$ in the limit $N\to\infty$ have led to 
impressive progress in quantitatively testing the $\AdS/\CFT$ 
correspondence \cite{Maldacena:1997re,Gubser:1998bc,Witten:1998qj}.
The correspondence conjectures a duality between these two theories
and, in particular, it predicts that, in the limit $N\to\infty$, with
the 't Hooft coupling $\lambda=g_\YM^2N$ fixed such that 
the gauge theory becomes planar, the energies of string states 
should match at any value of $\lambda$
the anomalous dimensions of gauge invariant composite 
operators with the same quantum numbers.

Tests of this prediction appeared impossible, since,
in both theories, the spectra can only be calculated perturbatively 
in incompatible regimes, i.e., to the first few orders in respective 
expansions at strong coupling $\lambda\gg1$ in the string theory and at 
weak coupling $\lambda\ll1$ in the gauge theory. 
Based on the assumption of all-order integrability, 
this obstacle has been overcome by a unification \cite{Beisert:2006ez}
of the Bethe ans\"atze of the string theory \cite{Arutyunov:2004vx} 
and of the gauge theory \cite{Beisert:2004hm,Beisert:2005fw}.
From these Bethe equations, an integral equation
for the so-called cusp anomalous dimension
was derived \cite{Eden:2006rx,Beisert:2006ez}. 
The found order-by-order solutions 
at strong and weak coupling
\cite{Benna:2006nd,Casteill:2007ct,Alday:2007qf,Basso:2007wd}
match with the strong coupling results from string theory
\cite{Gubser:2002tv,Frolov:2002av,Kruczenski:2002fb,Roiban:2007jf,Roiban:2007dq} and with the weak coupling results in the $\mathcal{N}=4$ SYM theory
\cite{Makeenko:2002qe,Kotikov:2003fb,Kotikov:2004er,Bern:2005iz,Bern:2006ew,Cachazo:2006az} to the known orders. 
This is an important quantitative test of the $\AdS/\CFT$ 
correspondence. It succeeds, since the cusp anomalous dimension 
is not affected by corrections due to the finite length of the string states 
and composite operators that, in general, are not captured by the 
Bethe ans\"atze. 
Based on the assumption of integrability,
proposals of how to incorporate 
these finite size corrections
were formulated
\cite{Ambjorn:2005wa,Bajnok:2008bm,Arutyunov:2009ur,Arutyunov:2009zu,Gromov:2009bc,Gromov:2009tv,Bombardelli:2009ns} in order to describe the full spectrum.
At weak coupling in the $\mathcal{N}=4$ SYM theory the finite size 
corrections are the so-called wrapping interactions 
\cite{Serban:2004jf,Beisert:2004hm,Sieg:2005kd}.
In Feynman diagram calculations 
\cite{Fiamberti:2007rj,Fiamberti:2008sh,Fiamberti:2009jw,Velizhanin:2008jd} 
and from integrability \cite{Bajnok:2008bm,Beccaria:2009eq}, some leading 
wrapping corrections were determined, and matching was found.
This rules out an earlier conjecture \cite{Rej:2005qt} that the
finite size effects might be captured by the Hubbard model.

A first sign of integrability in the $\AdS/\CFT$ correspondence
was discovered in a one-loop calculation
in the $\mathcal{N}=4$ SYM theory \cite{Minahan:2002ve}. The  
mixing under renormalization among different composite 
operators in which the scalar fields of the theory appear within 
a single trace over the gauge group indices
was mapped to the integrable Heisenberg spin chain. The spin chain 
states are thereby identified with the composite operators, and 
the Hamiltonian acting on these chains is determined by the 
UV divergences of the underlying Feynman diagrams.
The diagonalization of this system, e.g., by means of the Bethe ansatz, 
yields as eigenvalues the anomalous dimensions. 
The sums of bare dimensions and the anomalous dimensions are the 
conformal dimensions, and they are measured by the dilatation operator
of the underlying (super-)conformal algebra. 
The spin chain Hamiltonian is, therefore, also called (the quantum part
of) the dilatation operator.

At two loops, the renormalization of composite operators of BMN type
\cite{Berenstein:2002jq} was determined
by a calculation in component formalism \cite{Gross:2002su}. 
Thereby, a dilute gas approximation was used, i.e.,
interactions between the fields that are regarded as impurities
in the BMN operators were neglected. Moreover, only the Feynman diagrams 
that alter the relative positions of the impurities within 
the gauge trace were explicitly calculated.
The contribution from the remaining most complicated diagrams 
that leave the flavors unaffected was reconstructed 
from the condition that the BMN ground state has
vanishing anomalous dimension, since it saturates a BPS bound.

Supplementing the aforementioned results with the contributions from the
interactions between the impurities, the operator mixing problem in the 
planar limit is found to be integrable to two-loop order in the 
flavor $SU(2)$ subsector \cite{Beisert:2003tq}. 
Each operator in this subsector contains a
certain number of elementary fields of two different flavors that 
are given as two complex combinations of the six real scalar fields.
With the assumption of higher-loop integrability,
the dilatation operator of the flavor $SU(2)$ subsector was then constructed
first to three \cite{Beisert:2003tq} and then to higher loops
\cite{Beisert:2004hm}. By considering operator mixing in a bigger
subsector, it was then found that the symmetry algebra, together
with some assumptions and structural input from the underlying 
Feynman diagrams, fixes to three-loop order the dilatation 
operator of this subsector \cite{Beisert:2003ys}.

The predictions have been tested by various field theory 
calculations. The three-loop anomalous dimension of the Konishi 
operator matches the conjectured eigenvalue from integrability 
\cite{Kotikov:2004er,Eden:2004ua}.
At four loops and beyond, further field theory calculations
tested the structure and various eigenvalues of the dilatation 
operator 
\cite{Gross:2002su,Bern:2006ew,Beisert:2007hz,Fiamberti:2007rj,Fiamberti:2008sh,Fiamberti:2009jw,Velizhanin:2010cm}. 
The respective tests in 
\cite{Fiamberti:2007rj,Fiamberti:2008sh,Fiamberti:2009jw} 
were necessary in order to modify the dilatation operator 
such that some leading wrapping corrections could be included.
In this context, the five-loop result for the Konishi operator 
\cite{Bajnok:2009vm,Arutyunov:2010gb,Lukowski:2009ce} and
six-loop results for twist-three operators \cite{Velizhanin:2010cm},
as obtained from integrability, still have to be tested by direct 
Feynman diagram calculations.

Albeit the aforementioned tests of the structure and of 
some eigenvalues at higher loops, a direct field theoretical 
derivation of the three-loop dilatation operator in the flavor 
$SU(2)$ subsector is not yet available. 
In this paper, we perform this calculation in $\mathcal{N}=1$ 
superfield formalism.
Since our calculation yields the dilatation operator itself, it 
determines the three-loop planar spectrum of \emph{all} composite 
single-trace operators in the flavor $SU(2)$ subsector and 
goes beyond the existing
tests of some eigenvalues.
Our result implies three-loop integrability in the flavor $SU(2)$ 
subsector and it also fixes the coupling dependence in the magnon 
dispersion relation of the underlying Bethe equations to that order.

Another motivation for this work is to gain insight for 
similar perturbative calculations at higher orders.
We formulate and exploit finiteness conditions for the underlying 
supergraphs and uncover universal cancellation mechanisms
between overall UV divergences of entire classes of graphs.
This allows us to reduce the calculational effort significantly, and 
our findings should be of importance for extending 
the field theory calculations of 
the leading wrapping corrections for short operators along the lines 
of \cite{Sieg:2005kd,Fiamberti:2007rj,Fiamberti:2008sh,Fiamberti:2009jw},
also to the next-to-leading order.

\noindent The paper is organized as follows:

In Section \ref{sec:reno},
we introduce our notation and 
summarize some aspects of operator renormalization in 
the flavor $SU(2)$ subsector of $\mathcal{N}=4$ SYM theory.

In Section \ref{sec:integrability},
we reexamine the results from integrability in the flavor $SU(2)$
subsector and the existing tests from field theory to three-loop order. 
In particular, we argue that
our calculation determines to that order the magnon dispersion relation 
and integrability in this subsector directly from field theory.

In Section \ref{sec:finiteness},
we summarize some important implications of the finiteness 
conditions for the underlying Feynman diagrams which we 
derive in Appendix \ref{app:powercounting}.

In Section \ref{sec:onetwoloops},
we demonstrate the efficiency of our approach and
derive the one- and two-loop results in an instant.

In Section \ref{sec:threeloops}, we present the 
three-loop calculation, classifying the diagrams first according to  
the range of the interactions, i.e.,
the number of fields that are involved in the interactions, and then 
also according to the generated flavor manipulations. By employing the 
finiteness conditions, we reveal universal cancellations among the overall UV 
divergences of different Feynman diagrams. Then, we present the final result
of our calculation.

In Section \ref{sec:concl}, we draw our conclusions and comment on 
the implications our findings should have for
calculations of the next-to-leading wrapping corrections along the lines of
\cite{Sieg:2005kd,Fiamberti:2007rj,Fiamberti:2008sh,Fiamberti:2009jw}.

Several details of the calculation have been delegated to appendixes.
Based on the $\D$-algebra structure of the Feynman rules listed in 
Appendix \ref{app:Feynmanrules}, the finiteness conditions 
are derived in Appendix \ref{app:powercounting}. They are quite general and
also hold, e.g., for the $\beta$-deformation \cite{Leigh:1995ep}
of the $\mathcal{N}=4$ SYM theory and
for the $\mathcal{N}=6$ Chern-Simons theory 
\cite{Aharony:2008ug,Aharony:2008gk}.
In Appendix \ref{app:subdiag}, we determine the expressions of 
the one- and two-loop subdiagrams that appear in the calculation, and we 
derive the most complicated cancellation mechanism.
In Appendix \ref{app:hopolecanc}, as a consistency check of our result, 
we explicitly demonstrate the cancellation 
of higher-order poles in the logarithm of the renormalization constant.
Expressions for the relevant integrals and their overall UV divergences are
listed in Appendix \ref{app:integrals}.

\section{Renormalization in the flavor $SU(2)$ subsector}

\label{sec:reno}

In the following, we 
work in the $\mathcal{N}=1$ superfield formulation of $\mathcal{N}=4$ SYM 
theory in Fermi-Feynman gauge and use the conventions 
of \cite{Gates:1983nr}. 
An $SU(3)$ subgroup of the $SU(4)$ R-symmetry that we call flavor 
symmetry is manifest and it transforms the three chiral superfields 
$\phi^i=(\phi,\psi,Z)$ into each other.

The composite operators of the aforementioned flavor $SU(2)$ subsector 
appear as the lowest components of chiral superfields that themselves 
are products of chiral superfields,
using as building blocks only two of the three different chiral field 
flavors, e.g., $\phi$ and $Z$.
The length $L$ of such an operator is then defined as the number of 
its constituents $\phi$ and $Z$, and we call the appearing $\phi$
impurities. In order to obtain gauge invariant objects, the
color indices of the constituent fields have to be contracted with each other.
These contractions form a certain number of cycles that yields the 
number of gauge traces in the resulting so-called multi-trace operators.
Here, we will only consider the planar limit in which it is sufficient
to study the mixing of operators involving a single gauge trace only. 
They are denoted as single-trace operators.

The $\mathcal{N}=4$ SYM theory is finite 
\cite{Mandelstam:1982cb,Brink:1982wv,Howe:1982tm,Howe:1983sr}, and hence,
in terms of $\mathcal{N}=1$ superfields,
no infinities are encountered, apart from gauge artefacts 
\cite{Grisaru:1980nk,Grisaru:1980jc,Caswell:1980yi,Caswell:1980ru}. 
This is not the case when quantum corrections are considered for the 
correlation functions that also involve composite operators.
The appearing UV divergences from the loop integrals of the quantum 
corrections do not cancel and manifest themselves as poles in 
$\varepsilon$, where $\varepsilon$ is the regulator in dimensional reduction
\cite{Siegel:1979wq} in $D=4-2\varepsilon$ dimensions.
These poles have to be absorbed by the renormalization of the
composite operators as
\begin{equation}\label{opren}
\mathcal{O}_{a,\text{ren}}
=\mathcal{Z}_{a}{}^b(\lambda,\varepsilon)\mathcal{O}_{b,\text{bare}}
\col
\end{equation}
where $\mathcal{Z}$ is the matrix-valued renormalization constant
that is given as a power series in the 't Hooft coupling constant 
$\lambda=g_\YM^2N$.

The flavor $SU(2)$ subsector is closed under renormalization, at 
least perturbatively \cite{Minahan:2005jq}. 
Mixing within this subsector can only occur among
composite operators with the same length $L$ and number of 
impurities. 
For appropriately normalized operators,
the renormalization constant decomposes as
$\mathcal{Z}=\unitmatrix+\delta\mathcal{Z}$, where $\delta\mathcal{Z}$
can be brought to block-diagonal form. Each block acts within a subset of
operators that differ only by permutations of their field  
content within the gauge trace.
The permutations are generated from the nontrivial flavor structure
of the chiral and antichiral vertex of the $\mathcal{N}=4$ SYM theory.
These vertices and their connections within each Feynman diagram form 
its chiral structure that acts on the flavors of the interacting fields as
a fixed linear combination of products of permutations and of the identity
operation. The chiral structure of each diagram is 
captured by one of the chiral functions that were introduced in 
\cite{Fiamberti:2007rj}. 
In terms of the permutation $\perm_{ij}$ and identity $\unitmatrix_{ij}$ 
that act on the fields at sites $i$ and $j$ of a composite operator
of length $L$, they 
are defined as
\begin{equation}\label{chifuncdef}
\begin{aligned}
\chi(a_1,\dots,a_n)=\sum_{r=0}^{L-1}\prod_{i=1}^n(\perm-\unitmatrix)_{a_i+r\;a_i+r+1}
\col
\end{aligned}
\end{equation}
where $\chi()$ is the identity.
Periodicity with the period $L$ is understood.
The range of the interaction in flavor space, 
i.e., the number of 
nearest neighbors that are involved in flavor permutations,  
is extracted from the argument list $a_1,\dots,a_n$ of the chiral functions as
\begin{equation}\label{nneighborint}
\kappa=\max_{a_1,\dots, a_n}-\min_{a_1,\dots, a_n}{}+{}2\pnt
\end{equation}
It must not be confused with the range $R$ of the Feynman diagram itself, 
i.e., with the number of fields of the composite operators that are 
involved in the interaction. In fact, the range $R$
exceeds $\kappa$ if flavor-neutral 
vector fields establish interactions with further fields of the composite 
operators that are not themselves building up a nontrivial chiral structure.

According to the previous discussion, we can express $\delta\mathcal{Z}$ as 
a linear combination of chiral functions. The coefficient of 
each chiral function is determined from the Feynman graphs with the
respective chiral structure. It is the negative of the 
sum of the poles in $\varepsilon$ that capture the overall UV divergences 
of the individual graphs.
The result immediately determines the 
renormalization constant in \eqref{opren}, and the 
dilatation operator is then extracted from the latter as
\begin{equation}\label{DinZ}
\mathcal{D}
=\mu\frac{\de}{\de\mu}\ln\mathcal{Z}(\lambda\mu^{2\varepsilon},\varepsilon)
=\lim_{\varepsilon\rightarrow0}\left[2\varepsilon\lambda
\frac{\de}{\de \lambda}\ln\mathcal{Z}(\lambda,\varepsilon)\right]
\col
\end{equation}
where the second relation holds, since the logarithm cancels 
all higher-order poles in $\varepsilon$. 
This cancellation is an important consistency 
check for our calculation and it can be found in Appendix \ref{app:hopolecanc}.
In effect, the above description extracts the coefficient of the 
$\frac{1}{\varepsilon}$ pole of $\delta\mathcal{Z}$ and, at a given loop order 
$K$, multiplies it by a factor $2K$. This then yields the dilatation
operator (more precisely, the quantum part)
as a power series
\begin{equation}\label{Dex}
\mathcal{D}=\sum_{k\ge1}g^{2k}\mathcal{D}_k
\col\qquad
g=\frac{\sqrt{\lambda}}{4\pi}
\col
\end{equation}
where we have absorbed the powers of $4\pi$ that appear from the loop integrals
into a rescaled coupling constant $g$.

The expression of the dilatation operator in terms of chiral functions
allows for a general statement when considering the composite operators
\begin{equation}\label{groneimpstate}
\tr(Z^L)\col\qquad\tr(\phi Z^{L-1})
\end{equation}
that, for each length $L$, are the ground state and, respectively,
the first excited state in the flavor $SU(2)$ subsector.
All chiral functions \eqref{chifuncdef} with $n\ge1$ yield zero when 
they are applied to these states.
Only the identity $\chi()$ in flavor space yields the length $L$.
Since the operators \eqref{groneimpstate}
are protected and hence are not renormalized,
the matrix $\delta\mathcal{Z}$, and thus also $\mathcal{D}$, as 
defined in \eqref{DinZ}, have to vanish when applied to these states
and must not depend explicitly on $\chi()$.
We will come back to this statement in Section \ref{sec:finiteness}
and relate it to the preservation of conformal invariance on the quantum level.


\section{Three-loop integrability}
\label{sec:integrability}

To three-loop order, rewritten in the basis of chiral functions, 
the dilatation operator from integrability is predicted as 
\cite{Beisert:2003tq}
\begin{equation}\label{D1D2D3}
\begin{aligned}
\mathcal{D}_1
&={}-{}2\chi(1)
\col\\
\mathcal{D}_2
&={}-{}2[\chi(1,2)+\chi(2,1)]+4\chi(1)
\col\\
\mathcal{D}_3
&={}-{}4[\chi(1,2,3)+\chi(3,2,1)]+4i\epsilon_2[\chi(2,1,3)-\chi(1,3,2)]-4\chi(1,3)
\\
&\phantom{{}={}}
{}+{}16[\chi(1,2)+\chi(2,1)]
-24\chi(1)
\col
\end{aligned}
\end{equation}
where $\epsilon_2$ remains undetermined and does not enter the 
spectrum. It is associated with similarity transformations 
\cite{Beisert:2003tq,Beisert:2005wv}. 

The above expressions can be applied to an eigenstate of a single magnon 
with momentum $p$. This yields the respective coefficients in the 
weak coupling expansion of the magnon energy $E(p)$. At each loop order $K$,
the coefficient is given as a linear combination of $1\le l\le K$
individual contributions that are generated as
\begin{equation}\label{chiphaseshifts}
\frac{1}{2}[\chi(1,2,\dots,l)+\chi(l,\dots,2,1)]\to{}-{}4\cos(l-1)p\sin^2\tfrac{p}{2}
\end{equation}
by the chiral functions that can be associated with 
the magnon dispersion
relation of the all-order Bethe ansatz formulated in 
\cite{Beisert:2004hm,Beisert:2005fw}. The remaining chiral functions 
$\chi(2,1,3)$, $\chi(1,3,2)$, and $\chi(1,3)$ of \eqref{D1D2D3} 
only contribute when two magnons 
are present within their flavor interaction range $\kappa=4$
and hence are associated with  magnon scattering.

The basis of chiral functions is very convenient, since
the coefficients of all chiral functions of the form \eqref{chiphaseshifts}
are directly related to the magnon dispersion relation.
For a single magnon with momentum 
$p$, the dispersion relation is given by \cite{Beisert:2004hm,Beisert:2005fw}
\begin{equation}\label{magnondisp}
\begin{aligned}
E(p)&=\sqrt{1+4h^2(g)\sin^2\tfrac{p}{2}}-1
\col
\end{aligned}
\end{equation}
and it is determined by the underlying symmetry algebra up to an 
unknown function $h^2(g)$ of the coupling constant 
\cite{Beisert:2005tm}.
Results for the quantum corrections \cite{Minahan:2006bd,Papathanasiou:2007gd} 
of the giant magnon solution \cite{Hofman:2006xt}
at large $g$ and the field theory results for
two-loop anomalous dimensions of the 
Konishi operator \cite{Bianchi:2000hn,Arutyunov:2001mh,Bianchi:2001cm}
and of the BMN operators \cite{Gross:2002su}
suggest that $h^2(g)=4g^2$ is the exact result. This has also been argued
using S-duality \cite{Berenstein:2009qd}.

Setting $h^2(g)=4g^2$, the expansion of \eqref{magnondisp} immediately 
fixes the coefficients of all chiral functions of the form
\eqref{chiphaseshifts} not only in \eqref{D1D2D3} to three-loop order 
but in $\mathcal{D}_K$ at any loop order $K$. 
The respective terms in $\mathcal{D}_K$ are determined as
\begin{equation}
\begin{aligned}\label{DKphaseshiftcontrib}
E(p)|_{g^{2K}}&=
\frac{(-1)^{K+1}(2K)!}{(2K-1)K!^2}\big(4\sin^{2}\tfrac{p}{2}\big)^K
\to c_{K,1}\chi(1)
+\sum_{l=2}^Kc_{K,l}[\chi(1,2,\dots,l)+\chi(l,\dots,2,1)]
\col\\
c_{K,l}&=\frac{(-1)^{K+l+1}}{2K-1}\binom{2K}{K}\binom{2(K-1)}{K-l}
\pnt
\end{aligned}
\end{equation} 
In order to obtain the above expression from the $K$-loop 
coefficient in the expansion of \eqref{magnondisp}, we have kept one factor 
$\sin^2\tfrac{p}{2}$ and expressed the remaining powers
in terms of the cosine of integer multiples of $p$.
The respective trigonometric relation
can be found, e.g., in \cite{Gradshteyn:1980}.
Finally, we have used \eqref{chiphaseshifts} to replace the 
phase shifts by the respective chiral functions. 

Beyond explicit order-by-order evaluations, no guiding principle
from which one could determine $h^2(g)$ is presently known.
The function  $h^2(g)$
might even have a series expansion with coefficients of nonvanishing
transcendentality, as found in the case of $\mathcal{N}=6$ Chern-Simons theory 
\cite{Minahan:2009aq,Minahan:2009wg}.
We will, therefore, assume for a moment that $h^2(g)$ has
a generic nontrivial expansion at weak coupling as
\begin{equation}\label{hex}
h^2(\lambda)=4(g^2+g^4h_2+g^6h_3+\dots)\pnt
\end{equation}
Inserting this expansion into the magnon dispersion relation
and using \eqref{DKphaseshiftcontrib}, it deforms the dilatation operator 
in \eqref{D1D2D3} as
\begin{equation}
\begin{aligned}\label{D1D2D3def}
\mathcal{D}_{1,\text{def}}
&={}-{}2\chi(1)
\col\\
\mathcal{D}_{2,\text{def}}
&={}-{}2[\chi(1,2)+\chi(2,1)]+2(2-h_2)\chi(1)
\col\\
\mathcal{D}_{3,\text{def}}
&={}-{}4[\chi(1,2,3)+\chi(3,2,1)]+4i\epsilon_2[\chi(2,1,3)-\chi(1,3,2)]
-(4+s)\chi(1,3)\\
&\phantom{{}={}}
{}+{}4(4-h_2)[\chi(1,2)+\chi(2,1)]-2(12-4h_2+h_3)\chi(1)
\pnt
\end{aligned}
\end{equation}
We have thereby also introduced a deformation $s$ of the only magnon 
scattering term that affects the spectrum. If $s$ is 
nonvanishing, $\mathcal{D}_{\text{def}}$ no longer commutes with the 
higher local conserved charges, and integrability is lost. 

Note that the all-order resummation of the maximum shuffling 
terms \footnote{At each loop order $K$, the maximum shuffling terms are 
contained in the combinations \eqref{chiphaseshifts} with $l=K$.} 
in \cite{Gross:2002su} assumes that there are no deformations of the form 
\eqref{hex}. The maximum shuffling terms cannot provide any 
information on the $h_i$, $i\ge 2$, since, at each loop order $K$, the 
coefficients of the chiral functions \eqref{chiphaseshifts} with $l=K$ 
do not depend on these $h_i$.

Applying $\mathcal{D}_{\text{def}}$ to the Konishi descendant in the
flavor $SU(2)$ subsector yields the eigenvalue
\begin{equation}
\begin{aligned}
\gamma
&=12g^2-[48-12h_2]g^4+[336-12(8h_2-h_3+s)]g^6
\col
\end{aligned}
\end{equation}
and it should match with the anomalous dimension of the Konishi operator
to pass the existing tests from field theory.
The one-loop result 
\cite{Anselmi:1996mq,Anselmi:1996dd,Bianchi:1999ge} is reproduced, 
and the two-loop result
\cite{Bianchi:2000hn,Arutyunov:2001mh,Bianchi:2001cm,Gross:2002su} is found
for $h_2=0$, as mentioned earlier.
The three-loop eigenvalues for the Konishi operator and for another 
nonprotected operator obtained in \cite{Eden:2004ua} are sufficient
to fix $h_3=s=0$ but do not provide direct field theory results for
further operators. Such results would be very desirable, e.g., to 
test the anomalous dimensions of twist-two operators 
of the $SL(2)$ subsector \cite{Kotikov:2004er} that
 have been extracted as highest-transcendentality terms 
from a full three-loop calculation in QCD \cite{Moch:2004pa,Vogt:2004mw}, 
exploiting a relation of
BFKL and DGLAP evolution in $\mathcal{N}=4$ SYM theory \cite{Kotikov:2000pm}.
Our three-loop calculation of the dilatation operator itself 
determines \emph{all} coefficients 
of $\mathcal{D}_{3,\text{def}}$ in \eqref{D1D2D3def} and hence fixes 
$h_2$, $h_3$, and $s$ and tests integrability \emph{directly} from field theory.
In this way, it determines to three-loop order the planar anomalous 
dimensions of \emph{all} single-trace operators of the 
flavor $SU(2)$ subsector. 



\section{Finiteness conditions}
\label{sec:finiteness}

Based on power counting and structural properties of the Feynman rules
for the $\mathcal{N}=1$ superfields, in Appendix \ref{app:powercounting},
we derive finiteness conditions
for the diagrams that contribute to loop corrections of
a composite operator in the flavor $SU(2)$ subsector.
Diagrams in which at least two fields of the composite operators
are involved in the interaction (i.e., they have an interaction range $R\ge2$)
and in which all vertices appear in loops have no overall UV divergence 
at any loop order.
In the flavor $SU(2)$ subsector, 
vertices with vector fields and antichiral vertices 
can only appear in loops, and each chiral vertex that appears outside loops
leads to a flavor permutation. 
Diagrams with trivial chiral function $\chi()$ hence cannot have vertices 
outside their loops, and the finiteness conditions imply that, 
for $R\ge2$, they cannot have an overall UV divergence.
In other words, a contribution
of $\chi()$ to the dilatation operator could only come from a UV
divergent chiral self-energy at the respective order, 
but it is finite \cite{Grisaru:1980nk,Grisaru:1980jc,Caswell:1980yi,Caswell:1980ru,Mandelstam:1982cb,Brink:1982wv,Howe:1982tm,Howe:1983sr}, and the theory is conformal.
Since all chiral functions but $\chi()$ vanish when applied to the 
states \eqref{groneimpstate}, the above property relates the protection
of these states to the preservation of conformal invariance
at quantum level.

A further implication of the finiteness condition concerns the 
$\D$-algebra. The $\D$-algebra manipulations transform the supergraphs into
expressions which are local in the fermionic coordinates of
superspace. The considerations of Appendix \ref{app:powercounting} yield
restrictions for the $\D$-algebra manipulations, at the end of which
loop integrals with overall UV divergences are encountered: 
starting from an initial configuration, where the number of 
covariant spinor derivative $\D_\alpha$ inside loops is minimized by 
convenient choices of their positions at the chiral composite operator
and the chiral vertices, no further 
$\D_\alpha$ and only a limited number of 
$\barD_{\dot \alpha}$ must be transported outside loops by the $\D$-algebra 
manipulations.  At any loop order $K$, 
this restriction leads to complete cancellations among the overall 
UV divergences of diagrams with maximum range $R=K+1$, if
the interaction range in flavor space $\kappa$, as defined 
in \eqref{nneighborint}, is not maximal; i.e., it obeys $\kappa<R$
\cite{Fiamberti:2008sh}. 

Further details about the finiteness conditions and a discussion 
that includes also the case of $\mathcal{N}=6$ Chern-Simons (CS) theory in 
$\mathcal{N}=2$ superspace 
can be found in Appendix \ref{app:powercounting}.

\section{One-  and two-loop dilatation operator}
\label{sec:onetwoloops}

With the aforementioned finiteness conditions, we can immediately
derive the one- and two-loop dilatation operator. No reconstruction of 
parts of the Feynman diagrams from the BPS condition, as 
in the original one- and two-loop calculation in component fields 
in \cite{Berenstein:2002jq} and \cite{Gross:2002su}, is necessary here, but one 
obtains the full result.
At one loop, there is only a single UV divergent Feynman 
diagram. Its evaluation yields
\begin{equation}
\begin{aligned}\label{Z1}
\settoheight{\eqoff}{$\times$}%
\setlength{\eqoff}{0.5\eqoff}%
\addtolength{\eqoff}{-11\unitlength}%
\raisebox{\eqoff}{%
\fmfframe(-1,1)(-11,1){%
\begin{fmfchar*}(20,20)
\chione
\end{fmfchar*}}}
=\lambda I_1\chi(1)
\col\qquad
\mathcal{Z}_1=-\mathcal{I}_1\chi(1)
\col
\end{aligned}
\end{equation}
where, in the diagram, we have omitted all covariant spinor derivatives, and 
the bold horizontal line at the bottom represents the composite operator
(its further noninteracting fields are not drawn).
$\mathcal{I}_1$ denotes the pole part of the integral $I_1$ given in
\eqref{Integralexpr}.
Using \eqref{DinZ} and casting the result into the form \eqref{Dex}, we obtain
the expression for $\mathcal{D}_1$ in \eqref{D1D2D3}.

The two-loop calculation is reduced to the evaluation of three diagrams, 
when the finiteness conditions of Appendix \ref{app:powercounting} and
the finiteness of the two-loop chiral self-energy \eqref{ctwoloopse} are 
used. With the one-loop correction of the chiral vertex given in 
\eqref{ccconeloop},
the results for the diagrams with overall UV divergences 
are easily determined as
\begin{equation}
\begin{aligned}
\settoheight{\eqoff}{$\times$}%
\setlength{\eqoff}{0.5\eqoff}%
\addtolength{\eqoff}{-11\unitlength}%
\raisebox{\eqoff}{%
\fmfframe(-1,1)(-6,1){%
\begin{fmfchar*}(20,20)
\chionetwo
\end{fmfchar*}}}
=\lambda^2I_2\chi(1,2)
\col\qquad
\settoheight{\eqoff}{$\times$}%
\setlength{\eqoff}{0.5\eqoff}%
\addtolength{\eqoff}{-12\unitlength}%
\raisebox{\eqoff}{%
\fmfframe(-0.5,2)(-10.5,2){%
\begin{fmfchar*}(20,20)
\chione
\fmfcmd{fill fullcircle scaled 8 shifted vloc(__vc2) withcolor black ;}
\fmfiv{plain,label=$\scriptstyle\textcolor{white}{1}$,l.dist=0}{vloc(__vc2)}
\end{fmfchar*}}}
=
-2\lambda^2I_2\chi(1)
\col
\end{aligned}
\end{equation}
where the equalities hold up to finite terms.
Considering also the reflection of the first diagram,
the two-loop renormalization constant
becomes
\begin{equation}
\begin{aligned}\label{Z2}
\mathcal{Z}_2
&=-\mathcal{I}_2[\chi(1,2)+\chi(2,1)-2\chi(1)]
\col
\end{aligned}
\end{equation}
where $\mathcal{I}_2=\Kop\Rop(I_2)$ denotes the overall UV divergence of the 
integral $I_2$. Thereby, 
$\Rop$ subtracts the subdivergences, and $\Kop$ extracts the 
poles in $\varepsilon$. The integral $I_2$ and its
overall UV divergence $\mathcal{I}_2$ are listed in \eqref{Integralexpr}. 
Multiplying the $\frac{1}{\varepsilon}$ pole of \eqref{Z2} by $4$ 
yields the result for $\mathcal{D}_2$ in \eqref{D1D2D3}.

\section{Three-loop dilatation operator}

\label{sec:threeloops}

We organize the diagrams of the three-loop calculation according to
their interaction range $R$. This range must not be confused with the 
range of the flavor interactions $\kappa$ in \eqref{nneighborint} 
that is restricted as $\kappa\le R$.
At three loops, the maximum range diagrams have $R=4$. 
The next-to-maximum range diagrams have $R=3$, and the
diagrams of minimal range have $R=2$, since, according to the
finiteness of the three-loop chiral self-energy 
\cite{Grisaru:1980nk,Caswell:1980yi,Caswell:1980ru}, 
the $R=1$ diagrams are finite. Then, from the finiteness conditions,
as summarized in Section \ref{sec:finiteness}, we conclude that the
simplest chiral function that has to be considered is $\chi(1)$.
All equations involving three-loop diagrams are understood to hold up to 
irrelevant finite contributions.

\subsection{Maximum range diagrams}
\label{subsec:mrange}

At three loops the maximum number of fields of the composite operators
that can interact in a planar diagram is four. 
The respective diagrams only contain loops
that involve the interacting fields of the composite operator; i.e.,
if the composite operator is removed, the remaining interactions 
form a tree graph.

There are only four chiral maximum range diagrams.
They all give rise to loop integrals with simple poles in $\varepsilon$ 
and hence contribute to the dilatation operator.
They are determined as
\begin{equation}
\begin{aligned}
&
\settoheight{\eqoff}{$\times$}%
\setlength{\eqoff}{0.5\eqoff}%
\addtolength{\eqoff}{-12\unitlength}%
\raisebox{\eqoff}{%
\fmfframe(-1,2)(-1,2){%
\begin{fmfchar*}(20,20)
\chionetwothree
\end{fmfchar*}}}
=\lambda^3I_3\chi(1,2,3)
\col\quad
\settoheight{\eqoff}{$\times$}%
\setlength{\eqoff}{0.5\eqoff}%
\addtolength{\eqoff}{-12\unitlength}%
\raisebox{\eqoff}{%
\fmfframe(-1,2)(-1,2){%
\begin{fmfchar*}(20,20)
\chitwoonethree
\end{fmfchar*}}}
=\lambda^3I_{3\mathbf{bb}}\chi(2,1,3)
\col\quad
\settoheight{\eqoff}{$\times$}%
\setlength{\eqoff}{0.5\eqoff}%
\addtolength{\eqoff}{-12\unitlength}%
\raisebox{\eqoff}{%
\fmfframe(-1,2)(-1,2){%
\begin{fmfchar*}(20,20)
\chionethreetwo
\end{fmfchar*}}}
=\lambda^3I_{3\mathbf{b}}\chi(1,3,2)
\col
\end{aligned}
\end{equation}
where the not-displayed fourth diagram is obtained from the first one 
by reflection at the vertical axis.

In addition, there 
are maximum range diagrams which also contain vector fields. 
They are given by
\begin{equation}
\begin{aligned}\label{chi13diag}
&
\settoheight{\eqoff}{$\times$}%
\setlength{\eqoff}{0.5\eqoff}%
\addtolength{\eqoff}{-12\unitlength}%
\raisebox{\eqoff}{%
\fmfframe(-0.5,2)(-0.5,2){%
\begin{fmfchar*}(20,20)
\chionethree
\fmfi{photon}{vlm2--vrm1}
\end{fmfchar*}}}
=0
\col\qquad
\settoheight{\eqoff}{$\times$}%
\setlength{\eqoff}{0.5\eqoff}%
\addtolength{\eqoff}{-12\unitlength}%
\raisebox{\eqoff}{%
\fmfframe(-0.5,2)(-0.5,2){%
\begin{fmfchar*}(20,20)
\chionethree
\fmfi{photon}{vlm2--vrm3}
\end{fmfchar*}}}
+
\settoheight{\eqoff}{$\times$}%
\setlength{\eqoff}{0.5\eqoff}%
\addtolength{\eqoff}{-12\unitlength}%
\raisebox{\eqoff}{%
\fmfframe(-0.5,2)(-0.5,2){%
\begin{fmfchar*}(20,20)
\chionethree
\fmfi{photon}{vlm2--vrm4}
\end{fmfchar*}}}
=0
\col\qquad
\settoheight{\eqoff}{$\times$}%
\setlength{\eqoff}{0.5\eqoff}%
\addtolength{\eqoff}{-12\unitlength}%
\raisebox{\eqoff}{%
\fmfframe(-0.5,2)(-0.5,2){%
\begin{fmfchar*}(20,20)
\chionethree
\fmfi{photon}{vlm3--vrm3}
\end{fmfchar*}}}
=0
\col\\
&
\settoheight{\eqoff}{$\times$}%
\setlength{\eqoff}{0.5\eqoff}%
\addtolength{\eqoff}{-12\unitlength}%
\raisebox{\eqoff}{%
\fmfframe(-0.5,2)(-0.5,2){%
\begin{fmfchar*}(20,20)
\chionethree
\fmfi{photon}{vlm3--vrm4}
\end{fmfchar*}}}
=\lambda^3I_3\chi(1,3)
\col\qquad
\settoheight{\eqoff}{$\times$}%
\setlength{\eqoff}{0.5\eqoff}%
\addtolength{\eqoff}{-12\unitlength}%
\raisebox{\eqoff}{%
\fmfframe(-0.5,2)(-0.5,2){%
\begin{fmfchar*}(20,20)
\chionethree
\fmfi{photon}{vlm5--vrm4}
\end{fmfchar*}}}
=-2\lambda^3(I_3+I_{32\mathbf{t}})\chi(1,3)
\col\\
&
\settoheight{\eqoff}{$\times$}%
\setlength{\eqoff}{0.5\eqoff}%
\addtolength{\eqoff}{-12\unitlength}%
\raisebox{\eqoff}{%
\fmfframe(-0.5,2)(-0.5,2){%
\begin{fmfchar*}(20,20)
\chionethree
\fmfcmd{fill fullcircle scaled 8 shifted vloc(__vc2) withcolor black ;}
\fmfiv{plain,label=$\scriptstyle\textcolor{white}{1}$,l.dist=0}{vloc(__vc2)}
\end{fmfchar*}}}
=
-2\lambda^3I_1I_2\chi(1,3)
\col\qquad
\end{aligned}
\end{equation}
where the vanishing of the pole parts, as indicated by the first two equations,
is a consequence of the 
finiteness conditions that are derived in Appendix \ref{app:powercounting}.
For example, the finiteness of the first diagram follows immediately, 
since it matches the finiteness condition that all its vertices appear in 
loops. Because of this condition, we never have to consider graphs of this type
and disregard them in the following.

The last diagram in \eqref{chi13diag}
only yields higher-order poles in $\varepsilon$, since the interactions
occur in disconnected subdiagrams when the composite operator is 
removed. Although this diagram and its reflection do not contribute to the 
dilatation operator, we include them for the explicit check of 
the cancellation of all higher-order poles in Appendix \ref{app:hopolecanc}.
In addition to these diagrams, for a composite operator
of length $L>5$, there are similar diagrams which generate higher-order pole 
terms and have chiral functions $\chi(1,n)$, $n=4,\dots[\frac{L}{2}]+1$.
We disregard them, since, to three-loop order, there are no single-pole 
contributions to the coefficients of these chiral functions, and 
their cancellation in the logarithm of the renormalization constant
is straightforward.

The finiteness conditions of Appendix \ref{app:powercounting} 
restrict the possible 
$\D$-algebra manipulations which lead to contributions with 
overall UV divergences.
Starting from an initial configuration where a maximum number of 
covariant derivatives $\D_\alpha$ appears at propagators outside loops, 
no further $\D_\alpha$ must be brought outside the loops when performing
the $\D$-algebra manipulations. This restriction was found in 
\cite{Fiamberti:2008sh} and it simplified significantly the 
calculation of the leading wrapping corrections, also in the case of 
single-impurity operators in the $\beta$-deformed $\mathcal{N}=4$ SYM theory 
\cite{Fiamberti:2008sn}. 
The constraint on the $\D$-algebra
manipulations implies that, at a given loop order 
$K$, all diagrams with range $R=K+1$ that generate flavor interactions
of lower range $\kappa<K+1$ are finite, or their divergences cancel
against each other.
At three loops, we, therefore, need not consider 
all remaining range $R=4$ diagrams 
containing one of the chiral structures 
$\chi(1,2)$, $\chi(2,1)$, or $\chi(1)$.
As an example, we present in \eqref{r4chi1cancel}
the cancellations for some particular range $R=4$
diagrams involving the chiral structure $\chi(1)$ and two vector fields which
connect two further chiral fields to it. 
In the case when both these fields are not direct neighbors,
the cancellations are given by
\begin{equation}
\begin{aligned}
\label{r4chi1cancel}
&
\settoheight{\eqoff}{$\times$}%
\setlength{\eqoff}{0.5\eqoff}%
\addtolength{\eqoff}{-12\unitlength}%
\raisebox{\eqoff}{%
\fmfframe(-0.5,2)(-0.5,2){%
\begin{fmfchar*}(20,20)
\gchioneg
\fmfi{photon}{vm4--vglm5}
\fmfi{photon}{vm5--vgrm5}
\end{fmfchar*}}}
+
\settoheight{\eqoff}{$\times$}%
\setlength{\eqoff}{0.5\eqoff}%
\addtolength{\eqoff}{-12\unitlength}%
\raisebox{\eqoff}{%
\fmfframe(-0.5,2)(-0.5,2){%
\begin{fmfchar*}(20,20)
\gchioneg
\fmfi{photon}{vm3--vglm3}
\fmfi{photon}{vm3--vgrm3}
\end{fmfchar*}}}
&=0
\col\qquad
\settoheight{\eqoff}{$\times$}%
\setlength{\eqoff}{0.5\eqoff}%
\addtolength{\eqoff}{-12\unitlength}%
\raisebox{\eqoff}{%
\fmfframe(-0.5,2)(-0.5,2){%
\begin{fmfchar*}(20,20)
\gchioneg
\fmfi{photon}{vu3--vglu3}
\fmfi{photon}{vd3--vgrd3}
\end{fmfchar*}}}
+
\settoheight{\eqoff}{$\times$}%
\setlength{\eqoff}{0.5\eqoff}%
\addtolength{\eqoff}{-12\unitlength}%
\raisebox{\eqoff}{%
\fmfframe(-0.5,2)(-0.5,2){%
\begin{fmfchar*}(20,20)
\gchioneg
\fmfi{photon}{vm3--vglm3}
\fmfi{photon}{vm5--vgrm5}
\end{fmfchar*}}}
=0
\pnt
\end{aligned}
\end{equation}

Our analysis of the range four diagrams is now complete.
Including also the reflected diagrams where necessary, the contributions
of the range $R=4$ diagrams to the three-loop renormalization constant 
is the negative of the sum of all overall UV divergences.
We find
\begin{equation}
\begin{aligned}\label{Zr4}
\mathcal{Z}_{3,R=4}
&={}-{}\mathcal{I}_3(\chi(1,2,3)+\chi(3,2,1))
-\mathcal{I}_{3\mathbf{bb}}\chi(2,1,3)
-\mathcal{I}_{3\mathbf{b}}\chi(1,3,2)\\
&\phantom{{}={}}
{}-{}2(2\mathcal{I}_1\mathcal{I}_2-\mathcal{I}_{32\mathbf{t}})\chi(1,3)
\pnt
\end{aligned}
\end{equation}
In order to obtain the above result, we have used
the relation $\Kop\Rop(I_1I_2)=-\mathcal{I}_1\mathcal{I}_2$ for the overall 
divergence of a product of two integrals.

\subsection{Next-to-maximum range diagrams}
\label{subsec:nmrange}

At three loops, the next-to-maximum range diagrams involve $R=3$ 
neighboring fields of the composite operator: i.e., two
loops contain the three propagators that originate from the composite
operator, while one loop also remains in the diagram if the 
composite operator is removed by cutting the three connecting propagators.
Since the one-loop chiral self-energy is identically zero, 
at least three vertices must be involved in this loop.
One either obtains a box formed by only chiral field lines or 
a loop that obtains at least three vertices and that is built 
by using chiral and up to two vector fields.

The only chiral diagrams are
\begin{equation}
\begin{aligned}\label{crange3}
\settoheight{\eqoff}{$\times$}%
\setlength{\eqoff}{0.5\eqoff}%
\addtolength{\eqoff}{-12\unitlength}%
\raisebox{\eqoff}{%
\fmfframe(-0.5,2)(-5.5,2){%
\begin{fmfchar*}(20,20)
\chionetwoone
\end{fmfchar*}}}
&=\lambda^3I_3\chi(1,2,1)
\end{aligned}
\end{equation}
and its reflection that comes with the chiral function $\chi(2,1,2)$.
The appearing chiral functions $\chi(1,2,1)$ and $\chi(2,1,2)$
can be expressed in terms of a simpler one.
Using the definition \eqref{chifuncdef} in order to express them
in terms of products of permutations, 
then applying the rules found in \cite{Beisert:2005wv},
we obtain the identities
\begin{equation}\label{chiidentities}
\chi(1,2,1)=\chi(2,1,2)=\chi(1)\pnt
\end{equation}
However, 
in favour of a clear identification of the origin of the different
contributions, we will keep the original expressions and only make the 
identification at the very end.

At three loops, the nonvanishing range $R=3$ diagrams with 
chiral structure $\chi(1,2)$ or $\chi(2,1)$ contain one vector field line
which is attached to the chiral lines such that the formed loop 
involves at least three vertices. The finiteness
conditions of Appendix \ref{app:powercounting} thereby imply that,
in order to obtain a diagram with an overall UV divergence,
the chiral vertex which is not part of any loop 
must not become part of a loop when the vector field interaction is added.
Moreover, if the vector field yields a one-loop correction \eqref{ccconeloop}
of an (anti)chiral vertex which does not lead to the cancellation of 
a propagator inside a loop, the respective diagram is finite. 
The remaining diagrams are given by
\begin{equation}
\begin{aligned}\label{chi12r3}
\settoheight{\eqoff}{$\times$}%
\setlength{\eqoff}{0.5\eqoff}%
\addtolength{\eqoff}{-12\unitlength}%
\raisebox{\eqoff}{%
\fmfframe(-0.5,2)(-5.5,2){%
\begin{fmfchar*}(20,20)
\chionetwo
\fmfi{photon}{vm3{dir 180}..{dir -45}vm4}
\end{fmfchar*}}}
\col
\settoheight{\eqoff}{$\times$}%
\setlength{\eqoff}{0.5\eqoff}%
\addtolength{\eqoff}{-12\unitlength}%
\raisebox{\eqoff}{%
\fmfframe(-0.5,2)(-5.5,2){%
\begin{fmfchar*}(20,20)
\chionetwo
\fmfi{photon}{vm3{dir 0}..{dir -45}vm5}
\end{fmfchar*}}}
\col
\settoheight{\eqoff}{$\times$}%
\setlength{\eqoff}{0.5\eqoff}%
\addtolength{\eqoff}{-12\unitlength}%
\raisebox{\eqoff}{%
\fmfframe(-0.5,2)(-5.5,2){%
\begin{fmfchar*}(20,20)
\chionetwo
\fmfi{photon}{vm4{dir 0}..{dir 0}vm8}
\end{fmfchar*}}}
\col
\settoheight{\eqoff}{$\times$}%
\setlength{\eqoff}{0.5\eqoff}%
\addtolength{\eqoff}{-12\unitlength}%
\raisebox{\eqoff}{%
\fmfframe(-0.5,2)(-5.5,2){%
\begin{fmfchar*}(20,20)
\chionetwo
\fmfi{photon}{vm5{dir 120}..{dir 30}vm6}
\end{fmfchar*}}}
\col
\settoheight{\eqoff}{$\times$}%
\setlength{\eqoff}{0.5\eqoff}%
\addtolength{\eqoff}{-12\unitlength}%
\raisebox{\eqoff}{%
\fmfframe(-0.5,2)(-5.5,2){%
\begin{fmfchar*}(20,20)
\chionetwo
\fmfi{photon}{vm6{dir -45}..{dir 180}vm7}
\end{fmfchar*}}}
\col
\settoheight{\eqoff}{$\times$}%
\setlength{\eqoff}{0.5\eqoff}%
\addtolength{\eqoff}{-12\unitlength}%
\raisebox{\eqoff}{%
\fmfframe(-0.5,2)(-5.5,2){%
\begin{fmfchar*}(20,20)
\chionetwo
\fmfi{photon}{vm7{dir 225}..{dir -45}vm8}
\end{fmfchar*}}}
\col
\settoheight{\eqoff}{$\times$}%
\setlength{\eqoff}{0.5\eqoff}%
\addtolength{\eqoff}{-12\unitlength}%
\raisebox{\eqoff}{%
\fmfframe(-0.5,2)(-5.5,2){%
\begin{fmfchar*}(20,20)
\chionetwo
\fmfi{photon}{vm7{dir -45}..{dir -135}vm9}
\end{fmfchar*}}}
&=-\lambda^3I_3\chi(1,2)
\col\\
\settoheight{\eqoff}{$\times$}%
\setlength{\eqoff}{0.5\eqoff}%
\addtolength{\eqoff}{-12\unitlength}%
\raisebox{\eqoff}{%
\fmfframe(-0.5,2)(-5.5,2){%
\begin{fmfchar*}(20,20)
\chionetwo
\fmfi{photon}{vm3{dir 30}..{dir 15}vm6}
\end{fmfchar*}}}
\col
\settoheight{\eqoff}{$\times$}%
\setlength{\eqoff}{0.5\eqoff}%
\addtolength{\eqoff}{-12\unitlength}%
\raisebox{\eqoff}{%
\fmfframe(-0.5,2)(-5.5,2){%
\begin{fmfchar*}(20,20)
\chionetwo
\fmfi{photon}{vm4{dir 15}..{dir 30}vm7}
\end{fmfchar*}}}
\col
\settoheight{\eqoff}{$\times$}%
\setlength{\eqoff}{0.5\eqoff}%
\addtolength{\eqoff}{-12\unitlength}%
\raisebox{\eqoff}{%
\fmfframe(-0.5,2)(-5.5,2){%
\begin{fmfchar*}(20,20)
\chionetwo
\fmfi{photon}{vm6{dir -90}..{dir -90}vm9}
\end{fmfchar*}}}
&=\lambda^3I_3\chi(1,2)
\col\\
\settoheight{\eqoff}{$\times$}%
\setlength{\eqoff}{0.5\eqoff}%
\addtolength{\eqoff}{-12\unitlength}%
\raisebox{\eqoff}{%
\fmfframe(-0.5,2)(-5.5,2){%
\begin{fmfchar*}(20,20)
\chionetwo
\fmfi{photon}{vm4{dir 0}..{dir 90}vm5}
\end{fmfchar*}}}
\col
\settoheight{\eqoff}{$\times$}%
\setlength{\eqoff}{0.5\eqoff}%
\addtolength{\eqoff}{-12\unitlength}%
\raisebox{\eqoff}{%
\fmfframe(-0.5,2)(-5.5,2){%
\begin{fmfchar*}(20,20)
\chionetwo
\fmfi{photon}{vm5{dir -45}..{dir 0}vm7}
\end{fmfchar*}}}
&=0
\col\\
\settoheight{\eqoff}{$\times$}%
\setlength{\eqoff}{0.5\eqoff}%
\addtolength{\eqoff}{-12\unitlength}%
\raisebox{\eqoff}{%
\fmfframe(-0.5,2)(-5.5,2){%
\begin{fmfchar*}(20,20)
\chionetwo
\fmfi{photon}{vm5{dir -90}..{dir -90}vm8}
\end{fmfchar*}}}
&=\lambda^3I_{3\mathbf{t}}\chi(1,2)
\col\\
\settoheight{\eqoff}{$\times$}%
\setlength{\eqoff}{0.5\eqoff}%
\addtolength{\eqoff}{-12\unitlength}%
\raisebox{\eqoff}{%
\fmfframe(-0.5,2)(-5.5,2){%
\begin{fmfchar*}(20,20)
\chionetwo
\fmfi{photon}{vm8--vm9}
\end{fmfchar*}}}
&=-\lambda^3I_{3\mathbf{t}}\chi(1,2)
\end{aligned}
\end{equation}
and by their reflections.
It turns out that, due to the simplicity of the one-loop vertex correction 
\eqref{ccconeloop}
and due to the constraints on the $\D$-algebra manipulations, the effect
of the vector line is simply to add a triangle to the two-loop 
integral $I_2$ listed in \eqref{Integralexpr}. 
If the vector field interacts with one of the neighboring 
chiral lines of the composite operator that form the bubble in the lower-right 
corner of the chiral structure $\chi(1,2)$, this bubble is removed.
Moreover, after $\D$-algebra, there remains maximally one cubic 
vertex which is involved with only two lines in the loop integral. 
There are only two different three-loop integrals 
with an overall UV divergence that fulfill these restrictions and 
hence can be obtained after $\D$-algebra.
It can either be
$I_3$ or $I_{3\mathbf{t}}$ if a bubble is, respectively,
present or absent.
The sign of the individual contributions is determined by the color 
factor. The relative factor is negative for 
an odd number of (anti)chiral vertices which 
appear in the loop that involves the vector propagator and persists
when the composite operator is removed. This explains why, in \eqref{chi12r3},
there are only five possible results for the 
individual diagrams. The calculation essentially becomes a simple 
counting of their multiplicities.

Note that, in \eqref{chi12r3}, the contributions
which yield $I_{3\mathbf{t}}$ cancel against each other due to a relative 
sign from the color factors. Since, according to \eqref{Integralexpr},
this integral has a nonrational 
simple pole that is proportional to $\zeta(3)$,
a nonvanishing contribution would immediately require that the function
$h^2(g)$ in the magnon dispersion relation \eqref{magnondisp} received 
a transcendentality three contribution.

The range $R=3$ diagrams with chiral structure $\chi(1)$ contain up to two
flavor-neutral vector connections between $\chi(1)$ and one of its
neighboring field lines. Surprisingly, a partial evaluation of the 
$\D$-algebra 
reveals that the overall UV divergences of the
diagrams in which the vector fields interact with the additional
field line of the composite operator only via cubic vertices cancel 
among themselves.
The precise canceling combinations are 
\begin{equation}
\begin{aligned}\label{cancelr4}
&
\settoheight{\eqoff}{$\times$}%
\setlength{\eqoff}{0.5\eqoff}%
\addtolength{\eqoff}{-12\unitlength}%
\raisebox{\eqoff}{%
\fmfframe(-0.5,2)(-5.5,2){%
\begin{fmfchar*}(20,20)
\chioneg
\fmfi{photon}{vm3--vgm3}
\fmfi{photon}{vm5--vgm5}
\end{fmfchar*}}}
+
\settoheight{\eqoff}{$\times$}%
\setlength{\eqoff}{0.5\eqoff}%
\addtolength{\eqoff}{-12\unitlength}%
\raisebox{\eqoff}{%
\fmfframe(-0.5,2)(-5.5,2){%
\begin{fmfchar*}(20,20)
\chioneg
\fmfi{photon}{vu5--vgu5}
\fmfi{photon}{vd5--vgd5}
\end{fmfchar*}}}
=0
\col\qquad
\settoheight{\eqoff}{$\times$}%
\setlength{\eqoff}{0.5\eqoff}%
\addtolength{\eqoff}{-12\unitlength}%
\raisebox{\eqoff}{%
\fmfframe(-0.5,2)(-5.5,2){%
\begin{fmfchar*}(20,20)
\chioneg
\fmfi{photon}{vu3--vgu3}
\fmfi{photon}{vd3--vgd3}
\end{fmfchar*}}}
+
\settoheight{\eqoff}{$\times$}%
\setlength{\eqoff}{0.5\eqoff}%
\addtolength{\eqoff}{-12\unitlength}%
\raisebox{\eqoff}{%
\fmfframe(-0.5,2)(-5.5,2){%
\begin{fmfchar*}(20,20)
\chioneg
\fmfi{photon}{vm3--vgu3}
\fmfi{photon}{vm3--vgd3}
\end{fmfchar*}}}
+
\settoheight{\eqoff}{$\times$}%
\setlength{\eqoff}{0.5\eqoff}%
\addtolength{\eqoff}{-12\unitlength}%
\raisebox{\eqoff}{%
\fmfframe(-0.5,2)(-5.5,2){%
\begin{fmfchar*}(20,20)
\chioneg
\fmfi{photon}{vm5--vgu5}
\fmfi{photon}{vm5--vgd5}
\end{fmfchar*}}}
=0
\col\\
&
\settoheight{\eqoff}{$\times$}%
\setlength{\eqoff}{0.5\eqoff}%
\addtolength{\eqoff}{-12\unitlength}%
\raisebox{\eqoff}{%
\fmfframe(-0.5,2)(-5.5,2){%
\begin{fmfchar*}(20,20)
\chioneg
\fmfi{photon}{vm4{dir 135}..{dir 0}vu3}
\fmfi{photon}{vd3--vgd3}
\end{fmfchar*}}}
+
\settoheight{\eqoff}{$\times$}%
\setlength{\eqoff}{0.5\eqoff}%
\addtolength{\eqoff}{-12\unitlength}%
\raisebox{\eqoff}{%
\fmfframe(-0.5,2)(-5.5,2){%
\begin{fmfchar*}(20,20)
\chioneg
\fmfi{photon}{vm4{dir 135}..{dir 0}vm3}
\fmfi{photon}{vm5--vgm5}
\end{fmfchar*}}}
=0
\col\qquad
\settoheight{\eqoff}{$\times$}%
\setlength{\eqoff}{0.5\eqoff}%
\addtolength{\eqoff}{-12\unitlength}%
\raisebox{\eqoff}{%
\fmfframe(-0.5,2)(-5.5,2){%
\begin{fmfchar*}(20,20)
\chioneg
\fmfi{photon}{vm4{dir 135}..{dir 0}vd3}
\fmfi{photon}{vu3--vgu3}
\end{fmfchar*}}}
+
\settoheight{\eqoff}{$\times$}%
\setlength{\eqoff}{0.5\eqoff}%
\addtolength{\eqoff}{-12\unitlength}%
\raisebox{\eqoff}{%
\fmfframe(-0.5,2)(-5.5,2){%
\begin{fmfchar*}(20,20)
\chioneg
\fmfi{photon}{vm4{dir 135}..{dir 0}vm3}
\fmfi{photon}{vm3--vgm3}
\end{fmfchar*}}}
=0
\col\\
&
\settoheight{\eqoff}{$\times$}%
\setlength{\eqoff}{0.5\eqoff}%
\addtolength{\eqoff}{-12\unitlength}%
\raisebox{\eqoff}{%
\fmfframe(-0.5,2)(-5.5,2){%
\begin{fmfchar*}(20,20)
\chioneg
\fmfi{photon}{vm4--vm5}
\fmfi{photon}{vu3--vgu3}
\end{fmfchar*}}}
+
\settoheight{\eqoff}{$\times$}%
\setlength{\eqoff}{0.5\eqoff}%
\addtolength{\eqoff}{-12\unitlength}%
\raisebox{\eqoff}{%
\fmfframe(-0.5,2)(-5.5,2){%
\begin{fmfchar*}(20,20)
\chioneg
\fmfi{photon}{vd4--vd5}
\fmfi{photon}{vu5--vgu5}
\end{fmfchar*}}}
=0
\col\qquad
\settoheight{\eqoff}{$\times$}%
\setlength{\eqoff}{0.5\eqoff}%
\addtolength{\eqoff}{-12\unitlength}%
\raisebox{\eqoff}{%
\fmfframe(-0.5,2)(-5.5,2){%
\begin{fmfchar*}(20,20)
\chioneg
\fmfi{photon}{vu4--vu5}
\fmfi{photon}{vd5--vgd5}
\end{fmfchar*}}}
+
\settoheight{\eqoff}{$\times$}%
\setlength{\eqoff}{0.5\eqoff}%
\addtolength{\eqoff}{-12\unitlength}%
\raisebox{\eqoff}{%
\fmfframe(-0.5,2)(-5.5,2){%
\begin{fmfchar*}(20,20)
\chioneg
\fmfi{photon}{vm4--vm5}
\fmfi{photon}{vm5--vgm5}
\end{fmfchar*}}}
=0
\col\\
&
\settoheight{\eqoff}{$\times$}%
\setlength{\eqoff}{0.5\eqoff}%
\addtolength{\eqoff}{-12\unitlength}%
\raisebox{\eqoff}{%
\fmfframe(-0.5,2)(-5.5,2){%
\begin{fmfchar*}(20,20)
\chioneg
\fmfi{photon}{vm3--vgm3}
\fmfcmd{fill fullcircle scaled 8 shifted vgm3 withcolor black ;}
\fmfiv{label=$\scriptstyle\textcolor{white}{1}$,l.dist=0}{vgm3}
\end{fmfchar*}}}
+
\settoheight{\eqoff}{$\times$}%
\setlength{\eqoff}{0.5\eqoff}%
\addtolength{\eqoff}{-12\unitlength}%
\raisebox{\eqoff}{%
\fmfframe(-0.5,2)(-5.5,2){%
\begin{fmfchar*}(20,20)
\chioneg
\fmfi{photon}{vm5--vgm5}
\fmfcmd{fill fullcircle scaled 8 shifted vgm5 withcolor black ;}
\fmfiv{label=$\scriptstyle\textcolor{white}{1}$,l.dist=0}{vgm5}
\end{fmfchar*}}}
=0
\col\\
&
\settoheight{\eqoff}{$\times$}%
\setlength{\eqoff}{0.5\eqoff}%
\addtolength{\eqoff}{-12\unitlength}%
\raisebox{\eqoff}{%
\fmfframe(-0.5,2)(-5.5,2){%
\begin{fmfchar*}(20,20)
\chioneg
\fmfiset{p10}{vloc(__vc2)--vg2}
\svertex{vg3}{p10}
\fmfi{photon}{vm3--vg3}
\fmfi{photon}{vm5--vg3}
\fmfi{photon}{vg3--vg2}
\end{fmfchar*}}}
+
\settoheight{\eqoff}{$\times$}%
\setlength{\eqoff}{0.5\eqoff}%
\addtolength{\eqoff}{-12\unitlength}%
\raisebox{\eqoff}{%
\fmfframe(-0.5,2)(-5.5,2){%
\begin{fmfchar*}(20,20)
\chioneg
\fmfi{photon}{vm3--vgm3}
\fmfcmd{fill fullcircle scaled 8 shifted vm3 withcolor black ;}
\fmfiv{label=$\scriptstyle\textcolor{white}{1}$,l.dist=0}{vm3}
\end{fmfchar*}}}
+
\settoheight{\eqoff}{$\times$}%
\setlength{\eqoff}{0.5\eqoff}%
\addtolength{\eqoff}{-12\unitlength}%
\raisebox{\eqoff}{%
\fmfframe(-0.5,2)(-5.5,2){%
\begin{fmfchar*}(20,20)
\chioneg
\fmfi{photon}{vm5--vgm5}
\fmfcmd{fill fullcircle scaled 8 shifted vm5 withcolor black ;}
\fmfiv{label=$\scriptstyle\textcolor{white}{1}$,l.dist=0}{vm5}
\end{fmfchar*}}}%
+
\settoheight{\eqoff}{$\times$}%
\setlength{\eqoff}{0.5\eqoff}%
\addtolength{\eqoff}{-12\unitlength}%
\raisebox{\eqoff}{%
\fmfframe(-0.5,2)(-5.5,2){%
\begin{fmfchar*}(20,20)
\chioneg
\fmfi{photon}{vm5{dir 45}..{dir 180}vd3}
\fmfi{photon}{vu3--vgu3}
\end{fmfchar*}}}
+
\settoheight{\eqoff}{$\times$}%
\setlength{\eqoff}{0.5\eqoff}%
\addtolength{\eqoff}{-12\unitlength}%
\raisebox{\eqoff}{%
\fmfframe(-0.5,2)(-5.5,2){%
\begin{fmfchar*}(20,20)
\chioneg
\fmfi{photon}{vm5{dir 45}..{dir 180}vm3}
\fmfi{photon}{vm3--vgm3}
\end{fmfchar*}}}
+
\settoheight{\eqoff}{$\times$}%
\setlength{\eqoff}{0.5\eqoff}%
\addtolength{\eqoff}{-12\unitlength}%
\raisebox{\eqoff}{%
\fmfframe(-0.5,2)(-5.5,2){%
\begin{fmfchar*}(20,20)
\chioneg
\fmfi{photon}{vm5{dir 45}..{dir 180}vm3}
\fmfi{photon}{vm5--vgm5}
\end{fmfchar*}}}
+
\settoheight{\eqoff}{$\times$}%
\setlength{\eqoff}{0.5\eqoff}%
\addtolength{\eqoff}{-12\unitlength}%
\raisebox{\eqoff}{%
\fmfframe(-0.5,2)(-5.5,2){%
\begin{fmfchar*}(20,20)
\chioneg
\fmfi{photon}{vu5{dir 45}..{dir 180}vm3}
\fmfi{photon}{vd5--vgd5}
\end{fmfchar*}}}
&=0
\col
\end{aligned}
\end{equation}
where the equations hold up to finite parts, which 
do not enter the result for the dilatation operator.
In order to derive the above cancellations, we have used 
that a loop integral with overall UV divergence can only 
appear if by the $\D$-algebra manipulations,
no spinor derivative $\D_\alpha$ is brought outside the loops
\cite{Fiamberti:2007rj}. This constraint is part of the finiteness 
conditions derived in Appendix \ref{app:powercounting}. 
As an example, we derive the 
last relation of \eqref{cancelr4} in Appendix \ref{app:oneloopsubdiag}.
The cancellations in \eqref{cancelr4}
are universal and also hold if the chiral structure of the diagram
is different from $\chi(1)$, as long as the interactions of the 
neighboring chiral line with the remaining diagram 
are not altered.
At higher-loop orders, the next-to-maximum range diagrams with
two vector field lines involve more than a single chiral vertex.
It is then possible that also chiral vertices are involved in loops, 
and one should try to find cancellation 
patterns in these cases similar to the ones in \eqref{cancelr4}. 
This was not necessary 
here, since, at three loops, all next-to-maximum range diagrams 
have chiral structure $\chi(1)$ with only a single chiral vertex.
The diagrams in which this vertex appears in loops are all finite,
due to the finiteness conditions.

The cancellations of the overall UV divergences are not complete if two vector 
field connections with the chiral structure combine into a single 
quartic vector-matter interaction at the additional chiral field line.  
The relevant diagrams yield
\begin{equation}
\begin{aligned}\label{noncancelr4}
\settoheight{\eqoff}{$\times$}%
\setlength{\eqoff}{0.5\eqoff}%
\addtolength{\eqoff}{-12\unitlength}%
\raisebox{\eqoff}{%
\fmfframe(-0.5,2)(-5.5,2){%
\begin{fmfchar*}(20,20)
\chioneg
\fmfi{photon}{vu3--vgm3}
\fmfi{photon}{vd3--vgm3}
\end{fmfchar*}}}
&=-\frac{\lambda^3}{2}I_3\chi(1)
\col\qquad
\settoheight{\eqoff}{$\times$}%
\setlength{\eqoff}{0.5\eqoff}%
\addtolength{\eqoff}{-12\unitlength}%
\raisebox{\eqoff}{%
\fmfframe(-0.5,2)(-5.5,2){%
\begin{fmfchar*}(20,20)
\chioneg
\fmfi{photon}{vm3--vg2}
\fmfi{photon}{vm5--vg2}
\end{fmfchar*}}}
=
\frac{\lambda^3}{2}I_3\chi(1)
\col\qquad
\settoheight{\eqoff}{$\times$}%
\setlength{\eqoff}{0.5\eqoff}%
\addtolength{\eqoff}{-12\unitlength}%
\raisebox{\eqoff}{%
\fmfframe(-0.5,2)(-5.5,2){%
\begin{fmfchar*}(20,20)
\chioneg
\fmfi{photon}{vu5--vgm5}
\fmfi{photon}{vd5--vgm5}
\end{fmfchar*}}}
=-\frac{\lambda^3}{2}I_{3\mathbf{t}}\chi(1)
\col
\end{aligned}
\end{equation}
and we also have to consider their reflections.
There is an (accidental) cancellation between the first two contributions
in the three-loop graphs which does not seem to hold, in general, if 
further interactions or other chiral functions are involved at higher loops. 

Considering reflected diagrams where necessary, 
the contribution of all range $R=3$ diagrams to the renormalization constant 
is given by
\begin{equation}\label{Zr3}
\begin{aligned}
\mathcal{Z}_{3,R=3}
&{}={}-\mathcal{I}_3(\chi(1,2,1)+\chi(2,1,2))
-4\mathcal{I}_3(\chi(1,2)+\chi(2,1))
+\mathcal{I}_{3\mathbf{t}}\chi(1)
\pnt
\end{aligned}
\end{equation}

\subsection{Diagrams with nearest neighbor interactions}

The remaining diagrams which have to be calculated are corrections of  
$\chi(1)$ itself, involving no further fields of the composite operator.
As mentioned before, the interactions cannot include the chiral vertex
in a loop, since the respective graphs have no overall UV divergence, 
according to the finiteness conditions.
Because of the vanishing of the one-loop self-energy,
the remaining graphs can either be regarded as 
two-loop corrections of one of the propagators which connect
$\chi(1)$ to the composite operator, or as a
two-loop vertex correction of the antichiral vertex within $\chi(1)$.
With the chiral self-energy and
vertex correction calculated in Appendix \ref{app:twoloopsubdiag}, 
we obtain
\begin{equation}
\begin{aligned}
\settoheight{\eqoff}{$\times$}%
\setlength{\eqoff}{0.5\eqoff}%
\addtolength{\eqoff}{-12\unitlength}%
\raisebox{\eqoff}{%
\fmfframe(-0.5,2)(-10.5,2){%
\begin{fmfchar*}(20,20)
\chione
\fmfcmd{fill fullcircle scaled 8 shifted vm4 withcolor black ;}
\fmfiv{label=$\scriptstyle\textcolor{white}{2}$,l.dist=0}{vm4}
\end{fmfchar*}}}
=2\lambda^3I_{3\mathbf{t}}\chi(1)
\col\qquad
\settoheight{\eqoff}{$\times$}%
\setlength{\eqoff}{0.5\eqoff}%
\addtolength{\eqoff}{-12\unitlength}%
\raisebox{\eqoff}{%
\fmfframe(-0.5,2)(-10.5,2){%
\begin{fmfchar*}(20,20)
\chione
\fmfcmd{fill fullcircle scaled 8 shifted vloc(__vc2) withcolor black ;}
\fmfv{label=$\scriptstyle\textcolor{white}{2}$,l.dist=0}{vc2}
\end{fmfchar*}}}
=\lambda^3(4I_3-3I_{3\mathbf{t}})\chi(1)
\pnt
\end{aligned}
\end{equation}
We also have to consider the reflection of the first diagram.
Diagrams of the above type appear likewise
in the three-loop contribution to the chiral
self-energy. They are shown, respectively, 
in fig.\ 1 (f) and (b) of \cite{Caswell:1980yi}, 
and fig.\ 7 (f) and (b) of \cite{Caswell:1980ru},
and the expressions are given there in tab.\ 1.\ and 2.
Apart from a different normalization factor, our results coincide with these
expressions.
The nearest neighbor interactions contribute to the renormalization 
constant as
\begin{equation}
\begin{aligned}\label{Zr2}
\mathcal{Z}_{3,R=2}=-(4\mathcal{I}_3+I_{3\mathbf{t}})\chi(1)
\pnt
\end{aligned}
\end{equation}

\subsection{Final result}

Since the three-loop chiral self-energy is finite 
\cite{Grisaru:1980nk,Caswell:1980yi,Caswell:1980ru}, according to
the arguments in Section \ref{sec:finiteness}, no further graphs
have to be considered.
The complete three-loop contribution to the renormalization constant is then 
given by the sum of \eqref{Zr4}, \eqref{Zr3} and \eqref{Zr2} and it reads
\begin{equation}
\begin{aligned}\label{Z3res}
\mathcal{Z}_3
&={}-{}\mathcal{I}_3(\chi(1,2,3)+\chi(3,2,1)-4(\chi(1,2)+\chi(2,1))+4\chi(1)+\chi(1,2,1)+\chi(2,1,2))\\
&\phantom{{}={}}
{}+{}2(\mathcal{I}_{32\mathbf{t}}-2\mathcal{I}_1\mathcal{I}_2)\chi(1,3)
-\mathcal{I}_{3\mathbf{bb}}\chi(2,1,3)
-\mathcal{I}_{3\mathbf{b}}\chi(1,3,2)
\pnt
\end{aligned}
\end{equation}
We stress that the contributions with $\chi(1)$ 
which involve the integral $I_{3\mathbf{t}}$ with a nonrational 
simple pole in \eqref{Zr3} and \eqref{Zr2} cancel, as required 
by the simplest form of the magnon dispersion relation 
\eqref{magnondisp} with $h^2(g)=4g^2$.
As an important consistency check for \eqref{Z3res},
we explicitly demonstrate, in Appendix \ref{app:hopolecanc}, that 
to three-loop order
all higher-order poles cancel in the logarithm of the 
renormalization constant.

The three-loop dilatation operator is obtained by multiplying by $6$ the
$\frac{1}{\epsilon}$ pole of \eqref{Z3res} after the expressions
 of the poles of the integrals given in \eqref{Integralexpr}
have been inserted. The result reads
\begin{equation}
\begin{aligned}\label{D3}
\mathcal{D}_3
&={}-{}4(\chi(1,2,3)+\chi(3,2,1))+2(\chi(2,1,3)-\chi(1,3,2))-4\chi(1,3)\\
&\phantom{{}={}}
{}+{}16(\chi(1,2)+\chi(2,1))-16\chi(1)-4(\chi(1,2,1)+\chi(2,1,2))
\pnt
\end{aligned}
\end{equation}
It coincides with the prediction from integrability \eqref{D1D2D3}
if we insert the identities \eqref{chiidentities} that replace the
chiral functions $\chi(1,2,1)$ and $\chi(2,1,2)$ each by $\chi(1)$. 
The coefficient
parameterizing the similarity transformations 
\cite{Beisert:2003tq,Beisert:2005wv} is fixed to the value 
\begin{equation}\label{epsilon2}
\epsilon_2=-\frac{i}{2}
\end{equation}
in the scheme of Feynman diagrams in $\mathcal{N}=1$ superspace.

\section{Conclusions}
\label{sec:concl}

In this paper we have calculated to three-loop order the dilatation operator 
in the flavor $SU(2)$ subsector of $\mathcal{N}=4$ SYM theory
and confirmed the predictions from integrability. In particular, we have 
found cancellations of all contributions of transcendentality three, 
such that the result only contains rational numbers.
Since our calculation is based on Feynman diagrams 
in $\mathcal{N}=1$ superspace, it is a direct derivation of three-loop
integrability and of the planar three-loop spectrum of the flavor $SU(2)$ 
subsector from field theory. It also confirms to three-loop order
the expectation that there are no corrections to the function 
$h^2(g)=4g^2$ in the magnon dispersion relation.

From this work, we have also gained insight for 
future perturbative calculations at higher orders.
The finiteness conditions that we have derived in 
Appendix \ref{app:powercounting}
predict the finiteness of entire classes of graphs and
are quite universal. They also hold, e.g., for 
the $\beta$-deformed $\mathcal{N}=4$ SYM theory and
for $\mathcal{N}=6$ Chern-Simons theory in respective
formulations in $\mathcal{N}=1$ and $\mathcal{N}=2$ superspace.
Here, by making use of the finiteness conditions
in the case of $\mathcal{N}=4$ SYM theory, we have found universal 
cancellations among the overall UV divergences of graphs.
They reduced the calculational effort significantly.

Our findings should be of importance for future 
calculations of wrapping effects
for composite operators of length $L$ along the lines
of \cite{Sieg:2005kd,Fiamberti:2007rj,Fiamberti:2008sh,Fiamberti:2009jw}
beyond the presently known critical loop orders $K=L$.
Starting from the $K$-loop dilatation operator, as determined from the 
underlying integrability, one part of the procedure is
the subtraction of the contributions from 
all diagrams with range $R>L$.
At critical order $K=L$, such 
contributions only stem from maximum range $R=K+1$ diagrams
that also have a chiral structure of maximum range $\kappa=K+1$.
The remaining maximum range diagrams with chiral functions of lower range
$\kappa<R$ do not contribute, since their overall UV divergences
cancel. In our calculation, we have used this fact at two loops in 
Section \ref{sec:onetwoloops} and at three loops in
Subsection \ref{subsec:mrange}.
The subtraction of maximum range diagrams from the $K$-loop dilatation 
operator is performed simply by the omission of all terms with 
chiral functions of range $\kappa=K+1$. 
At the next order $K=L+1$, the contributions from all maximum and 
next-to-maximum range diagrams have to be identified and removed
from the dilatation operator.
At three loops, the next-to-maximum range 
diagrams show up for the first time and have been evaluated
in Subsection \ref{subsec:nmrange}. 
This analysis gives first hints of how the subtraction procedure
has to be extended at the order $K=L+1$.
Because of the identities \eqref{chiidentities},
the contribution of the chiral diagram \eqref{crange3} 
can be captured within a coefficient of a chiral function
of lower range. From this example, we conclude that
contributions from diagrams of nonmaximal range
$R<K+1$ are encoded within the coefficients of chiral 
functions of even lower range $\kappa<R$. After making 
use of relations like \eqref{chiidentities}, the simplified chiral 
functions no longer contain the information about the number of 
chiral and antichiral vertices of the underlying diagrams,
and, in particular, one cannot recover whether the underlying diagrams 
are chiral. Relations such as \eqref{chiidentities} only occur between certain
chiral functions. They clearly complicate 
the subtraction procedure at orders $K\ge L+1$, compared to the 
one at critical order $K=L$. But, there are also simplifications,
as, e.g., the cancellations \eqref{cancelr4} among the overall UV 
divergences of certain next-to-maximum range diagrams at three loops.
They are universal and hold for diagrams with
$R=K$ and $\kappa=R-1$ at any loop order. At three loops, where
these diagrams show up for the first time, they can only 
involve the simplest chiral 
structure $\chi(1)$ that contains a single chiral vertex.
Because of the generalized finiteness conditions, this vertex has to remain
outside loop in order not to yield a finite result.
At higher loops, the next-to-maximum range diagrams diagrams can have 
more complicated chiral structures with chiral vertices involved in loops,
such that the respective finiteness condition is not matched.
In these cases, one should try to find universal cancellations 
similar to the ones in \eqref{cancelr4}.
In any case, a complete cancellation of the overall UV divergences of
all diagrams with $R=K$ and $\kappa=R-1$ is excluded, since already
at three loops one finds a contribution from 
the diagrams in \eqref{noncancelr4}.

\section*{Acknowledgements}

\noindent 
I am very grateful to Francesco Fiamberti and Alberto Santambrogio
for valuable discussions. 

\appendix

\section{Feynman rules in $\mathcal{N}=1$ superspace}
\label{app:Feynmanrules}

The gauge fixed action $S=S_{\text{gauge}}+S_{\FP}+S_{\text{matter}}$
of $\mathcal{N}=4$ SYM theory in $\mathcal{N}=1$ superspace
contains one real vector superfield $V$, three chiral superfields 
$\phi_i$, $i=1,2,3$ and ghost fields $c$ and $c'$. 
In the conventions of \cite{Gates:1983nr}, it reads
\begin{equation}\label{action}
\begin{aligned}
S_{\text{gauge}}
&=\frac{1}{g_\YM^2}\Big[
\frac{1}{2}\int\de^4x\de^2\theta\,\tr\big(W^\alpha W_\alpha\big)
-\frac{1}{\alpha}\int\de^4x\de^4\theta\,\tr\big((\D^2V)(\barD^2V)\big)
\Big]
\col\\
S_{\FP}
&=
\int\de^4x\de^4\theta\tr\big((c'+\bar c')\Ld_{\frac{1}{2}g_\YM V}(c+\bar c+\coth\Ld_{\frac{1}{2}g_\YM V}(c-\bar c))\big)
\col\\
S_{\text{matter}}
&=
\int\de^4x\de^4\theta\,
\tr\big(\e^{-g_\YM V} \bar \phi_i \e^{g_\YM V}\phi^i\big)\\
&\phantom{{}={}}
+i\frac{g_\YM}{3!}\int\de^4x\de^2\theta\,
\epsilon_{ijk}\,\tr \big(\phi^i\comm{\phi^j\,}{\,\phi^k}\big)
+i\frac{g_\YM}{3!}\int\de^4x\de^2\bar\theta\,
\epsilon^{ijk}\,\tr \big(\bar\phi_i\comm{\bar\phi_j\,}{\,\bar\phi_k}\big)
\col
\end{aligned}
\end{equation}
where $W_\alpha = i\barD^2 \left(\e^{-g_\YM V}\D_\alpha\,\e^{g_\YM V}\right)$, 
$\Ld_VX=\comm{V\,}{\,X}$. The fields decompose as
$V=V_aT^a$, $\phi^i=\phi_a^iT^a$, $c=c_aT^a$, and $c'=c'_aT^a$, where the 
generators $T^a$ satisfy the $SU(N)$ algebra
\begin{equation}
\comm{T_a\,}{\,T_b} = i f_{abc} T_c\col
\end{equation}
and they are normalized as
\begin{equation}
\tr(T_a T_b) =\delta_{ab}
\pnt
\end{equation}

We use the Wick rotated Feynman rules, i.e., we have transformed
$\e^{-iS}\to\e^S$ in the path integral.
In supersymmetric Fermi-Feynman gauge, where $\alpha=1+\mathcal{O}(g_\YM^2)$, 
the vector, chiral, and ghost propagators are given 
by \footnote{The corrections from the gauge parameter $\alpha$ do not 
appear in the diagrams explicitly considered in this paper.}
\begin{equation}\label{propagators}
\begin{aligned}
\langle\, V_a\,V_b\,\rangle
&=
\settoheight{\eqoff}{$\times$}%
\setlength{\eqoff}{0.5\eqoff}%
\addtolength{\eqoff}{-3.75\unitlength}%
\raisebox{\eqoff}{%
\fmfframe(1,0)(1,0){%
\begin{fmfchar*}(15,7.5)
\fmfleft{v1}
\fmfright{v2}
\fmfforce{0.0625w,0.5h}{v1}
\fmfforce{0.9375w,0.5h}{v2}
\fmf{photon}{v1,v2}
\fmffreeze
\fmfposition
\fmfipath{pm[]}
\fmfiset{pm1}{vpath(__v1,__v2)}
\nvml{1}{$\scriptstyle p$}
\end{fmfchar*}}}
=-\frac{\delta_{ab}}{p^2}\delta^4(\theta_1-\theta_2)
\col\\
\langle\,\phi_a^i\,\bar\phi_{j\,b}\,\rangle
&=
\settoheight{\eqoff}{$\times$}%
\setlength{\eqoff}{0.5\eqoff}%
\addtolength{\eqoff}{-3.75\unitlength}%
\raisebox{\eqoff}{%
\fmfframe(1,0)(1,0){%
\begin{fmfchar*}(15,7.5)
\fmfleft{v1}
\fmfright{v2}
\fmfforce{0.0625w,0.5h}{v1}
\fmfforce{0.9375w,0.5h}{v2}
\fmf{plain}{v1,v2}
\fmffreeze
\fmfposition
\fmfipath{pm[]}
\fmfiset{pm1}{vpath(__v1,__v2)}
\nvml{1}{$\scriptstyle p$}
\end{fmfchar*}}}
=
\delta_j^i\frac{\delta_{ab}}{p^2}\delta^4(\theta_1-\theta_2)
\col\\
\langle\, \bar c'_a\,c_b\,\rangle
=-\langle\, c'_a\,\bar c_b\,\rangle
&=
\settoheight{\eqoff}{$\times$}%
\setlength{\eqoff}{0.5\eqoff}%
\addtolength{\eqoff}{-3.75\unitlength}%
\raisebox{\eqoff}{%
\fmfframe(1,0)(1,0){%
\begin{fmfchar*}(15,7.5)
\fmfleft{v1}
\fmfright{v2}
\fmfforce{0.0625w,0.5h}{v1}
\fmfforce{0.9375w,0.5h}{v2}
\fmf{dots}{v1,v2}
\fmffreeze
\fmfposition
\fmfipath{pm[]}
\fmfiset{pm1}{vpath(__v1,__v2)}
\nvml{1}{$\scriptstyle p$}
\end{fmfchar*}}}
=\frac{\delta_{ab}}{p^2}\delta^4(\theta_1-\theta_2)
\pnt
\end{aligned}
\end{equation}
The cubic gauge vertex is given by
\begin{equation}
\begin{aligned}\label{V3vertex}
V_{V^3}
&=
\Bigg(
\cvert{photon}{photon}{photon}{}{$\scriptstyle\D^\alpha$}{$\scriptstyle\barD^2\!\scriptstyle\D_{\!\alpha}$}
-
\cvert{photon}{photon}{photon}{}{$\scriptstyle\barD^2\!\scriptstyle\D_{\!\alpha}$}{$\scriptstyle\D^\alpha$}
+\cvert{photon}{photon}{photon}{$\scriptstyle\D_{\!\alpha}\!\scriptstyle\barD^2$}{}{$\scriptstyle\D^\alpha$}
-\cvert{photon}{photon}{photon}{$\scriptstyle\D_{\!\alpha}\!\scriptstyle\barD^2$}{$\scriptstyle\D^\alpha$}{}
+\cvert{photon}{photon}{photon}{$\scriptstyle\D^\alpha$}{$\scriptstyle\barD^2\!\scriptstyle\D_{\!\alpha}$}{}
-\cvert{photon}{photon}{photon}{$\scriptstyle\D^\alpha$}{}{$\scriptstyle\barD^2\!\scriptstyle\D_{\!\alpha}$}
\Bigg)
\frac{g_\YM }{2}\tr\big(T^a\comm{T^b\,}{\,T^c}\big)\col
\end{aligned}
\end{equation}
where the color indices are labeled $(a,b,c)$ clockwise.
The $\D$-algebra has to be performed for all six permutations of the structure 
of the covariant derivatives at its legs.
The only purpose of the vertices that appear on the right-hand side 
of the equation is to display this structure. They do not contain any other 
nontrivial factors.

The cubic gauge-matter vertices and gauge-ghost vertices
are given by
\begin{equation}\label{cvertices}
\begin{gathered}
\begin{aligned}
V_{\bar\phi_j V\phi^i}
&=
\cvert{photon}{plain}{plain}{}{$\scriptstyle\barD^2$}{$\scriptstyle\D^2$}
g_\YM\delta_i^j\tr\big(T^a\comm{T^b\,}{\,T^c}\big)
\col\\
V_{\phi^i\phi^j\phi^k}
&=\cvert{plain}{plain}{plain}{}{$\scriptstyle\barD^2$}{$\scriptstyle\barD^2$}
ig_\YM\epsilon_{ijk}\tr\big(T^a\comm{T^b\,}{\,T^c}\big)
\col\quad
&V_{\bar\phi_i\bar\phi_j\bar\phi_k}
&=\cvert{plain}{plain}{plain}{}{$\scriptstyle\D^2$}{$\scriptstyle\D^2$}
ig_\YM\epsilon^{ijk}\tr\big(T^a\comm{T^b\,}{\,T^c}\big)
\col\\
V_{Vcc'}
&=
\cvert{photon}{dots}{dots}{}{$\scriptstyle\barD^2$}{$\scriptstyle\barD^2$}\frac{g_\YM}{2}\tr\big(T^a\comm{T^b\,}{\,T^c}\big)
\col\qquad
&V_{V\bar c\bar c'}
&=
\cvert{photon}{dots}{dots}{}{$\scriptstyle\D^2$}{$\scriptstyle\D^2$}\frac{g_\YM}{2}\tr\big(T^a\comm{T^b\,}{\,T^c}\big)
\col\\
V_{Vc\bar c'}
&=
\cvert{photon}{dots}{dots}{}{$\scriptstyle\barD^2$}{$\scriptstyle\D^2$}\frac{g_\YM}{2}\tr\big(T^a\comm{T^b\,}{\,T^c}\big)
\col\qquad
&V_{V\bar cc'}
&=
\cvert{photon}{dots}{dots}{}{$\scriptstyle\D^2$}{$\scriptstyle\barD^2$}\frac{g_\YM}{2}\tr\big(T^a\comm{T^b\,}{\,T^c}\big)
\col
\end{aligned}
\end{gathered}
\end{equation}
where the color indices are labeled $(a,b,c)$ clockwise, 
starting with the leg to the lower-left.

For the three-loop renormalization in the flavor $SU(2)$ subsector, we only
need some of the quartic vertices. They read
\begin{equation}\label{qvertices}
\begin{aligned}
V_{V^2\bar\phi_i\phi^j}
&=
\qvert{photon}{photon}{plain}{plain}{}{}{$\scriptstyle\D^2$}{$\scriptstyle\barD^2$}\frac{g_\YM^2}{2}\delta_j^i\big[\tr\big(T^aT^bT^cT^d\big)+\tr\big(T^bT^aT^cT^d\big)\big]
\col\\
V_{V^2\phi^i\bar\phi_j}
&=
\qvert{photon}{photon}{plain}{plain}{}{}{$\scriptstyle\barD^2$}{$\scriptstyle\D^2$}\frac{g_\YM^2}{2}\delta_i^j\big[\tr\big(T^aT^bT^cT^d\big)+\tr\big(T^bT^aT^cT^d\big)\big]
\col\\
V_{V\bar\phi_iV\phi^j}
&=
\qvert{photon}{plain}{photon}{plain}{}{$\scriptstyle\D^2$}{}{$\scriptstyle\barD^2$}(-g_\YM^2)\delta_j^i\tr\big(T^aT^bT^cT^d\big)
\col
\end{aligned}
\end{equation}
where the color indices are labeled $(a,b,c,d)$ clockwise 
starting with the leg in the lower left corner.

\section{The power of power counting}
\label{app:powercounting}

In this Appendix, we derive conditions for the finiteness of 
superfield Feynman diagrams that yield loop corrections
for chiral composite operators.
Compared to \cite{Fiamberti:2008sh}, we simplify the derivation and 
generalize the result, predicting here 
the finiteness of larger classes of diagrams.
The considerations are based on power counting and general arguments 
and hold for $\mathcal{N}=4$ SYM theory and, e.g., also for its 
$\beta$-deformation, in terms of $\mathcal{N}=1$ superfields 
in Fermi-Feynman gauge. 
We discuss, in parallel, the case of $\mathcal{N}=6$ Chern-Simons theory 
in an $\mathcal{N}=2$ superfield formalism
\cite{Benna:2008zy,Akerblom:2009gx,Bianchi:2009ja} 
in supersymmetric Landau gauge. The resulting finiteness
conditions are the same, but, due to the different $\D$-algebra structure 
of the propagators, their implications slightly differ from the ones in the
SYM case.


In the considered theories, the order in the coupling constant $k$
can be obtained from the number $v_i$ of elementary vertices with $i$ legs as
\begin{equation}\label{kinv}
k=\sum_{i\ge3}(i-2)v_i
\pnt
\end{equation}
A nonamputated Feynman diagram of order $k$, with $l$ loops, $c$
connected pieces, $e$ external legs, $p$ propagators of which $e$ are external,
$v$ elementary vertices, and $n_{\D}$ and $n_{\barD}$ spinor derivatives
$\D_\alpha$ and $\barD_{\dot \alpha}$ in $\mathcal{N}=1$ superspace in four 
dimensions or $\D_\alpha$ and $\barD_{\alpha}$ in $\mathcal{N}=2$ superspace
in three dimensions, obeys the following relations:
\begin{equation}\label{subdiagrel}
c=v-p+e+l\col\qquad
e=k-2(l-c)\col\qquad 
n_{\D}+n_{\barD}
=\begin{cases}
4v\col & \mathcal{N}=4\text{ SYM}\\
k+4v+2e_V\col & \mathcal{N}=6\text{ CS}
\end{cases}
\col
\end{equation}
where $e_V$ is the number of external vector propagators (that contain
$\D^\alpha\bar\D_\alpha$ in Landau gauge).
The last equation reflects that, in $\mathcal{N}=4$ SYM theory,
each vertex contributes exactly four spinor derivatives. This holds
also for the vertices at higher order, involving increasing numbers of 
vector fields. In the case of $\mathcal{N}=6$ Chern-Simons theory, 
the last equation is obtained from the relations
\begin{equation}
\begin{aligned}
n_{\D}&=6v_{\bar\phi^4}+2\sum_{i\ge1}v_{\phi V^i\bar\phi}
+\sum_{i\ge3}v_{V^i}+p_V\col\\
n_{\barD}&=6v_{\phi^4}+2\sum_{i\ge1}v_{\phi V^i\bar\phi}
+\sum_{i\ge3}v_{V^i}+p_V\col\\
p_V&=\sum_{i\ge1}iv_{\phi V^i\bar\phi}+\sum_{i\ge3}iv_{V^i}+e_V
=k-2(v_{\phi^4}+v_{\bar\phi^4})+2\sum_{i\ge1}v_{\phi V^i\bar\phi}
+2\sum_{i\ge3}v_{V^i}+e_V
\col
\end{aligned}
\end{equation}
where the last relation for $p_V$ follows by using \eqref{kinv}.

The relations \eqref{subdiagrel} have to be modified if a composite operator is
part of the Feynman diagram, but they directly hold for the
subdiagram, which is obtained after cutting out the composite operator.
In the flavor $SU(2)$ subsector of $\mathcal{N}=4$ SYM and 
the flavor $SU(2)\times SU(2)$ subsector of $\mathcal{N}=6$ CS theory, 
a respective diagram should have even $e$ and $e_V=0$, and, moreover,
half of the external legs should be chiral and antichiral.
The number of chiral and antichiral vertices in the subdiagram
is then equal.
This implies $n_{\D}=n_{\barD}$.

As a next step, we write down the relations for the full diagram, including an
operator composed of $L$ chiral superfields, of which only $R=\frac{e}{2}\le L$
neighboring fields are contracted with the subdiagram. The remaining fields
in the composite operator do not interact but become additional external 
lines. They need not be considered in the power counting for the full 
diagram; i.e., we can replace the operator by one which has only $R$ legs. It
contains $2(R-1)$ spinor derivatives $\barD$ to impose the chirality 
constraint.
The following relations then hold
\begin{equation}
V=v+1\col\qquad P=p\col\qquad E=e-R\col\qquad 
N_{\D}=n_{\D}\col\qquad N_{\barD}=n_{\barD}+2(R-1)
\col
\end{equation}
where the capital variables refer to the full graph with the (shortened) 
composite operator included.

For the determination of the superficial degree of divergence, the number of 
propagators which appear in loops is relevant and not the total number of 
propagators. We hence first have to amputate the diagram and then also get 
rid of further propagators which do not appear in any loops. 
For this purpose, we introduce $v_0$ as the number of chiral 
vertices that do not appear in any
loops \footnote{For diagrams in the flavor $SU(2)$ or, respectively,
$SU(2)\times SU(2)$ subsector, all antichiral
  vertices have to appear in loops. The tree structures in which an
  antichiral vertex is not part of any loop cannot exist in these
  subsectors.} and denote by $r_{\D}$ and $r_{\barD}$ the number
of the respective spinor derivatives $\D$ and $\barD$ that, through the 
$\D$-algebra manipulations, are brought outside of the loops.
We then obtain for the number of propagators and spinor derivatives that
appear in loops, the respective relations
\begin{equation}
P_{\text{L}}=p-E-v_0
\col\qquad
N_{\text{L}\D}=n_{\D}-2v_0-r_{\D}\col\qquad 
N_{\text{L}\barD}=n_{\barD}-2-r_{\barD}
\col
\end{equation}
where the first equality is only true for a one-particle-irreducible
(1PI) connection to the
composite operator. In particular, it does not hold for the diagrams 
that contribute to the self-energy 
of the external lines. The second
equality relies on the fact that, at each antichiral vertex that is
involved in only a single loop, one can always place two covariant 
derivatives $\D_\alpha$ at the line that is not part of the loop.
The last equality follows from the fact that the chiral composite operator 
contains only $2(R-1)$ spinor derivatives $\barD$, but the initial 
configuration of derivatives for the full graph can be chosen such that $2R$ 
of the $\barD$ do not appear in loops. 
After $\D$-algebra, the spinor derivatives remaining inside the loops 
generate the following number of spacetime derivatives, 
i.e., factors of the loop momenta, in the numerators of the resulting loop 
integrals:
\begin{equation}
N_{\text{L}\partial}=\min(N_{\text{L}\D},N_{\text{L}\barD})-2L
\pnt
\end{equation}
Thereby, we have taken into account that two $\D$ and two $\barD$ in each loop 
are not transformed into spacetime derivatives but are required to 
obtain a nonzero result \cite{Gates:1983nr}.
A $K$-loop integral with $P_\text{L}$ (scalar) propagators and
$N_{\text{L}\partial}$ momenta 
in the numerators is superficially UV divergent if the following relation holds:
\begin{equation}
DK-2P_{\text{L}}+N_{\text{L}\partial}\ge0
\pnt
\end{equation}
With $e=2R$, $k=2K$, $n_{\D}=n_{\barD}$, and the first two relations in
\eqref{subdiagrel}, the above two equations can be combined and 
rephrased as
\begin{equation}
(D-4)K-2v+2v_0+n_{\D}-\max(2v_0+r_{\D},2+r_{\barD})\ge0\pnt
\end{equation}
Using that the value of $n_{\D}$ is one half of the sum given 
in the last relation in \eqref{subdiagrel}, we find, with 
$D=4$ for the $\mathcal{N}=4$ SYM theory and $D=3$ and $e_V=0$ for the 
$\mathcal{N}=6$ CS theory, the following necessary 
universal condition:
\begin{equation}
2v_0\ge\max(2v_0+r_{\D},2+r_{\barD})
\end{equation}
for obtaining a loop integral with overall UV divergence at the end of the 
$\D$-algebra manipulations.
It constrains the parameters as
\begin{equation}
\begin{aligned}
v_0\ge1\col\qquad r_{\D}=0\col\qquad r_{\barD}\le 2(v_0-1)\pnt
\end{aligned}
\end{equation}
This leads to the following finiteness conditions:\\
\emph{
Any Feynman diagram of $\mathcal{N}=4$ SYM theory in $\mathcal{N}=1$ 
superspace in Fermi-Feynman gauge
or of $\mathcal{N}=6$ CS theory in $\mathcal{N}=2$ superspace in Landau gauge
that could contribute to loop corrections of a chiral composite operator 
in the respective flavor $SU(2)$ or $SU(2)\times SU(2)$ subsectors
and with an interaction range $R\ge2$ has no overall UV divergence 
if at least one of the following criteria is matched
\footnote{$R\ge 2$ means that the composite operator is 1PI connected with the rest of the diagram, not including the noninteracting fields of the operator. This excludes diagrams in which one or more fields of the composite operator only involve self-energy corrections.}:
\begin{enumerate}
\item All of its chiral vertices are part of any loop.
\item One of its spinor derivatives $\D$ is brought outside the loops.
\item The number of its spinor derivatives $\barD$ brought outside 
loops becomes equal or larger than twice the number of chiral vertices 
that are not part of any loop.
\end{enumerate}
}
In the flavor $SU(2)$ or $SU(2)\times SU(2)$ subsectors, a chiral vertex that 
is not part of any loop always generates flavor permutations and thus a
nontrivial chiral structure of the diagram.
The above finiteness conditions hence imply the following rule:\\
\emph{All diagrams with an interaction range $R\ge2$ and trivial chiral structure
$\chi()$ have no overall UV divergence.}

In the case of $\mathcal{N}=6$ CS theory,
the finiteness conditions imply the following statement:\\
\emph{
A diagram with interaction range $R\ge2$ has no overall UV divergence
if it contains at least one cubic gauge-matter interaction at which the
chiral field line is not part of any loop. In particular,
if, in the diagram, exactly one of the chiral vertices appears outside
the loops, then it also has no overall UV divergence if the antichiral 
field of at least one cubic gauge-matter interaction is not part of any loop.}
\\
This statement relies on the fact that, in Landau gauge,
the vector field propagators carry $\D^\alpha\barD_{\alpha}$.
At the designated cubic gauge-matter vertices, at least one of 
them could only be moved outside the 
loops, matching at least one criterion of the previously formulated 
finiteness conditions.

\section{One- and two-loop subdiagrams}
\label{app:subdiag}

In this Appendix, we derive expressions for the one- and two-loop 
planar subdiagrams that appear in the three-loop calculation. 
We sum up all diagrams that contribute to the individual substructures
and partially perform the $\D$-algebra manipulations in order to obtain 
expressions that are local in the fermionic coordinates of superspace.
Locality in the fermionic coordinates is displayed by filling out gray the 
loops of the resulting integrals over the bosonic coordinates.
The prefactors of all vertices and propagators that are parts of the loops
are considered in the prefactors of the final results.
We omit factors of color traces that are identical to the color factors of the 
respective tree-level diagrams.

\subsection{One-loop subdiagrams}
\label{app:oneloopsubdiag}

The one-loop correction to the chiral vertex is easily evaluated as
\begin{equation}
\begin{aligned}\label{ccconeloop}
\cVat[\fmfcmd{fill fullcircle scaled 10 shifted vloc(__vg1) withcolor black ;}
\fmfiv{label=\small$\textcolor{white}{1}$,l.dist=0}{vloc(__vg1)}]
{plain}{plain}{plain}{phantom}{plain,ptext.in=$\scriptstyle\barD^2$}{plain,ptext.in=$\scriptstyle\barD^2$}
&=
\cVat
{plain,ptext.in=$\scriptstyle\barD^2$}{plain,ptext.in=$\scriptstyle\barD^2$,ptext.out=$\scriptstyle\D^2$}{plain,ptext.out=$\scriptstyle\D^2$}{photon}{plain,ptext.in=$\scriptstyle\barD^2$}{plain,ptext.in=$\scriptstyle\barD^2$}
+\dots
=
\left(\cVat[\fmfiset{p10}{p4--p6--p5--cycle}\fmfcmd{fill p10 withcolor \mympostgrey ;}]
{plain,l.side=left,l.dist=3,label=$\scriptstyle\Box$}{plain}{plain}{plain}{plain,ptext.in=$\scriptstyle\barD^2$}{plain,ptext.in=$\scriptstyle\barD^2$}
+\dots\right)i\lambda g_\YM\epsilon_{ijk}
\col
\end{aligned}
\end{equation}
where the ellipses denote the remaining two diagrams obtained by 
cyclic permutations of the external legs, and $\Box$ cancels the respective
propagator, thereby producing a minus.

The one-loop correction to the cubic gauge-matter vertex is
given by 
\begin{equation}
\begin{aligned}
\cVat[\fmfcmd{fill fullcircle scaled 10 shifted vloc(__vg1) withcolor black ;}
\fmfiv{label=\small$\textcolor{white}{1}$,l.dist=0}{vloc(__vg1)}]
{photon}{plain}{plain}{phantom}{plain}{plain}
&=
\cVat
{photon}{photon}{photon}{plain}{plain}{plain}
+
\cVat
{photon}{plain}{plain}{photon}{plain}{plain}
+
\cVat
{photon}{plain}{plain}{plain}{plain}{plain}
\\
&\phantom{{}={}}
+
\cVab{2}{-0.75}{}{$ $}{$ $}{ }
+
\cVab{2}{0.75}{}{$ $}{$ $}{ }\\
&\phantom{{}={}}
+
\cVab{3}{-0.75}{}{$ $}{$ $}{ }
+
\cVab{3}{0.75}{}{$ $}{$ $}{ }
%
\col
\end{aligned}
\end{equation}
where we have omitted the covariant derivatives. 
In the first diagram, we have to  consider the six configurations 
of the covariant derivatives at the cubic gauge vertex \eqref{V3vertex}.
Working out the $\D$-algebra for them, we find
\begin{equation}
\begin{aligned}
\cVat
{photon}{photon}{photon}{plain}{plain,ptext.in=$\scriptstyle\barD^2$}{plain,ptext.in=$\scriptstyle\D^2$}
&=\left(
\cVat[\fmfiset{p10}{p4--p6--p5--cycle}\fmfcmd{fill p10 withcolor \mympostgrey ;}]
{photon,ptext.in=$\hspace{0.4cm}\scriptstyle\D^\alpha\!\scriptstyle\barD^2\!\scriptstyle\D_{\!\alpha}$}{plain}{plain}{plain}{plain,ptext.in=$\scriptstyle\barD^2$}{plain,ptext.in=$\scriptstyle\D^2$}
+
\cVat[\fmfiset{p10}{p4--p6--p5--cycle}\fmfcmd{fill p10 withcolor \mympostgrey ;}]
{photon,ptext.out=$\scriptstyle\comm{\smash{\D_{\!\alpha}}}{\smash{\barD_{\!\dot\beta}}}\hspace{0.5cm}$}{plain}{plain}{derplain,label=$\scriptstyle l^{\alpha\dot\beta}$,l.side=left,l.dist=3}{plain,ptext.in=$\scriptstyle\barD^2$}{plain,ptext.in=$\scriptstyle\D^2$}
\right)\frac{\lambda}{2}g_\YM\delta_j^i
\pnt
\end{aligned}
\end{equation}
The covariant derivatives and also momenta are read off 
when leaving the vertices. The above graphical representation is, therefore, 
translated into the
following algebraic expression in the numerator of the respective 
loop integral:
\begin{equation}
\frac{\lambda}{2}\big(\D^\alpha\barD^2\D_{\alpha} 
+l^{\alpha\dot\beta}\comm{\smash{\barD_{\dot\beta}}\,}{\,\smash{D_\alpha}}\big)V(p_1)\barD^2\phi_i(p_2)\D^2\bar\phi^i(p_3)\col
\end{equation} 
where the covariant derivatives act to the right only on the first field 
that follows them,
and we have suppressed the dependence on the fermionic coordinates. 
In order to correctly apply the procedure of \cite{Gates:1983nr} that 
determines the sign coming from changing the order of the covariant 
derivatives, one
always has to start from an expression with all the indices in canonical 
order; i.e., one has to use $l^{\alpha\dot\beta}$ and $l^{\dot\beta\alpha}$
in the expressions coming, respectively, from the first and 
the second term of the commutator and then apply the procedure 
described in \cite{Gates:1983nr}.

The other two contributions involving only cubic vertices evaluate to
\begin{equation}
\begin{aligned}
\cVat
{photon}{plain,ptext.in=$\scriptstyle\barD^2$,ptext.out=$\scriptstyle\D^2$}{plain,ptext.in=$\scriptstyle\D^2$,ptext.out=$\scriptstyle\barD^2$}{photon}{plain,ptext.in=$\scriptstyle\barD^2$}{plain,ptext.in=$\scriptstyle\D^2$}
&=
\left(
\cVat[\fmfiset{p10}{p4--p6--p5--cycle}\fmfcmd{fill p10 withcolor \mympostgrey ;}]
{photon,ptext.in=$\hspace{0.4cm}\scriptstyle\barD^2\!\scriptstyle\D^2$}{plain}{plain}{plain}{plain,ptext.in=$\scriptstyle\barD^2$}{plain,ptext.in=$\scriptstyle\D^2$}
-
\cVat[\fmfiset{p10}{p4--p6--p5--cycle}\fmfcmd{fill p10 withcolor \mympostgrey ;}]
{photon,ptext.in=$\hspace{0.4cm}\scriptstyle\barD_{\!\dot\beta}\D_{\!\alpha}$}{plain}{derplain,label=$\scriptstyle (p_3-l)^{\alpha\dot\beta}$,l.side=right,l.dist=3}{plain}{plain,ptext.in=$\scriptstyle\barD^2$}{plain,ptext.in=$\scriptstyle\D^2$}
+
\cVat[\fmfiset{p10}{p4--p6--p5--cycle}\fmfcmd{fill p10 withcolor \mympostgrey ;}]
{photon}{plain}{plain,label=$\scriptstyle\Box$,l.side=right,l.dist=3}{plain}{plain,ptext.in=$\scriptstyle\barD^2$}{plain,ptext.in=$\scriptstyle\D^2$}
\right)(-\lambda) g_\YM\delta_j^i
\col\\
\cVat
{photon}{plain,ptext.in=$\scriptstyle\D^2$,ptext.out=$\scriptstyle\barD^2$}{plain,ptext.in=$\scriptstyle\barD^2$,ptext.out=$\scriptstyle\D^2$}{plain}{plain,ptext.in=$\scriptstyle\barD^2$}{plain,ptext.in=$\scriptstyle\D^2$}
&=
\left(\cVat[\fmfiset{p10}{p4--p6--p5--cycle}\fmfcmd{fill p10 withcolor \mympostgrey ;}]
{photon,ptext.in=$\hspace{0.4cm}\scriptstyle\barD^2\!\scriptstyle\D^2$}{plain}{plain}{plain}{plain,ptext.in=$\scriptstyle\barD^2$}{plain,ptext.in=$\scriptstyle\D^2$}
-
\cVat[\fmfiset{p10}{p4--p6--p5--cycle}\fmfcmd{fill p10 withcolor \mympostgrey ;}]
{photon,ptext.in=$\hspace{0.4cm}\scriptstyle\barD_{\!\dot\beta}\D_{\!\alpha}$}{derplain,label=$\scriptstyle (l+p_2)^{\alpha\dot\beta}$,l.side=left,l.dist=3}{plain}{plain}{plain,ptext.in=$\scriptstyle\barD^2$}{plain,ptext.in=$\scriptstyle\D^2$}
+\cVat[\fmfiset{p10}{p4--p6--p5--cycle}\fmfcmd{fill p10 withcolor \mympostgrey ;}]
{photon}{plain,label=$\scriptstyle\Box$,l.side=left,l.dist=3}{plain}{plain}{plain,ptext.in=$\scriptstyle\barD^2$}{plain,ptext.in=$\scriptstyle\D^2$}
\right)(-2\lambda) g_\YM\delta_j^i
\pnt
\end{aligned}
\end{equation}
The remaining contributions containing a quartic gauge-matter vertex 
are determined as
\begin{equation}
\begin{aligned}
\cVab{2}{-0.75}{}{$\scriptstyle\barD^2$}{$\scriptstyle\D^2$}{ptext.in=$\scriptstyle\barD^2$,ptext.out=$\scriptstyle\D^2$}
&=
\cVat[\fmfiset{p10}{p4--p6--p5--cycle}\fmfcmd{fill p10 withcolor \mympostgrey ;}]
{photon}{plain}{plain,label=$\scriptstyle\Box$,l.side=right,l.dist=3}{plain}{plain,ptext.in=$\scriptstyle\barD^2$}{plain,ptext.in=$\scriptstyle\D^2$}
\frac{\lambda}{2}g_\YM\delta_j^i
\col
&\cVab{2}{0.75}{}{$\scriptstyle\barD^2$}{$\scriptstyle\D^2$}{ptext.hin=2,ptext.hout=2,ptext.in=$\scriptstyle\barD^2$,ptext.out=$\scriptstyle\D^2$}
&=
\cVat[\fmfiset{p10}{p4--p6--p5--cycle}\fmfcmd{fill p10 withcolor \mympostgrey ;}]
{photon}{plain}{plain,label=$\scriptstyle\Box$,l.side=right,l.dist=3}{plain}{plain,ptext.in=$\scriptstyle\barD^2$}{plain,ptext.in=$\scriptstyle\D^2$}
\lambda g_\YM\delta_j^i\col\\
\cVab{3}{-0.75}{}{$\scriptstyle\barD^2$}{$\scriptstyle\D^2$}{ptext.in=$\scriptstyle\barD^2$,ptext.out=$\scriptstyle\D^2$}
&=
\cVat[\fmfiset{p10}{p4--p6--p5--cycle}\fmfcmd{fill p10 withcolor \mympostgrey ;}]
{photon}{plain,label=$\scriptstyle\Box$,l.side=left,l.dist=3}{plain}{plain}{plain,ptext.in=$\scriptstyle\barD^2$}{plain,ptext.in=$\scriptstyle\D^2$}
\frac{\lambda}{2}g_\YM\delta_j^i
\col
&\cVab{3}{0.75}{}{$\scriptstyle\barD^2$}{$\scriptstyle\D^2$}{ptext.hin=2,ptext.hout=2,ptext.in=$\scriptstyle\barD^2$,ptext.out=$\scriptstyle\D^2$}
&=
\cVat[\fmfiset{p10}{p4--p6--p5--cycle}\fmfcmd{fill p10 withcolor \mympostgrey ;}]
{photon}{plain,label=$\scriptstyle\Box$,l.side=left,l.dist=3}{plain}{plain}{plain,ptext.in=$\scriptstyle\barD^2$}{plain,ptext.in=$\scriptstyle\D^2$}
\lambda g_\YM\delta_j^i
\col
\end{aligned}
\end{equation}
where we have inserted $-\frac{\Box}{p^2}=1$ in order 
to obtain triangle integrals.
We sum up the above expressions and simplify the numerator. 
The terms in which the 
loop momentum $l$ is contracted with covariant derivatives 
combine and yield an 
anticommutator of these derivatives that can be replaced by the momentum $p_1=-p_2-p_3$ of the vector field they act on. After this step, the dependence on the loop momentum cancels out. The remaining terms simplify further by making 
use of the identities
\begin{equation}
\begin{aligned}
p_2^2-p_3^2=-\frac{1}{2}(p_2-p_3)^{\alpha\dot\beta}\acomm{\smash{\D_\alpha}}{\smash{\barD_{\dot\beta}}}\col\qquad
\D^\alpha\barD^2\D_\alpha=p_1^{\alpha\dot\beta}\D_\alpha\barD_{\dot\beta}+2\D^2\barD^2
\col
\end{aligned}
\end{equation}
where
$p_1$ is the momentum of the field the covariant derivatives act on.
The expression for the one-loop correction of the cubic gauge-matter vertex
can then be cast into the form
\begin{equation}
\begin{aligned}\label{cvaoneloop}
\cVat[\fmfcmd{fill fullcircle scaled 10 shifted vloc(__vg1) withcolor black ;}
\fmfiv{label=\small$\textcolor{white}{1}$,l.dist=0}{vloc(__vg1)}]
{photon}{plain}{plain}{phantom}{plain,ptext.in=$\scriptstyle\barD^2$}{plain,ptext.in=$\scriptstyle\D^2$}
&=\left(-
\cVat[\fmfiset{p10}{p4--p6--p5--cycle}\fmfcmd{fill p10 withcolor \mympostgrey ;}]
{photon,ptext.in=$\hspace{0.4cm}\scriptstyle\D^\alpha\!\scriptstyle\barD^2\!\scriptstyle\D_{\!\alpha}$}{plain}{plain}{plain}{plain,ptext.in=$\scriptstyle\barD^2$}{plain,ptext.in=$\scriptstyle\D^2$}
-\frac{1}{4}
\cVat[\fmfiset{p10}{p4--p6--p5--cycle}\fmfcmd{fill p10 withcolor \mympostgrey ;}]
{photon,ptext.out=$\scriptstyle\comm{\smash{\D_{\!\alpha}}}{\smash{\barD_{\!\dot\beta}}}\hspace{0.5cm}$}{plain}{plain}{plain}{derplain,ptext.in=$\scriptstyle\barD^2$,label=$\scriptstyle p_2^{\alpha\dot\beta}$,l.side=right,l.dist=0.5}{plain,ptext.in=$\scriptstyle\D^2$}
+\frac{1}{4}
\cVat[\fmfiset{p10}{p4--p6--p5--cycle}\fmfcmd{fill p10 withcolor \mympostgrey ;}]
{photon,ptext.out=$\scriptstyle\comm{\smash{\D_{\!\alpha}}}{\smash{\barD_{\!\dot\beta}}}\hspace{0.5cm}$}{plain}{plain}{plain}{plain,ptext.in=$\scriptstyle\barD^2$}{derplain,ptext.in=$\scriptstyle\D^2$,label=$\scriptstyle p_3^{\alpha\dot\beta}$,l.side=left,l.dist=2}
\,\,\right)\lambda g_\YM\delta_j^i
\pnt
\end{aligned}
\end{equation}

In the following, we will show that the sum of the diagrams in the last 
equation of \eqref{cancelr4} has no overall UV divergence. 
The finiteness conditions of Appendix
\ref{app:powercounting} guarantee that overall UV divergences can only appear
if all covariant derivatives $\D_{\alpha}$ remain within the loops of
these diagrams.
By partial integration, the factor $\D^2$ can then be transferred to act
on the vector field of the diagrams in \eqref{cancelr4}.
Here, we evaluate the appearing substructures with such a factor $\D^2$ 
acting on the vector field and show that their sum yields zero.
 
After $\D$-algebra, we find the following results for the 
individual contributions that contain the cubic gauge vertex \eqref{V3vertex}
or the one-loop correction of the cubic gauge-matter vertex
\eqref{cvaoneloop}:
\begin{equation}\label{cccVdiags1}
\begin{aligned}
\cccV
{plain,ptext.in=$\scriptstyle\D^2$}{plain,ptext.out=$\scriptstyle\D^2$}{photon,ptext.out=$\scriptstyle\D^2$}{plain,ptext.out=$\scriptstyle\D^2$}{plain,ptext.in=$\scriptstyle\D^2$,ptext.out=$\scriptstyle\barD^2$}{photon}{photon}{plain,ptext.out=$\scriptstyle\barD^2$}
&=\left(
\cccV[\fmfcmd{fill p5--p6--p7--p8--cycle withcolor \mympostgrey ;}]{plain,l.side=left,l.dist=3,label=$\scriptstyle\Box$}{plain,ptext.out=$\scriptstyle\D^2$}{photon,ptext.out=$\scriptstyle\D^2$}{plain,ptext.out=$\scriptstyle\D^2$}{plain}{plain}{plain,l.side=right,l.dist=2,label=$\scriptstyle\Box$}{plain}
-
\cccV[\fmfcmd{fill p5--p6--p7--p8--cycle withcolor \mympostgrey ;}]{plain,l.side=left,l.dist=3,label=$\scriptstyle\Box$}{plain,ptext.out=$\scriptstyle\D^2$}{photon,ptext.out=$\scriptstyle\D^2$}{plain,ptext.out=$\scriptstyle\D^2$}{plain}{plain,l.side=right,l.dist=2,label=$\scriptstyle\Box$}{plain}{plain}
+
\cccV[\fmfcmd{fill p5--p6--p7--p8--cycle withcolor \mympostgrey ;}]{plain}{plain,ptext.out=$\scriptstyle\D^2$}{photon,l.side=right,l.dist=3,label=$\scriptstyle\Box$,ptext.out=$\scriptstyle\D^2$}{plain,ptext.out=$\scriptstyle\D^2$}{plain,l.side=right,l.dist=2,label=$\scriptstyle\Box$}{plain}{plain}{plain}
-
\cccV[\fmfcmd{fill p5--p6--p7--p8--cycle withcolor \mympostgrey ;}]{plain}{plain,ptext.out=$\scriptstyle\D^2$}{photon,l.side=right,l.dist=3,label=$\scriptstyle\Box$,ptext.out=$\scriptstyle\D^2$}{plain,ptext.out=$\scriptstyle\D^2$}{plain}{plain}{plain}{plain,l.side=left,l.dist=2,label=$\scriptstyle\Box$}
\right.\\
&\hphantom{{}={}\Bigg({}\,}
\left.
-2
\cccV[\fmfcmd{fill p5--p6--p7--p8--cycle withcolor \mympostgrey ;}]{plain}{plain,ptext.out=$\scriptstyle\D^2$}{derphoton,ptext.out=$\scriptstyle\D^2$}{derplain,ptext.out=$\scriptstyle\D^2$}{plain,l.side=right,l.dist=3,label=$\scriptstyle\Box$}{plain}{plain}{plain}
+2
\cccV[\fmfcmd{fill p5--p6--p7--p8--cycle withcolor \mympostgrey ;}]{plain}{derplain,ptext.out=$\scriptstyle\D^2$}{derphoton,ptext.out=$\scriptstyle\D^2$}{plain,ptext.out=$\scriptstyle\D^2$}{plain}{plain}{plain}{plain,l.side=left,l.dist=2,label=$\scriptstyle\Box$}
\right)i\frac{\lambda}{2}g_\YM^2\epsilon_{ijk}
\col\\
\settoheight{\eqoff}{$\times$}%
\setlength{\eqoff}{0.5\eqoff}%
\addtolength{\eqoff}{-13\unitlength}%
\raisebox{\eqoff}{%
\fmfframe(2,1)(2,1){%
\begin{fmfchar*}(24,24)
\fmftop{v4}
\fmfleft{v1}
\fmfbottom{v2}
\fmfright{v3}
\fmfforce{(0.5w,h)}{v4}
\fmfforce{(0,0.5h)}{v1}
\fmfforce{(w,0.5h)}{v3}
\fmfforce{(0.5w,0)}{v2}
\fmffixed{(0,0.5h)}{vs2,vs3}
\fmffixed{(0,whatever)}{vs1,vs2}
\fmffixed{(0,whatever)}{vs2,v2}
\fmf{plain}{vs1,v1}
\fmf{plain}{vs2,v2}
\fmf{plain}{vs3,v4}
\fmf{plain}{vs1,vs3}
\fmf{plain}{vs1,vs2}
\fmffreeze
\fmf{phantom}{vs2,v3}
\fmf{photon}{vs3,v3}
\fmffreeze
\fmfposition
\fmfipath{p[],pca}
\fmfipair{vm[],vu[],vd[]}
\fmfiset{p1}{vpath(__vs1,__v1)}
\fmfiset{p2}{vpath(__vs2,__v2)}
\fmfiset{p3}{vpath(__vs2,__v3)}
\fmfiset{p4}{vpath(__vs3,__v4)}
\fmfiset{p5}{vpath(__vs1,__vs2)}
\fmfiset{p6}{vpath(__vs1,__vs3)}
\fmfiset{p7}{vpath(__vs3,__v3)}
\svertex{vm1}{p1}
\dvertex{vu1}{vd1}{p1}
\svertex{vm2}{p2}
\dvertex{vu2}{vd2}{p2}
\svertex{vm3}{p3}
\dvertex{vu3}{vd3}{p3}
\svertex{vm4}{p4}
\dvertex{vd4}{vu4}{p4}
\svertex{vm5}{p5}
\dvertex{vd5}{vu5}{p5}
\fmfcmd{fill fullcircle scaled 8 shifted vloc(__vs3) withcolor black ;}
\fmfiv{label=$\scriptstyle\textcolor{white}{1}$,l.dist=0}{vloc(__vs3)}
\fmfis{phantom,ptext.in=$\scriptstyle\D^2$,ptext.clen=7,ptext.hin=3,ptext.hout=3,ptext.oin=5,ptext.oout=3,ptext.sep=;}{reverse(p1)}
\fmfis{phantom,ptext.out=$\scriptstyle\D^2$,ptext.clen=7,ptext.hin=3,ptext.hout=3,ptext.oin=5,ptext.oout=3,ptext.sep=;}{p2}
\fmfis{phantom,ptext.in=$ $,ptext.clen=7,ptext.hin=3,ptext.hout=3,ptext.oin=5,ptext.oout=3,ptext.sep=;}{p5}
\fmfis{phantom,ptext.out=$\scriptstyle\D^2$,ptext.clen=7,ptext.hin=-8,ptext.hout=-8,ptext.oin=5,ptext.oout=3,ptext.sep=;}{p4}
\fmfis{phantom,ptext.in=$\scriptstyle\barD^2$,ptext.clen=7,ptext.hin=-8,ptext.hout=-8,ptext.oin=5,ptext.oout=3,ptext.sep=;}{p6}
\fmfis{phantom,ptext.out=$\scriptstyle\D^2$,ptext.clen=7,ptext.hin=3,ptext.hout=3,ptext.oin=5,ptext.oout=3,ptext.sep=;}{p7}
\end{fmfchar*}}}
&=\left(
2\cccV[\fmf{phantom,l.side=right,l.dist=2,label=$\scriptstyle\Box$}{vs2,vc}\fmf{phantom}{vc,vs4}\fmffreeze\fmfposition\fmfcmd{fill vloc(__vs3)--vloc(__vc)--vloc(__vs4)--vloc(__vs3)--cycle withcolor \mympostgrey ;}\fmf{plain}{vs3,vc}\fmf{plain}{vs2,vs4}\fmf{plain}{vs2,v1}]{phantom}{plain,ptext.in=$\scriptstyle\D^2$}{derphoton,ptext.out=$\scriptstyle\D^2$}{derplain,ptext.out=$\scriptstyle\D^2$}{phantom}{phantom}{plain}{phantom}
-
\cccV[\fmf{phantom,l.side=right,l.dist=2,label=$\scriptstyle\Box$}{vs2,vc}\fmf{phantom}{vc,vs4}\fmffreeze\fmfposition\fmfcmd{fill vloc(__vs3)--vloc(__vc)--vloc(__vs4)--vloc(__vs3)--cycle withcolor \mympostgrey ;}\fmf{plain}{vs3,vc}\fmf{plain}{vs2,vs4}\fmf{plain}{vs2,v1}]{phantom}{plain,ptext.out=$\scriptstyle\D^2$}{photon,l.side=right,l.dist=3,label=$\scriptstyle\Box$,ptext.out=$\scriptstyle\D^2$}{plain,ptext.out=$\scriptstyle\D^2$}{phantom}{phantom}{plain}{phantom}
\right)i\frac{\lambda}{2}g_\YM^2\epsilon_{ijk}\\
&=
\left(
2
\cccV[\fmfcmd{fill p5--p6--p7--p8--cycle withcolor \mympostgrey ;}]{plain}{plain,ptext.out=$\scriptstyle\D^2$}{derphoton,ptext.out=$\scriptstyle\D^2$}{derplain,ptext.out=$\scriptstyle\D^2$}{plain,l.side=right,l.dist=3,label=$\scriptstyle\Box$}{plain}{plain}{plain}
-
\cccV[\fmfcmd{fill p5--p6--p7--p8--cycle withcolor \mympostgrey ;}]{plain}{plain,ptext.out=$\scriptstyle\D^2$}{photon,l.side=right,l.dist=3,label=$\scriptstyle\Box$,ptext.out=$\scriptstyle\D^2$}{plain,ptext.out=$\scriptstyle\D^2$}{plain,l.side=right,l.dist=3,label=$\scriptstyle\Box$}{plain}{plain}{plain}
\right)i\frac{\lambda}{2}g_\YM^2\epsilon_{ijk}
\col\\
\settoheight{\eqoff}{$\times$}%
\setlength{\eqoff}{0.5\eqoff}%
\addtolength{\eqoff}{-13\unitlength}%
\raisebox{\eqoff}{%
\fmfframe(2,1)(2,1){%
\begin{fmfchar*}(24,24)
\fmftop{v4}
\fmfleft{v1}
\fmfbottom{v2}
\fmfright{v2}
\fmfforce{(0.5w,h)}{v4}
\fmfforce{(0,0.5h)}{v1}
\fmfforce{(w,0.5h)}{v3}
\fmfforce{(0.5w,0)}{v2}
\fmffixed{(0,0.5h)}{vs2,vs3}
\fmffixed{(0,whatever)}{vs1,vs2}
\fmffixed{(0,whatever)}{vs2,v2}
\fmf{plain}{vs1,v1}
\fmf{plain}{vs2,v2}
\fmf{plain}{vs3,v4}
\fmf{plain}{vs1,vs2}
\fmf{plain}{vs1,vs3}
\fmffreeze
\fmf{photon}{vs2,v3}
\fmffreeze
\fmfposition
\fmfipath{p[],pca}
\fmfipair{vm[],vu[],vd[]}
\fmfiset{p1}{vpath(__vs1,__v1)}
\fmfiset{p2}{vpath(__vs2,__v2)}
\fmfiset{p3}{vpath(__vs2,__v3)}
\fmfiset{p4}{vpath(__vs3,__v4)}
\fmfiset{p5}{vpath(__vs1,__vs2)}
\fmfiset{p6}{vpath(__vs1,__vs3)}
\svertex{vm1}{p1}
\dvertex{vu1}{vd1}{p1}
\svertex{vm2}{p2}
\dvertex{vu2}{vd2}{p2}
\svertex{vm3}{p3}
\dvertex{vu3}{vd3}{p3}
\svertex{vm4}{p4}
\dvertex{vd4}{vu4}{p4}
\svertex{vm5}{p5}
\dvertex{vd5}{vu5}{p5}
\fmfcmd{fill fullcircle scaled 8 shifted vloc(__vs2) withcolor black ;}
\fmfiv{label=$\scriptstyle\textcolor{white}{1}$,l.dist=0}{vloc(__vs2)}
\fmfis{phantom,ptext.in=$\scriptstyle\D^2$,ptext.clen=7,ptext.hin=3,ptext.hout=3,ptext.oin=5,ptext.oout=3,ptext.sep=;}{reverse(p1)}
\fmfis{phantom,ptext.out=$\scriptstyle\D^2$,ptext.clen=7,ptext.hin=3,ptext.hout=3,ptext.oin=5,ptext.oout=3,ptext.sep=;}{p2}
\fmfis{phantom,ptext.out=$\scriptstyle\D^2$,ptext.clen=7,ptext.hin=3,ptext.hout=3,ptext.oin=5,ptext.oout=3,ptext.sep=;}{p3}
\fmfis{phantom,ptext.out=$\scriptstyle\D^2$,ptext.clen=7,ptext.hin=-8,ptext.hout=-8,ptext.oin=5,ptext.oout=3,ptext.sep=;}{p4}
\fmfis{phantom,ptext.in=$\scriptstyle\barD^2$,ptext.clen=7,ptext.hin=3,ptext.hout=3,ptext.oin=5,ptext.oout=3,ptext.sep=;}{p5}
\end{fmfchar*}}}
&=\left(
2\cccV[\fmf{phantom}{vs2,vc}\fmf{phantom,l.side=right,l.dist=2,label=$\scriptstyle\Box$}{vc,vs4}\fmffreeze\fmfposition\fmfcmd{fill vloc(__vs3)--vloc(__vc)--vloc(__vs2)--vloc(__vs3)--cycle withcolor \mympostgrey ;}\fmf{plain}{vs3,vc}\fmf{plain}{vs2,vs4}\fmf{plain}{vs4,v1}]{phantom}{derplain,ptext.out=$\scriptstyle\D^2$}{derphoton,ptext.out=$\scriptstyle\D^2$}{plain,ptext.out=$\scriptstyle\D^2$}{phantom}{plain}{phantom}{phantom}
-
\cccV[\fmf{phantom}{vs2,vc}\fmf{phantom,l.side=right,l.dist=2,label=$\scriptstyle\Box$}{vc,vs4}\fmffreeze\fmfposition\fmfcmd{fill vloc(__vs3)--vloc(__vc)--vloc(__vs2)--vloc(__vs3)--cycle withcolor \mympostgrey ;}\fmf{plain}{vs3,vc}\fmf{plain}{vs2,vs4}\fmf{plain}{vs4,v1}]{phantom}{plain,ptext.in=$\scriptstyle\D^2$}{photon,l.side=right,l.dist=3,label=$\scriptstyle\Box$,ptext.out=$\scriptstyle\D^2$}{plain,ptext.out=$\scriptstyle\D^2$}{phantom}{plain}{phantom}{phantom}
\right)(-i)\frac{\lambda}{2}g_\YM^2\epsilon_{ijk}\\
&=
\left(
2
\cccV[\fmfcmd{fill p5--p6--p7--p8--cycle withcolor \mympostgrey ;}]{plain}{derplain,ptext.out=$\scriptstyle\D^2$}{derphoton,ptext.out=$\scriptstyle\D^2$}{plain,ptext.out=$\scriptstyle\D^2$}{plain}{plain}{plain}{plain,l.side=left,l.dist=2,label=$\scriptstyle\Box$}
-
\cccV[\fmfcmd{fill p5--p6--p7--p8--cycle withcolor \mympostgrey ;}]{plain}{plain,ptext.out=$\scriptstyle\D^2$}{photon,l.side=right,l.dist=3,label=$\scriptstyle\Box$,ptext.out=$\scriptstyle\D^2$}{plain,ptext.out=$\scriptstyle\D^2$}{plain}{plain}{plain}{plain,l.side=left,l.dist=3,label=$\scriptstyle\Box$}
\right)(-i)\frac{\lambda}{2}g_\YM^2\epsilon_{ijk}
\pnt
\end{aligned}
\end{equation}
In order to obtain the above results, we made use of the relation 
$\D^\alpha\barD^2\D_\alpha=-\Box+\D^2\barD^2+\barD^2\D^2$.
Furthermore, we consider the following contributions:
\begin{equation}\label{cccVdiags2}
\begin{aligned}
\cccV[\fmf{photon}{vs2,vs4}\fmf{photon}{vs4,v3}\fmffreeze\fmfposition\fmfis{phantom,ptext.out=$\scriptstyle\D^2$,ptext.clen=7,ptext.hin=3,ptext.hout=3,ptext.oin=5,ptext.oout=3,ptext.sep=;}{vpath(__vs4,__v3)}]{plain,ptext.in=$\scriptstyle\D^2$}{plain,ptext.out=$\scriptstyle\D^2$}{phantom}{plain,ptext.out=$\scriptstyle\D^2$}{plain,ptext.in=$\scriptstyle\D^2$,ptext.out=$\scriptstyle\barD^2$}{phantom}{phantom}{plain,ptext.out=$\scriptstyle\barD^2$}
&=
\cccV[\fmfcmd{fill p5--p6--p7--p8--cycle withcolor \mympostgrey ;}]{plain,l.side=left,l.dist=3,label=$\scriptstyle\Box$}{plain,ptext.out=$\scriptstyle\D^2$}{photon,ptext.out=$\scriptstyle\D^2$}{plain,ptext.out=$\scriptstyle\D^2$}{plain}{plain}{plain,l.side=right,l.dist=2,label=$\scriptstyle\Box$}{plain}
i\frac{\lambda}{2}g_\YM^2\epsilon_{ijk}
\col
\cccV[\fmf{phantom}{vs2,vc}\fmf{phantom}{vc,vs4}\fmffreeze\fmf{photon}{vs2,vc}\fmf{photon}{vs4,v3}\fmffreeze\fmfposition\fmfis{phantom,ptext.out=$\scriptstyle\D^2$,ptext.clen=7,ptext.hin=3,ptext.hout=3,ptext.oin=5,ptext.oout=3,ptext.sep=;}{vpath(__vs4,__v3)}\fmfis{plain,ptext.out=$\scriptstyle\barD^2$,ptext.clen=7,ptext.hin=3,ptext.hout=3,ptext.oin=5,ptext.oout=5,ptext.sep=;}{vloc(__vs1)--vloc(__vc)}\fmfis{plain,ptext.in=$\scriptstyle\D^2$,ptext.out=$\scriptstyle\barD^2$,ptext.clen=7,ptext.hin=-8,ptext.hout=3,ptext.oin=5,ptext.oout=0,ptext.sep=;}{vloc(__vc)--vloc(__vs4)}]{plain,ptext.in=$\scriptstyle\D^2$}{plain,ptext.out=$\scriptstyle\D^2$}{phantom}{plain,ptext.out=$\scriptstyle\D^2$}{plain,ptext.in=$\scriptstyle\D^2$,ptext.out=$\scriptstyle\barD^2$}{phantom}{phantom}{phantom}
=
\cccV[\fmfcmd{fill p5--p6--p7--p8--cycle withcolor \mympostgrey ;}]{plain,l.side=left,l.dist=3,label=$\scriptstyle\Box$}{plain,ptext.out=$\scriptstyle\D^2$}{photon,ptext.out=$\scriptstyle\D^2$}{plain,ptext.out=$\scriptstyle\D^2$}{plain}{plain}{plain,l.side=right,l.dist=2,label=$\scriptstyle\Box$}{plain}
(-i)\lambda g_\YM^2\epsilon_{ijk}
\col\\
\cccV[\fmf{photon}{vs2,vs4}\fmf{photon}{vs2,v3}\fmffreeze\fmfposition\fmfis{phantom,ptext.out=$\scriptstyle\D^2$,ptext.clen=7,ptext.hin=3,ptext.hout=3,ptext.oin=5,ptext.oout=3,ptext.sep=;}{vpath(__vs2,__v3)}]{plain,ptext.in=$\scriptstyle\D^2$}{plain,ptext.out=$\scriptstyle\D^2$}{phantom}{plain,ptext.out=$\scriptstyle\D^2$}{plain,ptext.in=$\scriptstyle\D^2$,ptext.out=$\scriptstyle\barD^2$}{phantom}{phantom}{plain,ptext.out=$\scriptstyle\barD^2$}
&=
\cccV[\fmfcmd{fill p5--p6--p7--p8--cycle withcolor \mympostgrey ;}]{plain,l.side=left,l.dist=3,label=$\scriptstyle\Box$}{plain,ptext.out=$\scriptstyle\D^2$}{photon,ptext.out=$\scriptstyle\D^2$}{plain,ptext.out=$\scriptstyle\D^2$}{plain}{plain,l.side=right,l.dist=2,label=$\scriptstyle\Box$}{plain}{plain}
(-i)\frac{\lambda}{2}g_\YM^2\epsilon_{ijk}
\col
\cccV[\fmf{phantom}{vs2,vc}\fmf{phantom}{vc,vs4}\fmffreeze\fmf{photon}{vs4,vc}\fmf{photon}{vs2,v3}\fmffreeze\fmfposition\fmfis{phantom,ptext.out=$\scriptstyle\D^2$,ptext.clen=7,ptext.hin=3,ptext.hout=3,ptext.oin=5,ptext.oout=3,ptext.sep=;}{vpath(__vs2,__v3)}\fmfis{plain,ptext.out=$\scriptstyle\barD^2$,ptext.clen=7,ptext.hin=3,ptext.hout=-8,ptext.oin=5,ptext.oout=5,ptext.sep=;}{vloc(__vs1)--vloc(__vc)}\fmfis{plain,ptext.in=$\scriptstyle\D^2$,ptext.out=$\scriptstyle\barD^2$,ptext.clen=7,ptext.hin=3,ptext.hout=-8,ptext.oin=5,ptext.oout=0,ptext.sep=;}{vloc(__vc)--vloc(__vs2)}]{plain,ptext.in=$\scriptstyle\D^2$}{plain,ptext.out=$\scriptstyle\D^2$}{phantom}{plain,ptext.out=$\scriptstyle\D^2$}{phantom}{phantom}{phantom}{plain,ptext.in=$\scriptstyle\D^2$,ptext.out=$\scriptstyle\barD^2$}
=
\cccV[\fmfcmd{fill p5--p6--p7--p8--cycle withcolor \mympostgrey ;}]{plain,l.side=left,l.dist=3,label=$\scriptstyle\Box$}{plain,ptext.out=$\scriptstyle\D^2$}{photon,ptext.out=$\scriptstyle\D^2$}{plain,ptext.out=$\scriptstyle\D^2$}{plain}{plain,l.side=right,l.dist=2,label=$\scriptstyle\Box$}{plain}{plain}
i\lambda g_\YM^2\epsilon_{ijk}
\pnt
\end{aligned}
\end{equation}
In the above expression,  
we have removed and inserted contracted propagators as 
$-\frac{\Box}{p^2}=1$, where $p$ is the respective momentum.
This does not affect the result and allows us 
to transform all integrals into box integrals with different numerators. 
It is then easy to see that the contributions in 
\eqref{cccVdiags1} and \eqref{cccVdiags2} sum up to zero.
This demonstrates the cancellation of the overall UV divergences
described by the last relation in \eqref{cancelr4}.

\subsection{Two-loop subdiagrams}
\label{app:twoloopsubdiag}

The finite two-loop chiral self-energy and two-loop 
chiral vertex correction appear as subdiagrams in three-loop diagrams.
Here, we derive the results for these subdiagrams. 
We use that several diagrams are a priori vanishing
because the one-loop self-energies of the 
chiral and vector fields and certain color contractions are zero.

The two-loop chiral self-energy contains the following nonvanishing
contributions:
\begin{equation}
\begin{aligned}
S_1=\swftwoone
&=-2\lambda^2I_{2\mathbf{t}}
\col\\
S_2=\swftwotwo{1}{1}
&=\frac{\lambda^2}{2}I_2
\col\qquad
S_3=\swftwotwo{1}{-1}=\lambda^2I_2
\col\\
S_4=\swftwofour{1}{1}
&=\frac{\lambda^2}{2}I_1^2
\col\qquad
S_5=\swftwofour{1}{-1}
=\lambda^2I_1^2
\col\\
S_6=\swftwofive{1}
&=\lambda^2(-I_1^2-I_{2\mathbf{t}})
\col\\
S_7=\swftwosix{plain}{photon}{plain}
&=-2\lambda^2I_2
\col\\
S_8=\swftwosix{photon}{photon}{photon}
&=\frac{\lambda^2}{2}(-I_1^2+2I_2+2I_{2\mathbf{t}})
\col\\
\end{aligned}
\end{equation}
where we have omitted a factor $p^{2(D-3)}$, and the covariant derivatives
$\D^2$ and $\barD^2$ at the external legs after $\D$-algebra.
Expressions for the integrals are given in \eqref{Integralexpr} and
\eqref{I2t}.
Summing up the above contributions, thereby including also the reflected 
diagrams where required, all divergences cancel out, and we find
\begin{equation}
\begin{aligned}
\Sigma_2
&=S_1+4S_2+4S_3+2S_4+2S_5+2S_6+4S_7+2S_8
=-2\lambda^2I_{2\mathbf{t}}\pnt
\end{aligned}
\end{equation}
Restoring the covariant derivatives and the correct 
proportionality to the external momentum $p$, 
the two-loop chiral self-energy can be written as
\begin{equation}\label{ctwoloopse}
\settoheight{\eqoff}{$\times$}%
\setlength{\eqoff}{0.5\eqoff}%
\addtolength{\eqoff}{-7.5\unitlength}%
\raisebox{\eqoff}{%
\fmfframe(1,0)(1,0){%
\begin{fmfchar*}(20,15)
\fmfleft{v1}
\fmfright{v2}
\fmf{plain}{v1,v2}
\fmffreeze
\fmfposition
\fmfipair{vm[]}
\fmfipath{p[]}
\fmfiset{p1}{vpath(__v1,__v2)}
\svertex{vm1}{p1}
\fmfcmd{fill fullcircle scaled 10 shifted vm1 withcolor black ;}
\fmfiv{label=\small$\textcolor{white}{2}$,l.dist=0}{vm1}
\end{fmfchar*}}}
={}
-2\lambda^2p^{2(D-3)}
\settoheight{\eqoff}{$\times$}%
\setlength{\eqoff}{0.5\eqoff}%
\addtolength{\eqoff}{-7.5\unitlength}%
\raisebox{\eqoff}{%
\fmfframe(1,0)(1,0){%
\begin{fmfchar*}(20,15)
\fmfleft{v1}
\fmfright{v2}
\fmffixed{(0.66w,0)}{vc1,vc2}
\fmf{plain}{v1,vc1}
\fmf{plain}{vc2,v2}
\fmf{plain,left=0.5}{vc1,vc2}
\fmf{plain,left=0.5}{vc2,vc1}
\fmffreeze
\fmfposition
\fmfipair{vm[]}
\fmfipath{p[]}
\fmfiset{p1}{vpath(__v1,__vc1)}
\fmfiset{p2}{vpath(__v2,__vc2)}
\fmfiset{p3}{vpath(__vc1,__vc2)}
\fmfiset{p4}{vpath(__vc2,__vc1)}
\fmfcmd{fill p3--p4--cycle withcolor \mympostgrey;}
\svertex{vm1}{p3}
\svertex{vm2}{p4}
\fmfi{plain}{vm1--vm2}
\fmfis{phantom,ptext.in=$\scriptstyle\D^2$,ptext.clen=7,ptext.hin=3,ptext.hout=3,ptext.oin=4,ptext.oout=3,ptext.sep=;}{p1}
\fmfis{phantom
,ptext.out=$\scriptstyle\barD^2$,ptext.clen=7,ptext.hin=3,ptext.hout=3,ptext.oin=4,ptext.oout=3,ptext.sep=;}{p2}
\end{fmfchar*}}}
\pnt
\end{equation}
The gray scaled part of the graph is identified as the integral 
$I_{2\mathbf{t}}$ given in \eqref{I2t}.


The two-loop correction of the chiral vertex is given as a sum of the 
following nonvanishing contributions:
\begin{equation}
\begin{aligned}
\cVat[\fmfcmd{fill fullcircle scaled 10 shifted vloc(__vg1) withcolor black ;}
\fmfiv{label=\small$\textcolor{white}{2}$,l.dist=0}{vloc(__vg1)}]
{plain}{plain}{plain}{phantom}{plain}{plain}
&=
\cVatg[\fmfcmd{fill fullcircle scaled 8 shifted vloc(__vg1) withcolor black ;}\fmfiv{label=$\scriptstyle\textcolor{white}{1}$,l.dist=0}{vloc(__vg1)}\fmfi{photon}{vm2--vm3}]{plain}{plain}{plain}{plain}{plain}
+
\cVatg[\fmfi{photon}{vm1{dir 90}..{dir -45}vi3}\fmfi{phantom}{vo2--vo3}\fmfcmd{fill fullcircle scaled 8 shifted vm1 withcolor black ;}\fmfiv{label=$\scriptstyle\textcolor{white}{1}$,l.dist=0}{vm1}]{plain}{plain}{plain}{plain}{plain}
+
\cVatg[\fmfi{photon}{vm1{dir -90}..{dir 45}vi2}\fmfi{phantom}{vo2--vo3}\fmfcmd{fill fullcircle scaled 8 shifted vm1 withcolor black ;}\fmfiv{label=$\scriptstyle\textcolor{white}{1}$,l.dist=0}{vm1}]{plain}{plain}{plain}{plain}{plain}
+\cVatg[\fmfi{photon}{vm2--vi3}\fmfi{photon}{vm2--vo3}]{plain}{plain}{plain}{plain}{plain}
+\cVatg[\fmfi{photon}{vi2--vm3}\fmfi{photon}{vo2--vm3}]{plain}{plain}{plain}{plain}{plain}
\\
&\phantom{{}={}}
+\dots\col
\end{aligned}
\end{equation}
where we have omitted the covariant derivatives.
The first and the next two contributions, respectively, contain 
the one-loop corrections of the chiral vertex in \eqref{ccconeloop} and 
of the cubic gauge-matter vertex in \eqref{cvaoneloop}. The ellipsis denotes
the two contributions obtained by cyclic permutations of all displayed 
diagrams.
After $\D$-algebra, the final result can be cast into the form
\begin{equation}
\begin{aligned}
\cVat[\fmfcmd{fill fullcircle scaled 10 shifted vloc(__vg1) withcolor black ;}
\fmfiv{label=\small$\textcolor{white}{2}$,l.dist=0}{vloc(__vg1)}]
{plain}{plain}{plain}{phantom}{plain,ptext.in=$\scriptstyle\barD^2$}{plain,ptext.in=$\scriptstyle\barD^2$}
&=\left(
2
\cVatt[\fmfcmd{fill p7--p8--p5--p6--p9--cycle withcolor \mympostgrey ;}]{plain,label=$\scriptstyle\Box^2$,l.side=left,l.dist=3}{plain}{plain}{plain}{plain}{plain}{plain,ptext.in=$\scriptstyle\barD^2$}{plain,ptext.in=$\scriptstyle\barD^2$}
-
\cVatt[\fmfcmd{fill p7--p8--p5--p6--p9--cycle withcolor \mympostgrey ;}]{plain,label=$\scriptstyle\Box$,l.side=left,l.dist=3}{plain}{plain}{plain}{plain}{plain,label=$\scriptstyle\Box$,l.side=right,l.dist=3}{plain,ptext.in=$\scriptstyle\barD^2$}{plain,ptext.in=$\scriptstyle\barD^2$}
-
\cVatt[\fmfcmd{fill p7--p8--p5--p6--p9--cycle withcolor \mympostgrey ;}]{plain,label=$\scriptstyle\Box$,l.side=left,l.dist=3}{plain}{plain}{plain}{plain,label=$\scriptstyle\Box$,l.side=left,l.dist=3}{plain}{plain,ptext.in=$\scriptstyle\barD^2$}{plain,ptext.in=$\scriptstyle\barD^2$}
-2
\cVatt[\fmfcmd{fill p7--p8--p5--p6--p9--cycle withcolor \mympostgrey ;}]{plain,label=$\scriptstyle\Box $,l.side=left,l.dist=3}{plain}{plain}{plain,label=$\scriptstyle\Box$,l.dist=3,l.side=left}{plain}{plain}{plain,ptext.in=$\scriptstyle\barD^2$}{plain,ptext.in=$\scriptstyle\barD^2$}\right.\\
&\phantom{{}={}\Bigg({}\,}\left.
+
\cVatt[\fmfcmd{fill p7--p8--p5--p6--p9--cycle withcolor \mympostgrey ;}]{plain,label=$\scriptstyle\Box$,l.side=left,l.dist=3}{plain}{plain}{plain}{plain}{plain}{plain,ptext.in=$\scriptstyle\barD^2$}{plain,ptext.in=$\scriptstyle\barD^2$,label.side=left,label=$\scriptstyle\Box$,l.dist=3}
+
\cVatt[\fmfcmd{fill p7--p8--p5--p6--p9--cycle withcolor \mympostgrey ;}]{plain,label=$\scriptstyle\Box$,l.side=left,l.dist=3}{plain}{plain}{plain}{plain}{plain}{plain,ptext.in=$\scriptstyle\barD^2$,label.side=right,label=$\scriptstyle\Box$,l.dist=3}{plain,ptext.in=$\scriptstyle\barD^2$}
+
\cVatt[\fmfcmd{fill p7--p8--p5--p6--p9--cycle withcolor \mympostgrey ;}]{plain}{plain}{plain}{plain}{plain}{plain,l.side=right,label=$\scriptstyle\Box$,l.dist=3}{plain,ptext.in=$\scriptstyle\barD^2$}{plain,ptext.in=$\scriptstyle\barD^2$,label.side=left,label=$\scriptstyle\Box$,l.dist=3}
+
\cVatt[\fmfcmd{fill p7--p8--p5--p6--p9--cycle withcolor \mympostgrey ;}]{plain}{plain}{plain}{plain}{plain,l.side=left,label=$\scriptstyle\Box$,l.dist=3}{plain}{plain,ptext.in=$\scriptstyle\barD^2$,l.side=right,label=$\scriptstyle\Box$,l.dist=3}{plain,ptext.in=$\scriptstyle\barD^2$}
\right.\\
&\phantom{{}={}\Bigg({}\,}\left.
+\dots\vphantom{
\settoheight{\eqoff}{$\times$}%
\setlength{\eqoff}{0.5\eqoff}%
\addtolength{\eqoff}{-14\unitlength}%
\raisebox{\eqoff}{%
\fmfframe(12.5,14)(12.5,14){%
}}
}\right)\frac{\lambda^2}{2}ig_\YM\epsilon_{ijk}
\pnt
\end{aligned}
\end{equation}

\section{Cancellation of higher-order poles}
\label{app:hopolecanc}

In this Appendix, we demonstrate the cancellation of all higher-order poles
in the logarithm of the renormalization constant. The cancellations 
are required in a consistent renormalization procedure.
We expand the renormalization constant, as introduced 
in \eqref{opren} and \eqref{DinZ}, to three-loop order as
\begin{equation}
\mathcal{Z}=\unitmatrix+\lambda\mathcal{Z}_1
+\lambda^2\mathcal{Z}_2
+\lambda^3\mathcal{Z}_3
+\mathcal{O}(\lambda^4)\pnt
\end{equation}
Its logarithm then has the series expansion
\begin{equation}\label{lnZ}
\ln\mathcal{Z}=
\lambda\mathcal{Z}_1
+\lambda^2\Big(\mathcal{Z}_2-\frac{1}{2}\mathcal{Z}_1^2\Big)
+\lambda^3\Big(\mathcal{Z}_3-\frac{1}{2}(\mathcal{Z}_1\mathcal{Z}_2+\mathcal{Z}_2\mathcal{Z}_1)+\frac{1}{3}\mathcal{Z}_1^3\Big)
+\mathcal{O}(\lambda^4)
\col
\end{equation}
where the one- and two-loop contributions to the renormalization constant
were obtained in \eqref{Z1} and \eqref{Z2} and
are given by
\begin{equation}\label{Z1Z2}
\mathcal{Z}_1=-\mathcal{I}_1\chi(1)\col\qquad
\mathcal{Z}_2=-\mathcal{I}_2(\chi(1,2)+\chi(2,1)-2\chi(1))+\mathcal{I}_1^2\chi(1,3)+\dots\pnt
\end{equation}
In the case of $\mathcal{Z}_2$, we have restored one contribution that we 
neglected in \eqref{Z2} because it only contains
a quadratic pole in $\varepsilon$ and hence does not contribute to the 
dilatation operator. 
Here, this term is required, since its multiplication by the one-loop 
contribution 
in \eqref{lnZ} generates chiral functions that also appear in three-loop 
diagrams which have simple poles. The respective contributions hence cancel 
higher-order poles coming from these diagrams. 
The ellipsis denotes further terms that only involve quadratic poles 
in $\varepsilon$ and chiral functions with range $\kappa\ge5$ that 
cannot appear in the three-loop dilatation operator.
The higher-order poles from these contributions cancel  
separately in a straightforward way, and we have neglected the respective 
Feynman diagrams from the very beginning.

When \eqref{Z1Z2} is inserted into \eqref{lnZ}, one
encounters products of chiral functions. They can be expanded 
in terms of simple chiral functions, thereby taking care of 
factors of two coming from flavor contractions and minus signs 
from the color factors.
The results read
\begin{equation}
\begin{aligned}\label{chiprodex}
\chi(1)^2&=\chi(1,2)+\chi(2,1)+2\chi(1,3)-2\chi(1)+\dots\col\\
\chi(1)\chi(1,2)&=\chi(1,2,3)-2\chi(1,2)+\chi(2,1,2)+\chi(1,3,2)
+\dots
\col\\
\chi(1)\chi(2,1)&=\chi(1,3,2)+\chi(1,2,1)-2\chi(2,1)+\chi(3,2,1)
+\dots
\col\\
\chi(1,2)\chi(1)&=\chi(1,2,3)-2\chi(1,2)+\chi(1,2,1)+\chi(2,1,3)
+\dots
\col\\
\chi(2,1)\chi(1)&=\chi(2,1,3)+\chi(2,1,2)-2\chi(2,1)+\chi(3,2,1)
+\dots
\col\\
\chi(1)\chi(1,3)&=\chi(2,1,3)-4\chi(1,3)
+\dots
\col\\
\chi(1,3)\chi(1)&=\chi(1,3,2)-4\chi(1,3)
+\dots
\col\\
\chi(1)^3
&=\chi(1,2,3)+\chi(3,2,1)+2(\chi(1,3,2)+\chi(2,1,3))-12\chi(1,3)\\
&\phantom{{}={}}{}+{}\chi(2,1,2)+\chi(1,2,1)
-4(\chi(1,2)-\chi(2,1))
+4\chi(1)+\dots
\col
\end{aligned}
\end{equation}
where the ellipses denote chiral functions with range $\kappa\ge5$.

The above products are used to reexpress the one- and two-loop 
renormalization constants in 
\eqref{Z1Z2} and $\mathcal{Z}_3$ in \eqref{Z3res} in a very convenient
form
\begin{equation}
\begin{aligned}\label{Z1Z2Z3}
\mathcal{Z}_1&={}-\mathcal{I}_1\chi(1)\col\\\
\mathcal{Z}_2&={}-\mathcal{I}_2\chi(1)^2+(2\mathcal{I}_2+\mathcal{I}_1^2)\chi(1,3)+\dots\col\\
\mathcal{Z}_3
&={}-\mathcal{I}_3\chi(1)^3
+2(-6I_3-2\mathcal{I}_1\mathcal{I}_2+\mathcal{I}_{32\mathbf{t}})\chi(1,3)\\
&\phantom{{}={}}
+(2\mathcal{I}_3-\mathcal{I}_{3\mathbf{bb}})\chi(2,1,3)
+(2\mathcal{I}_3-\mathcal{I}_{3\mathbf{b}})\chi(1,3,2)+\dots
\col
\end{aligned}
\end{equation}
With these expressions, 
the combination appearing at two loops in \eqref{lnZ} is then 
given by
\begin{equation}
\begin{aligned}
\mathcal{Z}_2-\frac{1}{2}\mathcal{Z}_1^2
&=
-\Big(\mathcal{I}_2+\frac{1}{2}\mathcal{I}_1^2\Big)
(\chi(1)^2-2\chi(1,3))
\pnt
\end{aligned}
\end{equation}
Inserting the explicit expressions for the poles listed in 
\eqref{Integralexpr}, we find that the quadratic poles cancel in the 
above combination.
Moreover, the linear 
combination of chiral functions which appears on the right-hand side
does not contain $\chi(1,3)$ after expanding the product of chiral
functions as in \eqref{chiprodex}.
This is a consequence of the fact that, 
at two-loop order, all Feynman diagrams which generate $\chi(1,3)$ 
only have double poles. 

Inserting the expressions \eqref{Z1Z2Z3} into the combination appearing at 
three loops in \eqref{lnZ},
we find
\begin{equation}
\begin{aligned}
&\mathcal{Z}_3
-\frac{1}{2}(\mathcal{Z}_1\mathcal{Z}_2+\mathcal{Z}_2\mathcal{Z}_1)+\frac{1}{3}\mathcal{Z}_1^3\\
&={}-{}\Big(\mathcal{I}_3+\mathcal{I}_1\mathcal{I}_2+\frac{1}{3}\mathcal{I}_1^3\Big)\chi(1)^3
-12\Big(I_3+\mathcal{I}_1\mathcal{I}_2+\frac{1}{3}\mathcal{I}_3^2
+\mathcal{I}_{32\mathbf{t}}\Big)\chi(1,3)\\
&\phantom{{}={}}
+\frac{1}{2}\Big(4\mathcal{I}_3-\mathcal{I}_{3\mathbf{b}}-\mathcal{I}_{3\mathbf{bb}}-2\mathcal{I}_1\mathcal{I}_2+\mathcal{I}_1^3\Big)(\chi(1,3,2)+\chi(2,1,3))\\
&\phantom{{}={}}
-\frac{1}{2}(\mathcal{I}_{3\mathbf{b}}-\mathcal{I}_{3\mathbf{bb}})(\chi(1,3,2)-\chi(2,1,3))
\pnt
\end{aligned}
\end{equation}
With the explicit results for the integrals \eqref{Integralexpr}, 
one verifies that the coefficients of $\chi(1)^3$, $\chi(1,3)$ and of 
the symmetric combination $\chi(1,3,2)+\chi(2,1,3)$ are free of cubic 
and quadratic poles in $\varepsilon$ but that the coefficient of the 
antisymmetric combination is not free of them. This is not an inconsistency.
The two chiral functions $\chi(1,3,2)$ and
$\chi(2,1,3)$ are different but they yield the same result whenever 
applied to a any state of the flavor $SU(2)$ subsector. The last contribution,
therefore, yields zero whenever we consider the full Feynman diagrams
with the composite operator included. 
In fact, the coefficient 
of an antisymmetric combination of $\chi(1,3,2)$ and
$\chi(2,1,3)$ is given by $\epsilon_2$ and it is  
associated to an ambiguity in fixing a scheme 
\cite{Beisert:2003tq,Beisert:2005wv}. It
does not alter the anomalous dimensions, and, as can be seen from
\eqref{D1D2D3}, integrability 
makes no prediction for it, while we have found the value 
given in \eqref{epsilon2}.

\section{Integrals}
\label{app:integrals}

In $D$-dimensional Euclidean space,
the scalar $G$-function is defined as
\begin{equation}
G(\alpha,\beta)=
\frac{\Gamma(\tfrac{D}{2}-\alpha)\Gamma(\tfrac{D}{2}-\beta)\Gamma(\alpha+\beta-\tfrac{D}{2})}{(4\pi)^{\frac{D}{2}}\Gamma(\alpha)\Gamma(\beta)\Gamma(D-\alpha-\beta)}\col
\end{equation}
and it describes the simple loop integral that involves two propagators
of massless fields with respective weights $\alpha$ and $\beta$ evaluated 
with external momentum $p^2=1$. 
The $G$-function with one momentum or, respectively, two momenta 
in the numerators of the integrals is defined as
\begin{equation}
\begin{aligned}
G_1(\alpha,\beta)
&=\frac{1}{2}(-G(\alpha,\beta-1)+G(\alpha-1,\beta)+G(\alpha,\beta))\col\\
G_2(\alpha,\beta)
&=\frac{1}{2}(-G(\alpha,\beta-1)-G(\alpha-1,\beta)+G(\alpha,\beta))
\pnt
\end{aligned}
\end{equation}
To three-loop order, we need the following integrals and their overall
UV divergences:
\begin{equation}\label{Integralexpr}
\begin{gathered}
\begin{aligned}
I_1&=
\settoheight{\eqoff}{$\times$}%
\setlength{\eqoff}{0.5\eqoff}%
\addtolength{\eqoff}{-4.5\unitlength}%
\raisebox{\eqoff}{%
\fmfframe(-1,-3)(-1,-3){%
\begin{fmfchar*}(15,15)
  \fmfleft{in}
  \fmfright{out1}
\fmf{phantom}{in,v1}
\fmf{phantom}{out,v2}
\fmfforce{(0,0.5h)}{in}
\fmfforce{(w,0.5h)}{out}
\fmffixed{(0.75w,0)}{v1,v2}
\fmf{plain,right=0.5}{v1,v2}
\fmf{plain,left=0.5}{v1,v2}
\end{fmfchar*}}}
=G(1,1)
\col\quad
&\mathcal{I}_1&=\frac{1}{(4\pi)^2}\frac{1}{\varepsilon} \col\\
I_2&=
\settoheight{\eqoff}{$\times$}%
\setlength{\eqoff}{0.5\eqoff}%
\addtolength{\eqoff}{-6.75\unitlength}%
\raisebox{\eqoff}{%
\fmfframe(-0.5,-5.5)(1,4){%
\begin{fmfchar*}(15,15)
  \fmfleft{in}
  \fmfright{out1}
\fmf{phantom}{in,v1}
\fmf{phantom}{out,v2}
\fmfforce{(0,0h)}{in}
\fmfforce{(w,0h)}{out}
\fmffixed{(0,0.75h)}{v2,v3}
\fmfpoly{phantom}{v1,v2,v3}
\fmf{plain}{v1,v2}
\fmf{plain}{v1,v3}
\fmf{plain,right=0.25}{v2,v3}
\fmf{plain,left=0.25}{v2,v3}
\end{fmfchar*}}}
=G(1,1)G(3-\tfrac{D}{2},1)
\col\quad
&\mathcal{I}_2&=\frac{1}{(4\pi)^4}\Big(-\frac{1}{2\varepsilon^2}+\frac{1}{2\varepsilon}\Big) 
\col\\
I_3&=
\settoheight{\eqoff}{$\times$}%
\setlength{\eqoff}{0.5\eqoff}%
\addtolength{\eqoff}{-6.75\unitlength}%
\raisebox{\eqoff}{%
\fmfframe(-0.5,-5.5)(9,4){%
\begin{fmfchar*}(15,15)
  \fmfleft{in}
  \fmfright{out1}
\fmf{phantom}{in,v1}
\fmf{phantom}{out,v2}
\fmfforce{(0,0h)}{in}
\fmfforce{(w,0h)}{out}
\fmffixed{(0,0.75h)}{v2,v3}
\fmfpoly{phantom}{v1,v2,v3}
\fmfpoly{phantom}{v3,v2,v4}
\fmf{plain}{v1,v2}
\fmf{plain}{v1,v3}
\fmf{plain}{v2,v3}
\fmf{plain}{v3,v4}
\fmf{plain,right=0.25}{v2,v4}
\fmf{plain,left=0.25}{v2,v4}
\end{fmfchar*}}}
=G(1,1)G(3-\tfrac{D}{2},1)G(5-D,1)
\col
&\mathcal{I}_3&=\frac{1}{(4\pi)^6}\Big(\frac{1}{6\varepsilon^3}-\frac{1}{2\varepsilon^2}+\frac{2}{3\varepsilon}\Big)
\col\\
I_{3\mathbf{t}}&=
\settoheight{\eqoff}{$\times$}%
\setlength{\eqoff}{0.5\eqoff}%
\addtolength{\eqoff}{-9\unitlength}%
\raisebox{\eqoff}{%
\fmfframe(-0.5,-5.5)(9,8.5){%
\begin{fmfchar*}(15,15)
  \fmfleft{in}
  \fmfright{out1}
\fmf{phantom}{in,v1}
\fmf{phantom}{out,v2}
\fmfforce{(0,0h)}{in}
\fmfforce{(w,0h)}{out}
\fmffixed{(0,0.75h)}{v2,v3}
\fmfpoly{phantom}{v1,v2,v3}
\fmfpoly{phantom}{v3,v2,v4}
\fmf{plain}{v1,v2}
\fmf{plain}{v1,v3}
\fmf{plain}{v2,v3}
\fmf{plain}{v3,v4}
\fmf{plain}{v2,v4}
\fmf{plain,right=1}{v1,v4}
\end{fmfchar*}}}
=I_{2\mathbf{t}}G(5-D,1)
\col\quad
&\mathcal{I}_{3\mathbf{t}}&=\frac{1}{(4\pi)^6}\frac{1}{\varepsilon}2\zeta(3)
\col\\
I_{3\mathbf{b}}&=
\settoheight{\eqoff}{$\times$}%
\setlength{\eqoff}{0.5\eqoff}%
\addtolength{\eqoff}{-6.75\unitlength}%
\raisebox{\eqoff}{%
\fmfframe(-0.5,-5.5)(9,4){%
\begin{fmfchar*}(15,15)
  \fmfleft{in}
  \fmfright{out1}
\fmf{phantom}{in,v1}
\fmf{phantom}{out,v2}
\fmfforce{(0,0h)}{in}
\fmfforce{(w,0h)}{out}
\fmffixed{(0,0.75h)}{v2,v3}
\fmfpoly{phantom}{v1,v2,v3}
\fmfpoly{phantom}{v3,v2,v4}
\fmf{plain}{v1,v2}
\fmf{plain}{v1,v3}
\fmf{plain}{v2,v4}
\fmf{plain}{v3,v4}
\fmf{plain,right=0.25}{v2,v3}
\fmf{plain,left=0.25}{v2,v3}
\end{fmfchar*}}}
\col\quad
&\mathcal{I}_{3\mathbf{b}}&=\frac{1}{(4\pi)^6}
\Big(\frac{1}{3\varepsilon^3}-\frac{2}{3\varepsilon^2}+\frac{1}{3\varepsilon}\Big)\col
&\\
I_{3\mathbf{bb}}&=
\settoheight{\eqoff}{$\times$}%
\setlength{\eqoff}{0.5\eqoff}%
\addtolength{\eqoff}{-6.75\unitlength}%
\raisebox{\eqoff}{%
\fmfframe(-0.5,-5.5)(9,4){%
\begin{fmfchar*}(15,15)
  \fmfleft{in}
  \fmfright{out1}
\fmf{phantom}{in,v1}
\fmf{phantom}{out,v2}
\fmfforce{(0,0h)}{in}
\fmfforce{(w,0h)}{out}
\fmffixed{(0,0.75h)}{v2,v3}
\fmfpoly{phantom}{v1,v2,v3}
\fmfpoly{phantom}{v3,v2,v4}
\fmf{plain,right=0.25}{v1,v2}
\fmf{plain,left=0.25}{v1,v2}
\fmf{plain}{v1,v3}
\fmf{plain}{v3,v4}
\fmf{plain,right=0.25}{v2,v4}
\fmf{plain,left=0.25}{v2,v4}
\end{fmfchar*}}}
=G(1,1)^2G(3-\tfrac{D}{2},3-\tfrac{D}{2})
\col\quad
&\mathcal{I}_{3\mathbf{bb}}&=\frac{1}{(4\pi)^6}
\Big(\frac{1}{3\varepsilon^3}-\frac{1}{3\varepsilon^2}-\frac{1}{3\varepsilon}\Big)
\col\\
\end{aligned}\\
\begin{aligned}
\hphantom{I_{3\mathbf{bb}}}&\hfill\\[-\baselineskip]
I_{32\mathbf{t}}&=
\settoheight{\eqoff}{$\times$}%
\setlength{\eqoff}{0.5\eqoff}%
\addtolength{\eqoff}{-6.75\unitlength}%
\raisebox{\eqoff}{%
\fmfframe(-0.5,-3.5)(10.5,2){%
\begin{fmfchar*}(15,15)
  \fmfleft{in}
  \fmfright{out1}
\fmf{phantom}{in,v1}
\fmf{phantom}{out,v2}
\fmfforce{(0,0h)}{in}
\fmfforce{(w,0h)}{out}
\fmffixed{(0.75w,0)}{v1,v2}
\fmfpoly{phantom}{v1,v2,v3}
\fmfpoly{phantom}{v3,v2,v4}
\fmfpoly{phantom}{v4,v2,v5}
\fmf{plain}{v1,v2}
\fmf{derplain}{v3,v1}
\fmf{plain}{v2,v3}
\fmf{plain}{v2,v4}
\fmf{plain}{v3,v4}
\fmf{plain}{v2,v5}
\fmf{derplain}{v4,v5}
\end{fmfchar*}}}
=G_1(2,1)G_1(4-\tfrac{D}{2},1)G_2(6-\tfrac{D}{2},1)
\col\qquad
\mathcal{I}_{32\mathbf{t}}=\frac{1}{(4\pi)^6}
\Big(-\frac{1}{3\varepsilon}\Big)
\col\qquad\quad\!\!\!\!
\end{aligned}
\end{gathered}
\end{equation}
where the integral $I_{2\mathbf{t}}$ that appears as substructure in 
$I_{3\mathbf{t}}$ and in the final expression for the two-loop chiral 
self-energy \eqref{ctwoloopse} is finite and given by
\begin{equation}
\begin{aligned}\label{I2t}
I_{2\mathbf{t}}=
\settoheight{\eqoff}{$\times$}%
\setlength{\eqoff}{0.5\eqoff}%
\addtolength{\eqoff}{-6.75\unitlength}%
\raisebox{\eqoff}{%
\fmfframe(-0.5,-5.5)(9,4){%
\begin{fmfchar*}(15,15)
  \fmfleft{in}
  \fmfright{out1}
\fmf{phantom}{in,v1}
\fmf{phantom}{out,v2}
\fmfforce{(0,0h)}{in}
\fmfforce{(w,0h)}{out}
\fmffixed{(0,0.75h)}{v2,v3}
\fmfpoly{phantom}{v1,v2,v3}
\fmfpoly{phantom}{v3,v2,v4}
\fmf{plain}{v1,v2}
\fmf{plain}{v1,v3}
\fmf{plain}{v2,v3}
\fmf{plain}{v3,v4}
\fmf{plain}{v2,v4}
\end{fmfchar*}}}
=\frac{2}{D-4}G(1,1)(G(1,2)+G(3-\tfrac{D}{2},2))
=\frac{1}{(4\pi)^4}6\zeta(3)+\mathcal{O}(\epsilon)\pnt
\end{aligned}
\end{equation}

\unitlength=0.875mm

\end{fmffile}

\footnotesize
\bibliographystyle{JHEP}
\bibliography{references}

\end{document}